\documentclass[aps,prd,reprint,nofootinbib]{revtex4-2}

\usepackage[utf8]{inputenc}

\usepackage{graphicx}
\usepackage{hyperref}
\usepackage{latexsym}
\usepackage{amsmath}
\usepackage{amssymb}
\usepackage{bbm}
\usepackage{ulem}
\usepackage{pdfsync}
\usepackage{epsfig}
\usepackage{epstopdf}
\usepackage{xcolor}
\usepackage{comment}
\usepackage{footnote}
\usepackage{slashed}
\usepackage{multirow}
\usepackage{xfrac}
\usepackage{adjustbox}
\usepackage{caption, subcaption} 
\usepackage{placeins}
\usepackage{ulem}
\usepackage{cancel}

\usepackage{color} 
\definecolor{orange}{rgb}{0.9,0.2,0}
\definecolor{brown}{rgb}{0.7,0.3,0.2} \definecolor{fuxia}{rgb}{1,0,1}
\definecolor{skyblue}{rgb}{0,0.1,0.9}
\definecolor{violetred}{rgb}{0.8,0.13,0.56}
\definecolor{deeppink}{rgb}{1.00,0.08,0.5}
\definecolor{pink}{rgb}{1.00,0.75,0.80}
\definecolor{orchid}{rgb}{0.85,0.44,0.84}
\definecolor{lightpink}{rgb}{1.00,0.71,0.76}
\definecolor{bluish}{rgb}{0,0.6,0.8}
\definecolor{lightgray}{rgb}{0.95,0.95,0.95}

\begin{document}

\title{Theoretical constraints on models with vector-like fermions}

\date{\today}

\author{Amit Adhikary} \email{Amit.Adhikary@cpt.univ-mrs.fr}
\affiliation{Aix Marseille Univ, Université de Toulon, CNRS, CPT,
  IPhU, Marseille, France}

\author{Marek Olechowski} \email{Marek.Olechowski@fuw.edu.pl}
\affiliation{Institute of Theoretical Physics, Faculty of Physics,
  University of Warsaw, Pasteura 5, PL 02-093, Warsaw, Poland}

\author{Janusz Rosiek} \email{Janusz.Rosiek@fuw.edu.pl}
\affiliation{Institute of Theoretical Physics, Faculty of Physics,
  University of Warsaw, Pasteura 5, PL 02-093, Warsaw, Poland}
  
\author{Micha{\l} Ryczkowski} \email{Michal.Ryczkowski@fuw.edu.pl}
\affiliation{Institute of Theoretical Physics, Faculty of Physics,
  University of Warsaw, Pasteura 5, PL 02-093, Warsaw, Poland}

\begin{abstract}
We provide a set of theoretical constraints on models in which the
Standard Model field content is extended by vector-like fermions and
in some cases also by a real scalar singlet. Our approach is based on
the study of electroweak vacuum stability, perturbativity of model
couplings and gauge couplings unification with the use of
renormalization group equations.  We show that careful analysis of
these issues leads to strong constraints on the parameter space of the
considered models. This, in turn, has important implications for
phenomenology, as we show using examples of
the double Higgs boson production, electroweak precision observables,
and the electroweak phase transition. 
\end{abstract}

\maketitle

\section{Introduction}
\label{sec:intro}

In spite of the enormous success of the Standard Model (SM) of the
elementary interactions, crowned with the Higgs boson discovery at the
LHC~\cite{CMS:2012qbp, ATLAS:2012yve}, there is a number of
theoretical and experimental issues which cannot be addressed by the
SM itself.  To name a few, SM does not provide a mechanism for
explaining the problem of scalar and fermion mass hierarchy, nature of
dark matter, flavor structure of the theory or the source of CP
violation and the related problem of observed baryon asymmetry in the
Universe.

Numerous extensions of the SM have been considered over the years to
tackle the problems mentioned above.  The wide class of such Beyond
the SM (BSM) models contains so called vector-like fermions (VLF),
i.e.~additional heavy fermion multiplets characterised by the unique
feature that the left-handed and right-handed components of these
states transform identically under the SM gauge group, distinguishing
them from the SM chiral fermions (see e.g.~Refs.~\cite{Cheng:1999bg,
  Arkani-Hamed:2002ikv, Han:2003gf, Cheng:2005as, Kang:2007ib,
  Cacciapaglia:2018qep, Cacciapaglia:2011fx, Aguilar-Saavedra:2013qpa,
  Ellis:2014dza, Angelescu:2015uiz, Arhrib:2016rlj, Barducci:2017xtw,
  Arhrib:2018pdi, Song:2019aav, delAguila:1989rq,
  Bhattacharya:2021ltd, CarcamoHernandez:2023wzf}).  Therefore, their
mass terms of the form $(\overline{\Psi}_L \Psi_R+\text{h.c.})$ are
gauge invariant and remain unbounded as they lack connection to gauge
symmetry breaking mechanism.

Models with vectorlike quarks (VLQ) or leptons (VLL), have been
studied in various contexts, in particular: stability of the Higgs
vacuum~\cite{Blum:2015rpa, Gopalakrishna:2018uxn, Arsenault:2022xty,
  Hiller:2022rla, Hiller:2023bdb}, possible enhancement in the Higgs
pair production rate~\cite{Cacciapaglia:2017gzh, Cheung:2020xij},
electroweak phase transition (EWPT) and
baryogenesis~\cite{Egana-Ugrinovic:2017jib, Bell:2019mbn,
  Davoudiasl:2012tu, Fairbairn:2013xaa, Angelescu:2018dkk,
  Chao:2014dpa, Cao:2021yau, Matsedonskyi:2020mlz, Baldes:2016rqn,
  Carena:2004ha}, observed anomaly in the measurement of the muon
anomalous magnetic moment~\cite{Poh:2017tfo, Crivellin:2018qmi,
  Athron:2021iuf, Muong-2:2021ojo, Hiller:2019mou, Hiller:2020fbu,
  Hiller:2019tvg}, the gauge coupling
unification~\cite{Dermisek:2012ke, Bhattacherjee:2017cxh,
  Emmanuel-Costa:2005qsv, Barger:2006fm, Dorsner:2014wva,
  Kowalska:2019qxm, Olivas:2021nft} or the flavor
physics~\cite{delAguila:2000rc, Bobeth:2016llm, Crivellin:2021bkd,
  Alves:2023ufm}, impact on the electroweak precision
observables~\cite{Lavoura:1992np,Kearney:2012zi,
  Arsenault:2022xty,Abouabid:2023mbu}, and dark matter
properties~\cite{Fan:2015sza, Belyaev:2022shr, Barman:2019tuo}.  Their
ability of providing the proper solution to these issues depends on
the details of a specific VLF model, its field content and assumed
constraints on their masses and couplings.  Allowed ranges of such
parameters can be derived both from the analysis of the current
experimental results and applying consistency conditions on the model
structure.

The current lower limit on the masses of vector-like fermions is given
by the direct searches for VLF from ATLAS \cite{ATLAS:2022ozf,
  ATLAS:2022hnn, ATLAS:2022tla, ATLAS:2023sbu, ATLAS:2023pja,
  ATLAS:2023bfh} and CMS \cite{CMS:2022yxp, CMS:2022tdo, CMS:2022cpe,
  CMS:2022fck, CMS:2023agg}.  Actual values of such masses assumed in
a given model cannot be too high and simultaneously their couplings
(especially VLF Yukawa interactions) cannot be too small in order to
have consequences significant enough to explain interesting
phenomenological effects.  However, they cannot be considered free
parameters constrained only by the experimental results -- as
mentioned above, one should consider also bounds following from
theoretical consistency of a given model.  In particular, it turns out
that strong limits are provided by the requirements of model
perturbativity and its vacuum stability, eventually further tightened
by additional conditions related to the gauge couplings unification.

In this paper, we reexamine such constraints in a more detailed way,
providing stronger than known before, independent theoretical
constraint on the parameter space of models featuring VLF multiplets.
As we show, such new bounds seriously limit the ability of simple VLF
models to have significant impact on phenomena which were often
considered in their context, such as double Higgs boson production,
electroweak precision tests or electroweak phase transition.

In order to investigate possibilities to weaken the constrains on
purely VLF scenarios, we also analyse models containing an extra
scalar singlet in addition to VLF fields.  The phenomenological
applications of the pure scalar singlet extension of the SM have also
been widely studied in the context of EWPT and collider
phenomenology~\cite{Curtin:2014jma, Profumo:2007wc, Noble:2007kk,
  Espinosa:2011ax, Espinosa:1993bs, Fernandez-Martinez:2022stj,
  No:2013wsa, Barger:2007im, Huang:2017jws, Profumo:2014opa,
  Chen:2017qcz}, EWPT and gravitational waves
spectra~\cite{Hashino:2016xoj, Ellis:2022lft} or dark
matter~\cite{Gonderinger:2009jp, Cline:2013gha, He:2009yd}.  We apply
our theoretical constraints also to the scalar singlet model itself,
finding its parameter space more limited than in the known literature.

In order to limit the number of free parameters in the considered
models while understanding their most important features, we derive
new bounds assuming certain simplified relations between the VLF
parameters, like uniform values of masses and Yukawa couplings for all
added VLF multiplet generations. Therefore, our analysis provides
insight into the most generic properties of VLF models, not depending
on details of the eventual flavor-like structure of new fermion
sector.  Relaxing assumptions on VLF parameters may to some extent
weaken the constraints which we are discussing, but at the cost of
increasing the complexity of the VLF models and/or fine tuning between
its parameters, thus making them less natural and less appealing in
explaining the desired phenomenological effects.

To illustrate the effects of the theoretical bounds on VLF and real
scalar couplings which we derive, we analyse their impact on two
problems often considered in the context of such models - double Higgs
boson production and EWPT.  We show that, after taking into account
the correlation between the single and double Higgs boson production
rates and the experimental constraints of the former, the enhancement
in the latter can be at most $\sim 15\%$, far below the current
experimental limits~\cite{ATLAS:2022jtk, CMS:2022dwd}, and likely also
below the sensitivity expected in future colliders.  Similarly, we
show that 1- or 2-step EWPT can occur only for a limited range of
scalar singlet parameters, with negligible impact of VLF interactions.
Therefore, more thorough inclusion of the consistency conditions in
VLF and real scalar models suggests that the phenomenological
implications of such models can be more limited than it was previously
thought.

Still, this does not mean that all VLF models can be rejected just on
the basis of consistency conditions we are exploiting -- in the
literature there are examples of theories with more complicated BSM
sectors, containing in addition to VLF multiplets also (pseudo)scalar
and/or vector particles (e.g.~\cite{Cheung:2020xij, Bobeth:2016llm, Hieu:2020hti,
  Fan:2015sza, Belyaev:2022shr, Barman:2019tuo}), or with more
complicated structure of VLF parameters (e.g.~flavor
non-diagonal and interacting with the SM fermions~\cite{delAguila:2000rc, 
Bobeth:2016llm, Crivellin:2021bkd, Alves:2023ufm}).
These conditions should be then
applied in a case by case manner to determine allowed parameter space
for a given scenario.
  
Breaking of EW vacuum stability in VLF models may also indicate
possibility that they themselves are an effective SM extension, and
the full ultraviolet completion of the SM, valid to very high scales,
requires existence of even more heavy particles. In such scenario
breaking of the conditions used in this work (see
\eqref{eq:condition1}--\eqref{eq:condition3}) does not mean that a
given VLF model should be automatically rejected, but that it needs to
be modified with unknown (before such modification is fully specified)
effects on the phenomenology.

The paper is organised as follows.  In Section~\ref{sec:model} we
present the model of interest along with relevant notation.
Section~\ref{sec:constraints} introduces a set of conditions utilised
for constraining the parameter space of considered model scenarios.
In Section~\ref{sec:parspace} we present our main results, outlining
the constraints on parameter space for models featuring VLF alone,
real scalar only, and the combined models with VLF and real scalar.
In Section~\ref{sec:pheno}, we consider the phenomenological
consequences of our approach, exploring the impact of the derived
constraints on double Higgs production, electroweak precision
observables and electroweak phase transition.  Finally, we conclude
and outline future directions in Section~\ref{sec:concl}. Expressions
for 1-loop RGE equations, oblique electroweak $\mathbb{S}, \mathbb{T}$
parameters and for effective scalar potential at finite temperature
are collected in Appendices.

\begin{table}[htb!]
\begin{center}
\vskip 3mm
\begin{tabular}{|c|c|c|c|c|}
\hline
$\psi$ & $S U(2)_L$& $Y_{W}$ & $T_3$ & $Q_{E M}$ \\
\hline
$Q^d_{L, R}=\left(\begin{array}{c}U^d_{L, R} \\ D^d_{L,
    R}\end{array}\right)$ & $2$& $+1 / 6$ & $\begin{array}{c}+1/2
  \\ -1/2\end{array}$ & $\begin{array}{c} +2/3 \\ -1/3 \end{array}$ \\
\hline
$U^s_{L,  R}$& $1$&$+2/3$ & $0$ & $+2/3$ \\
\hline
$D^s_{L,  R}$& $1$&$-1/3$& $0$ & $-1/3$ \\
\hline
$L^d_{L, R}=\left(\begin{array}{c}N^d_{L, R} \\ E^d_{L,
    R}\end{array}\right)$ & $2$ &$-1/2$ & $\begin{array}{c}+1/2
  \\ -1/2\end{array}$ & $\begin{array}{c}0 \\ -1\end{array}$ \\
\hline
$N^s_{L,  R}$& $1$&$0$ & $0$ & $0$ \\
\hline
$E^s_{L,  R}$& $1$&$-1$ & $0$ & $-1$ \\
\hline
\end{tabular}
\end{center}
\caption{VLF multiplets extending the SM field content.  The
  superscripts $d$ and $s$ denote $SU(2)$ doublets and singlets,
  respectively.}
\label{tab:Charges}
\end{table}

\section{Models with vector-like fermions}
\label{sec:model}

In this paper we consider a class of models where the SM particle
content is extended by adding multiplets of heavy vector-like fermions
and/or scalar field.  Thus, we consider Lagrangian which contains, in
addition to the SM fields, vector-like quarks and/or leptons in
doublet and singlet representations of the $SU(2)$ weak isospin gauge
group (see Table~\ref{tab:Charges}) and optionally also a new gauge
singlet scalar field.

We assume that the only interactions of new vector-like fermions with
the SM particles occur through their Yukawa couplings with the SM
Higgs doublet.  As we will discuss in Section~\ref{sec:IV:C}, though
mixed Yukawa couplings between SM and new vector-like fermions can be
written down (see
Eq.~\eqref{eq:LSMSVLF:Coupled:1}~\eqref{eq:LSMSVLF:Coupled:2}), these
terms would further have a negative influence on EW vacuum stability
and stability of perturbative expansion, further shrinking allowed
parameter space of models studied in this work. We therefore neglect
such terms in our analysis. The SM scalar potential reads:
\begin{equation}
V_{SM}(\Phi)=-\mu^2 \Phi^\dagger\Phi + \lambda (\Phi^\dagger\Phi
)^2\,,
\label{eq:V:SM}
\end{equation}
with $\Phi$ being an $SU(2)$ complex doublet with weak hypercharge
$Y_W=1/2$ of the following form:
\begin{equation}
\Phi=
\begin{pmatrix}
    G^+ \\
    \frac{1}{\sqrt{2}}\left(v+H+iG^0\right)  \\
\end{pmatrix}\,,
\label{eq:Higgs:doublet}
\end{equation} 
where $H$ is the physical Higgs field, $G^0$ and $G^+$ are Goldstone
fields and $v$ is the vacuum expectation value (VEV).

New terms in the Lagrangian involving VLF, corresponding to their
Dirac masses and Yukawa interactions with the SM Higgs field, have the
following form:
\begin{widetext}
\begin{equation}
\begin{aligned}
\mathcal{L}\supset-\sum_{i,j=1}^{n_{Q}} &M_{Q^d}^{ij} \bar{Q}^d_i
Q^d_j - \sum_{i,j=1}^{n_{U}}\biggl(M_{U^s}^{ij}\bar{U}^s_iU^s_j +
(y_U^{ij} \bar{Q}^d_i\Tilde{\Phi} U^s_j + \text{h.c.})\biggr) -
\sum_{i,j=1}^{n_{D}}\biggl(M_{D^s}^{ij}\bar{D}^s_iD^s_j +(y_D^{ij}
\bar{Q}^d_i\Phi D^s_j + \text{h.c.})\biggr)\\
-\sum_{i,j=1}^{n_L} &M_{L^d}^{ij} \bar{L}^d_i L^d_j -
\sum_{i,j=1}^{n_N}\biggl(M_{N^s}^{ij}\bar{N}^s_iN^s_j + (y_N^{ij}
\bar{L}^d_i\Tilde{\Phi} N^s_j + \text{h.c.}) \biggr) -
\sum_{i,j=1}^{n_E}\biggl(M_{E^s}^{ij}\bar{E}^s_iE^s_j + (y_E^{ij}
\bar{L}^d_i\Phi E^s_j+ \text{h.c.})\biggr),
\label{eq:L:VLF:INT:1}
\end{aligned}
\end{equation}
\end{widetext}
where $\Tilde{\Phi}$ is defined as:
\begin{equation}
\Tilde{\Phi}=i\tau_2\Phi^*\,,
\end{equation} 
Eq.~\eqref{eq:L:VLF:INT:1} describes the whole class of models - upper
summation limits denote the number of VLF doublets and singlets added
to the SM Lagrangian, e.g.~$n_Q=1$, $n_U=1$ and $n_D=0$ indicates a
model with one vector-like quark $SU(2)$ doublet, one ``up-type'' VLQ
$SU(2)$ singlet and zero ``down-type'' VLQ $SU(2)$ singlets (note that
$\sum_{1}^{0}\equiv 0$).

As mentioned in the Introduction, in order to understand the most
common features of VLF models, in our analysis we limit the discussion
to the simplified case with minimal number of free parameters,
assuming for the mass and coupling matrices in~\eqref{eq:L:VLF:INT:1}
to be VLF-flavor diagonal and identical for each vector-like
generation.  Then the VLF mass matrices in the $(F^d, F^s)$
interaction basis, where $F^d \in \{U^d, D^d, N^d, E^d\}$ and $F^s \in
\{U^s, D^s, N^s, E^s\}$, after the spontaneous gauge symmetry
breaking, have the following form:
\begin{equation}
\widetilde{\mathbf{M}}_F=
\begin{pmatrix}
 M_{F^d}   & \frac{1}{\sqrt{2}} \, v \, y_F\\
 \frac{1}{\sqrt{2}} \, v \, y_F   & M_{F^s}\\
\end{pmatrix}.
\end{equation}
The eigenvalues of this mass matrix are equal
\begin{equation}
\begin{aligned}
M_{F_1}&= \frac{1}{2} \left(M_{F^d} + M_{F^s} +
\sqrt{(M_{F^d}-M_{F^s})^2 + 2 v^2 y_F^2}\right), \\
M_{F_2}&= \frac{1}{2} \left(M_{F^d} + M_{F^s} -
\sqrt{(M_{F^d}-M_{F^s})^2 + 2 v^2 y_F^2}\right),
\label{eq:Mass:VLQ}
\end{aligned}
\end{equation}
and the mixing angle $\gamma_F$ relating the interaction and mass
bases is given by the condition $\tan{2\gamma_{F}}={\sqrt{2}\, v\,
  y_F}/\left({M_{F^d}-M_{F^s}}\right)$.

In some cases (models of classes B and C defined below) we will add to
the SM one real scalar singlet $S$ with an unbroken $\mathbb{Z}_2$
symmetry\footnote{This $\mathbb{Z}_2$ symmetry is assumed to simplify
some formulae but it is not crucial for the present work because it
forbids terms, linear and cubic in $S$, which anyway do not alter
renormalization group $\beta$ functions used to obtain the main
results of our analysis.}  under which $S\rightarrow -S$.  The full
scalar potential in such cases reads:
\begin{equation}
V(\Phi, S)=V_{SM}(\Phi) +\frac{1}{2} \mu_S^2 \,
S^2+\frac{1}{2}\lambda_{H S} \, \Phi^\dagger \Phi \, S^2+\frac{1}{4}
\lambda_S \, S^4\, ,
\label{eq:VSMS}
\end{equation}
which, for vanishing VEV of $S$, gives the following tree level masses
of the scalar particles:
\begin{equation}
\begin{aligned}
M_H^2&=2\lambda v^2 = 2 \mu^2,  \\
M_S^2&=\mu_S^2+\frac{1}{2}\lambda_{H S} v^2\,.
\end{aligned}
\end{equation}

Phenomenological consequences of a given model depend on type and
number of additional fields and on their masses and couplings.  In
order to illustrate the effects of extending SM by various types of
fields, in the following we consider three specific classes of models:
\begin{itemize}
\item Class A: models with SM extended by vector-like fermions only,
\item Class B: models with SM extended by the real scalar only,
\item Class C: models extended by both vector-like fermions and real
  scalar.
\end{itemize}
Within each class we also vary the number of added VLF multiplets of
each kind following the notation introduced in
Eq.~\eqref{eq:L:VLF:INT:1}.  As a benchmark scenarios, we consider
three representative scenarios of VLF models:
\begin{itemize}
\item Scenario I - $n$ VLF doublets, $2\times n$ VLF singlets ($n$ of
  ``up-type'' and $n$ of ``down-type''),
\item Scenario II - $n$ VLF doublets, $n$ ``up-type'' VLF singlets,
  $0$ ``down-type'' VLF singlets,
\item Scenario III - $n$ VLF doublets, $0$ ``up-type'' VLF singlets,
  $n$ ``down-type'' VLF singlets.
\end{itemize}
We will refer to $n$ as the number of VLF families (composition of one
family depends on the Scenario chosen from the list above).  For
simplicity we consider SM extensions with vector-like quarks and
vector-like leptons separately.
For example: Scenario I with VLQ and $n=1$ is obtained by choosing
$n_Q=n_U=n_D=1$, $n_L=n_E=n_N=0$; Scenario III with VLL and $n=2$
corresponds to $n_L=n_E=2$, $n_Q=n_U=n_D=n_N=0$ etc.

Finally, SM input parameters used in our analysis are set in
accordance with Ref.~\cite{Buttazzo:2013uya}.  Updated experimental
input parameters are taken as central values from
Ref.~\cite{ParticleDataGroup:2022pth}:
\begin{equation}
\begin{aligned}
\label{eq:rge_input}
&M_t= 172.83 \text{ GeV},\\ 
&g_{1}\left(M_{\mathrm{t}}\right) = \sqrt{5/3}\times g^\prime =
\sqrt{5/3}\times 0.358144, \\
&g_{2}\left(M_{\mathrm{t}}\right) = 0.64772, \quad
g_{3}\left(M_{\mathrm{t}}\right) = 1.1646, \\
&y_{\mathrm{t}}\left(M_{\mathrm{t}}\right) = 0.93436, \quad
\lambda\left(M_{\mathrm{t}}\right) = 0.12637.
\end{aligned}
\end{equation}
As we checked, the theoretical and experimental uncertainties of
the experimental input parameters do not significantly affect our
results and conclusions.

\section{Theoretical constraints}
\label{sec:constraints}

In our analysis we consider energy scale dependence of the model
parameters defined in the previous Section, dictated by the
renormalization group equations (RGEs) including 1- and 2-loop
contributions.  Such evolution turns out to impose important
theoretical constraints on the parameter space.

We assume that model scenarios introduced in the previous Section are
valid up to a given cut-off energy scale $\Lambda$.  We treat them as
effective models which at some scale above $\Lambda$ may be embedded
in a more fundamental theory.  So, $\Lambda$ is the upper energy limit
up to which we demand the relevant constraints to be satisfied.  In
particular, we take into account bounds based on the following
conditions.

\begin{enumerate}
\item Stability of the vacuum.  We demand that the scalar potential is
  bounded from below.  In models of class A this corresponds to the
  condition that the Higgs self-coupling $\lambda$ is positive up to
  the cut-off scale:
\begin{equation}
\lambda(\mu)>0 \quad \text{for} \quad \mu\le\Lambda\,.
\label{eq:condition1}
\end{equation}
In models with the singlet scalar (classes B and C) stability of the
scalar potential \eqref{eq:VSMS} requires two additional conditions:
$\lambda_{HS}(\mu)>-2\sqrt{\lambda(\mu)\lambda_S(\mu)}$ and
$\lambda_{S}(\mu)>0$ for $\mu\le\Lambda$. Both are always fulfilled in
models considered in this paper. Both couplings, $\lambda_S$ and
$\lambda_{HS}$, are positive at all relevant scales if they are
positive at low scale $\mu=M_t$ because the leading contributions
\eqref{eq:RGE:S:1} to their $\beta$ functions are positive.
\item Perturbativity of the model couplings up to the scale $\Lambda$:
\begin{equation}
\kappa_i(\mu)\leq 4\pi \quad \text{for} \quad \mu\le\Lambda\,,
\label{eq:condition2}
\end{equation}
where $\kappa_i=(\lambda, y_t^2, g_1^2, g_2^2, g_3^2, y_F^2,
\lambda_{HS}, \lambda_S)$.
\item
Stability of the perturbative expansion:
\begin{equation}	
\min_{[\mu,{\mu\times 10^\delta}]}\left|
\frac{\beta^{(2)}_{\kappa_i}(\mu)}{\beta^{(1)}_{\kappa_i}(\mu)}
\right|
\leq \Delta \quad \text{for} \quad \mu\le\Lambda\,.
\label{eq:condition3}
\end{equation}
where (1) and (2) superscripts indicate, respectively, 1-loop and
2-loop contribution to a $\beta$-function for a given coupling
evolution.  We consider the ratio of 2- and 1-loop contributions
minimised over some range of $\mu$ scale, in order to avoid its
spurious large values around points where a given 1-loop term
vanishes.  $\Delta$ is a maximal allowed value of such regularised
ratio. For our numerical analysis we choose $\delta = 1$ and $\Delta =
0.4$.  We checked that the constraints on the model parameters
following from the condition \eqref{eq:condition3} depend weakly on
the precise values of the $\delta$ and $\Delta$ -- e.g. using higher
value of $\Delta=0.6$ leads to increase of maximal allowed values of
couplings by $\mathcal{O}(20\%)$.
\end{enumerate}

When considering phenomenological applications of the conditions
listed above, we also take into account additional information related
to the issues under discussion.  This includes available experimental
limits on the triple Higgs boson interaction, relevant for the
prediction of double Higgs production at the LHC, and existing bounds
on masses of vector-like fermions and extra real scalar field, and
electroweak precision observables.  All conditions combined lead to
bounds on the maximal allowed number of VLF multiplets and values of
their couplings.

For our numerical analysis we use the 2-loop RGE evolution.  The full
2-loop RGEs (obtained with SARAH~\cite{Staub:2013tta} and
cross-checked with RGBeta~\cite{Thomsen:2021ncy}) are lengthy and we
do not display them in the paper, for reference and easier analytical
discussion of various effects we collected the 1-loop RGEs in
Appendix~\ref{app:rge}.  We include contributions from new particles
to the RGE $\beta$-functions only for (renormalization) energy scale
$\mu$ above their respective mass scales, i.e. $\mu\geq M_F$ for
vector-like fermions and $\mu\geq M_S$ for the scalar singlet.

\section{Allowed parameters space of VLF models}
\label{sec:parspace}

We study the impact of
conditions~\eqref{eq:condition1}--\eqref{eq:condition3} defined in
Section~\ref{sec:constraints} on the parameters of the VLF sector
i.e.~on the number of VLF multiplets, their masses and couplings
(Eq.~\eqref{eq:L:VLF:INT:1} and Table~\ref{tab:Charges}).  As a
result, we obtain maximal allowed values of VLF Yukawa couplings for
which those conditions are satisfied up to a given cut-off energy
scale $\Lambda$.  As our analysis will show, in models were SM is
extended by the vector-like fermions only, such couplings are limited
to relatively small values.  In order to alleviate these stringent
constraints and enlarge the allowed parameter space, we later consider
also models where, in addition to extra VLF multiplets, a new real
scalar singlet field is present.

\begin{figure*}[htb!]
 \centering
\begin{tabular}{ccc}
\includegraphics[width=0.3\linewidth]{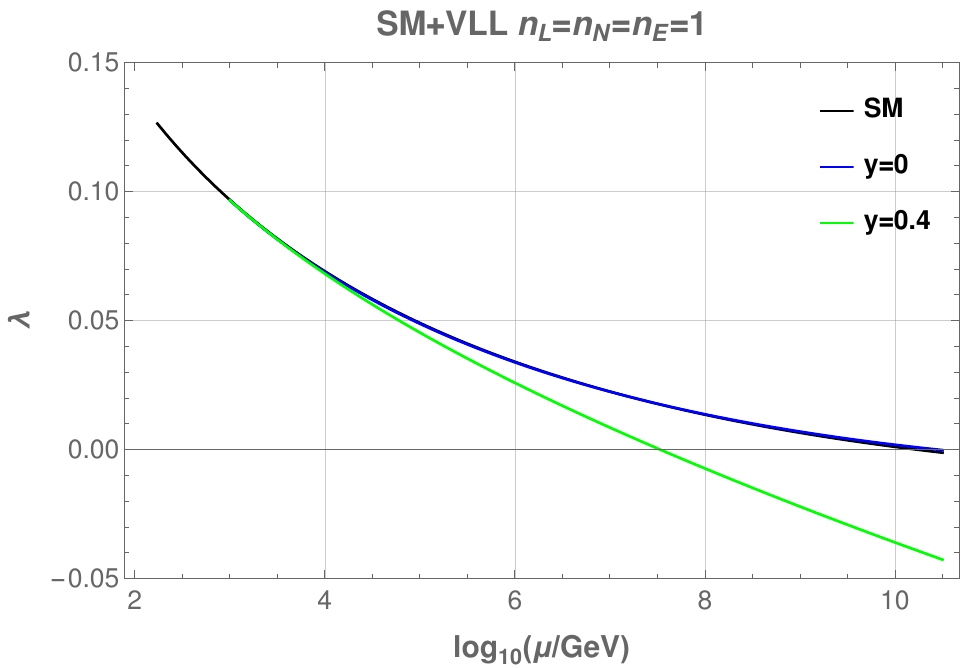}
&
\includegraphics[width=0.3\linewidth]{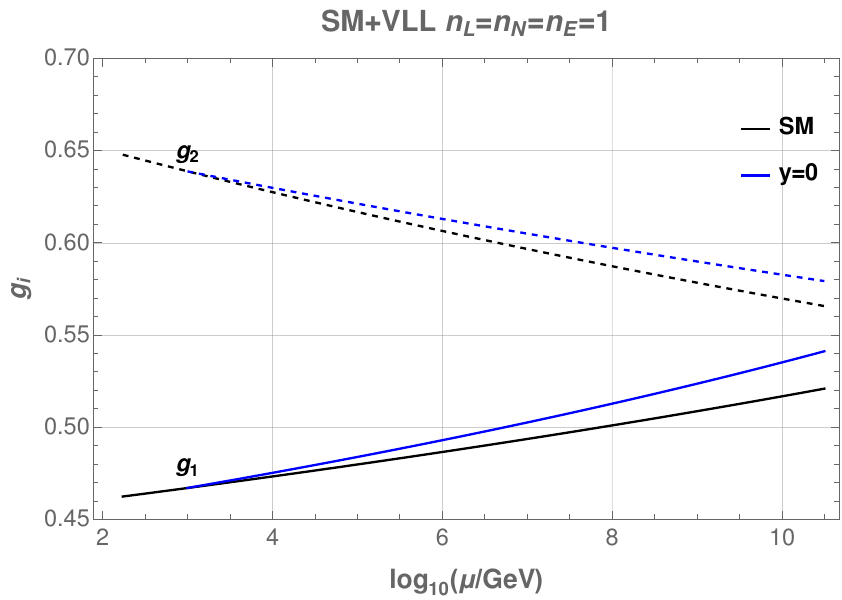}
&
\includegraphics[width=0.3\linewidth]{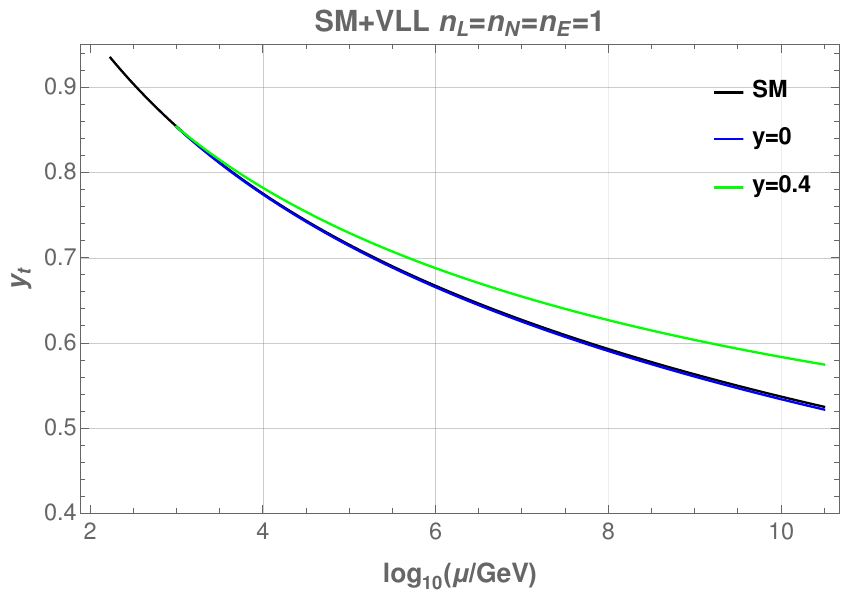}
\\
\includegraphics[width=0.3\linewidth]{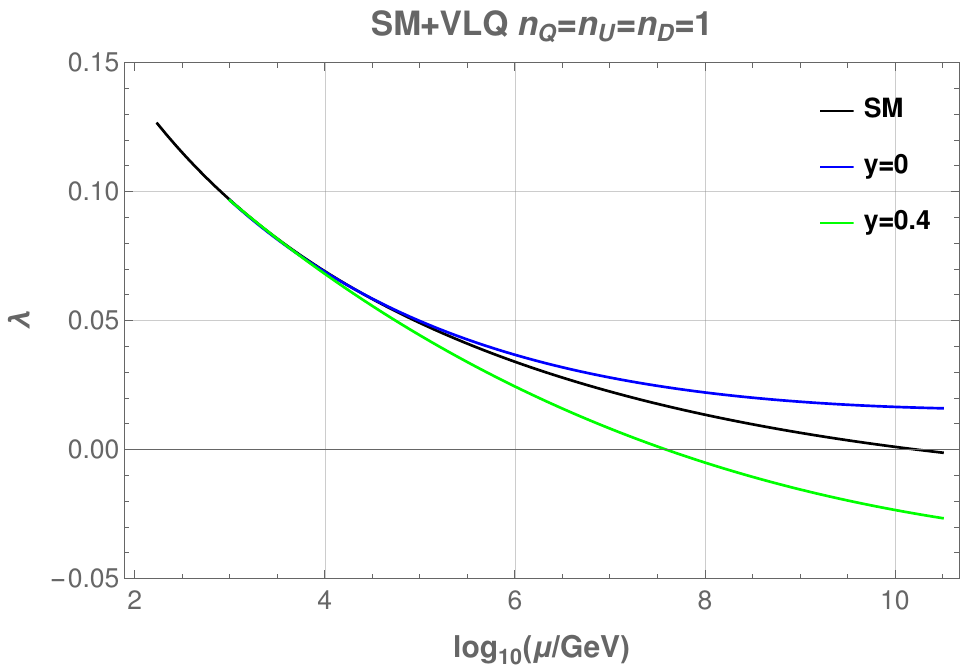}
&
\includegraphics[width=0.3\linewidth]{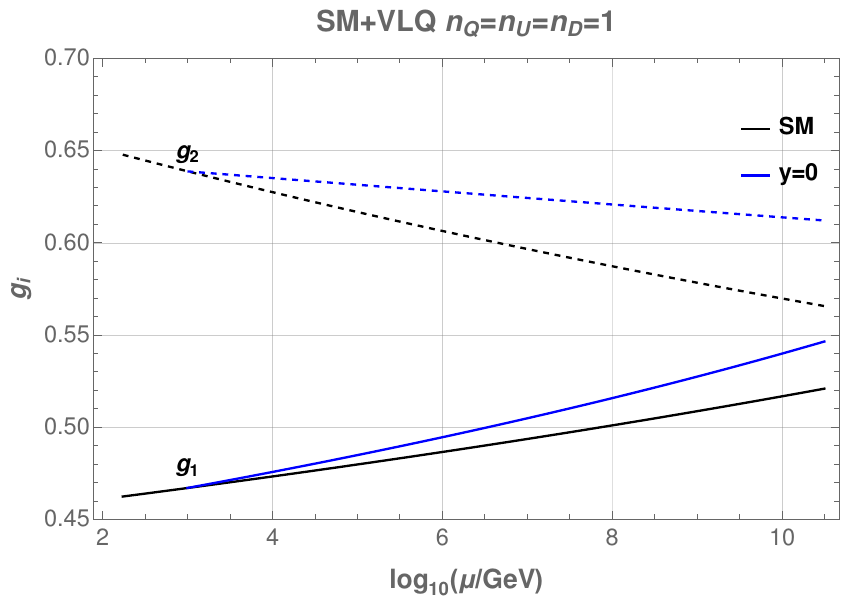}
&
\includegraphics[width=0.3\linewidth]{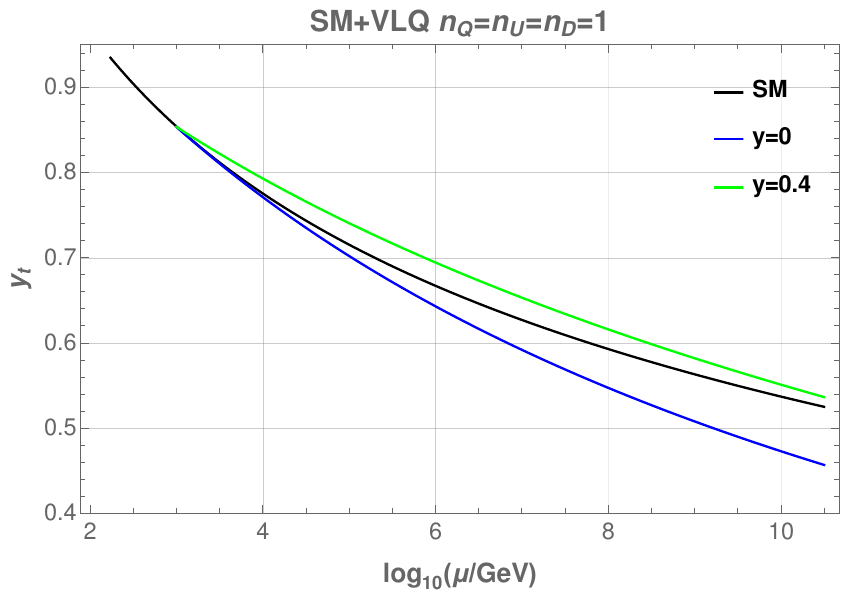}
\end{tabular}
\caption{Running of $\lambda$ (left column), $g_1$ and $g_2$
  (continuous and dashed lines respectively, middle column) and $y_t$
  (right column) at 2-loop in the SM extended by VLF assuming $M_F=1$
  TeV and varying VLF Yukawa coupling $y$.}
\label{fig:lambda:VLF:1}
\end{figure*}

In order to limit the number of free parameters, in our numerical
analysis we make further simplifying assumptions, imposing that all
Dirac masses of singlet and doublet VLF are the identical,
$M_{F^d}=M_{F^s}=M_{F}$.  Furthermore, we assume a single universal
value of all VLF Yukawa couplings: $y_F=y$ for all new fermion
multiplets.
For the VLF masses, relying on the results of recent direct searches
from ATLAS \cite{ATLAS:2022ozf, ATLAS:2022hnn, ATLAS:2022tla,
  ATLAS:2023sbu, ATLAS:2023pja, ATLAS:2023bfh} and CMS
\cite{CMS:2022yxp, CMS:2022tdo, CMS:2022cpe, CMS:2022fck,
  CMS:2023agg}, we adopt the lower limit $M_F\gtrsim 1$ TeV.  Too
heavy VLF would lead to their effective decoupling and negligible
modifications of the SM particles interactions (e.g.~the triple Higgs
coupling) and of observable phenomenology.  Moreover, the cut-off
scale $\Lambda$ must be larger than the heaviest particle in a given
model, otherwise there would be no energy scale at which such model
could be consistently defined.  Thus, we consider masses of new
particles, VLF and/or the scalar $S$, to be of the order
$\mathcal{O}(1 \div 10)$~TeV.  Pushing $\Lambda$ to very large values
significantly shrinks allowed model parameter space.  For this reasons
we consider $\Lambda=100 \div 1000$~TeV as typical interesting range
of the cut-off scale which we will use in most of our further
considerations.

\subsection{SM extended with vector-like fermions only -- case A}
\label{sec:SMVLF}

We illustrate the impact of VLF on the running of the Higgs quartic
coupling $\lambda$ and EW vacuum stability given by
\eqref{eq:condition1} assuming $n=1$ (as we checked, behaviour of
models with larger numbers of VLF families is qualitatively very
similar, see also discussion in Section~\ref{sec:IV:gauge}).
  
Plots in Fig.~\ref{fig:lambda:VLF:1} show how varying VLF Yukawa
couplings influences the running of $\lambda$, the gauge couplings
$g_1$ and $g_2$ and the top Yukawa coupling $y_t$.  Two main regimes
can be distinguished depending on the magnitude of $y$:

\smallskip

\noindent {\bf 1.  VLF with small (or vanishing) Yukawa couplings.}

\smallskip

They have positive impact on the stability of the EW vacuum, as
compared to the SM, eventually even leading to the stability up to the
Planck scale.  This can be understood at the 1-loop level as follows.
For $y=0$, VLF contributions affect directly the running of $g_1$ and
$g_2$ gauge couplings (see Eq.~\eqref{eq:RGE:VLF:1}), which in turn
leads to bigger values of $\beta_\lambda$ (see
Eq.~\eqref{eq:RGE:SM:1}).  Adding VLQ multiplets leads to larger
effects than adding VLL ones due to larger number of degrees of
freedom, affecting RGE running through a bigger colour factor
$N_c^\prime$ in Eq.~\eqref{eq:RGE:VLF:1}. This effect in the case of
$g_1$ is more model-dependent due to differences in hypercharges of
quarks and leptons.

\smallskip

\noindent {\bf 2. VLF with larger Yukawa couplings. }

\smallskip

Increasing the VLF Yukawa couplings typically has negative impact on
the stability of the EW vacuum.  For these couplings bigger than some
critical value the EW vacuum becomes unstable.  This follows again
from the 1-loop RGEs: the leading direct VLF contribution to the
running of $\lambda$ is proportional to $ (2\lambda-y_F^2) y_F^2 $
(see the first equation in~\eqref{eq:RGE:VLF:1}) which is positive for
small values of $y_F^2$ but becomes negative for $y_F^2>2\lambda$.
The negative contribution grows with increasing $y_F^2$ and very soon
overcomes the indirect positive contribution mentioned in point 1
above.  This effect is further strengthened by the positive
contribution of VLF to $\beta_{y_t}$ which indirectly leads to bigger
negative term in $\beta_\lambda$ proportional to $y_t^4$ (see the
first equation in \eqref{eq:RGE:SM:1}).

\begin{figure}[tb!]
\centering
\includegraphics[width=1\linewidth]{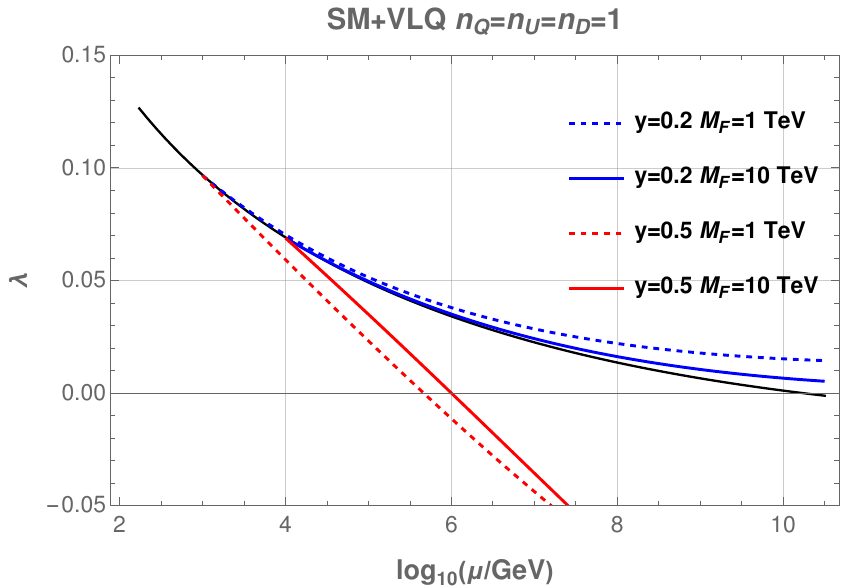}
\caption{Running of $\lambda$ coupling for chosen values of $M_F$ and
  $y$.  Black line indicates $\lambda$ behaviour in the SM.}
\label{fig:lambda:VLF:M}
\end{figure}

The maximal value of the Yukawa coupling consistent with the stability
condition \eqref{eq:condition1} depends not only on the cut-off scale
$\Lambda$ but also on the masses of VLF.  Any effects from VLF are
stronger when their masses are smaller because lighter particles
modify RGEs starting from lower energy scales. This applies to both
types of effects: those which improve and those which worsen stability
of the EW vacuum.  In Fig.~\ref{fig:lambda:VLF:M} we illustrate this
feature for two values of $y$ in scenario I with VLQ.

\begin{figure*}[htb!]
\centering
\begin{tabular}{cc}
\includegraphics[width=0.45\linewidth]{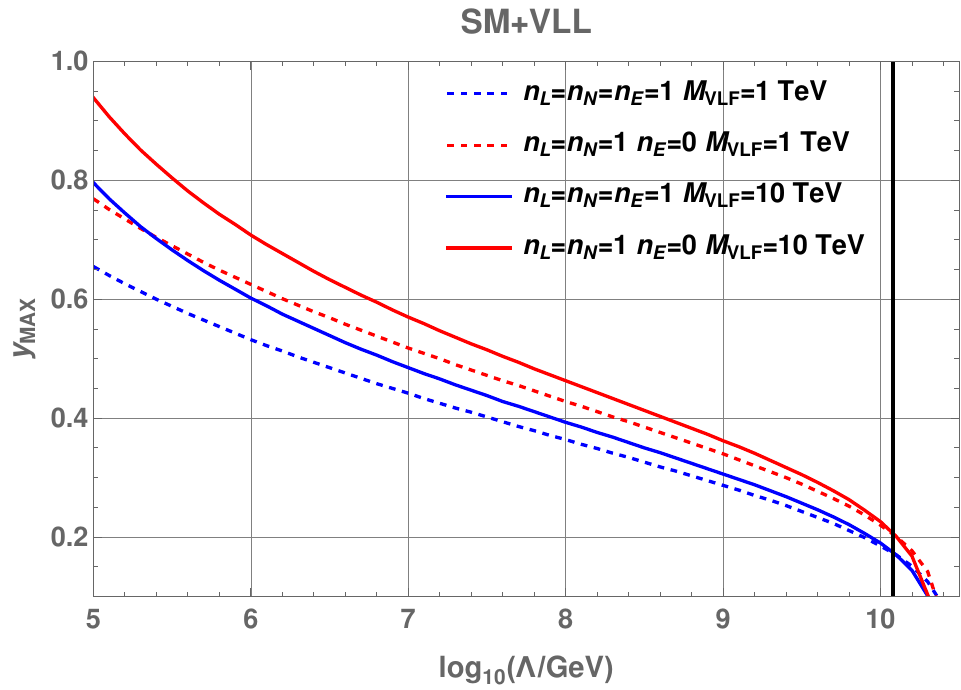} & 
\includegraphics[width=0.45\linewidth]{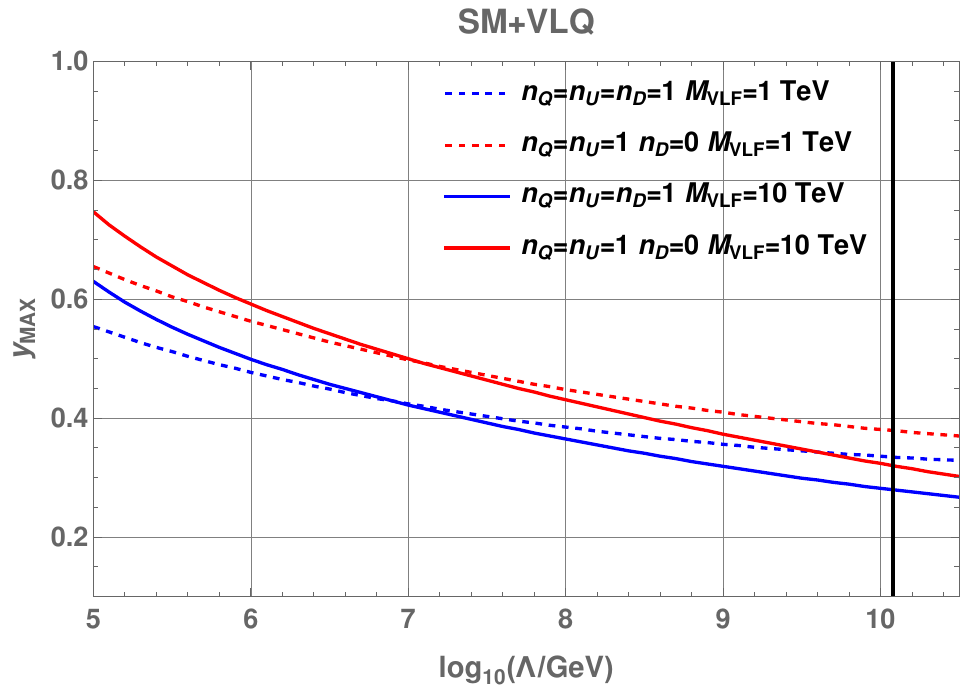}
\end{tabular}
\caption{Maximal values of VLF Yukawa coupling for which stability
  condition is satisfied up to the given scale $\mu$.  $M_F=1,\space
  10$ TeV.  The vertical lines indicate the scale of the SM stability
  breakdown.  }
\label{fig:Y:VLF:1}
\end{figure*}

As we argued above, VLF with Yukawa couplings larger than some
critical values have negative impact on the EW vacuum stability.  The
bigger are such couplings the lower is the energy scale at which
$\lambda$ becomes negative.  Thus, demanding $\lambda$ to be positive
up to some chosen cut-off scale $\Lambda$ results in an upper limit on
the Yukawa coupling, $y_{MAX}$, in a given model\footnote{We have
assumed here that all VLF's Yukawa couplings are equal. In more
general models the corresponding upper bound applies to some
combination of different Yukawa couplings.}
(we remind that by couplings without explicit scale-dependence we
denote these couplings renormalized at the scale $\mu=M_t$).

Negative contributions to $\beta_\lambda$ from the VLF increase with
the corresponding Yukawa couplings. Thus, the maximal value of the
Yukawa coupling in a given model, $y_{MAX}$, is a decreasing functions
of the cut-off scale $\Lambda$. Such dependence for scales above the
$10^{9\div 10}$~GeV (close to the scale at which $\lambda$ becomes
negative in the SM) is very different for VLL and VLQ models.  In the
case of VLL models $y_{MAX}$ drops very quickly to zero, while in VLQ
models $y_{MAX}$ decreases very slowly.  The reason for this
difference follows from the impact of VLF with small Yukawa couplings
on the RGE evolution of $\lambda$ discussed in the point 1~above (and
illustrated in the upper row of plots in
Fig.~\ref{fig:lambda:VLF:1}). Namely, VLL may improve the EW vacuum
stability only sightly and only for very small Yukawa couplings, while
addition of VLQ with not too large Yukawa couplings may stabilise the
EW vacuum even up to the Planck scale.

A few examples of $y_{MAX}(\Lambda)$ for VLL and VLQ models are shown
in Fig.~\ref{fig:Y:VLF:1}.  The biggest possible value of $y_{MAX}$,
corresponding to the lowest cut-off scale considered in this paper,
$\Lambda=100$~TeV, in scenarios I and II for VLL and VLQ models and
for two values of $M_F$, are collected in Table~\ref{tab:SM+VLF}.

\begin{table}[tb!]
\centering
\begin{tabular}{|c|c|c|c|}
\hline
& \multirow{3}{*}{Scenario} & \multicolumn{2}{|c|}{$y_{MAX}$} \\
\cline{3-4}
& & $M_F=1$ TeV &$M_F=10$ TeV \\
\hline 
\multirow{2}{*}{VLQ} & $n_Q=n_U=n_D=1$	& 0.55	& 0.63	 \\ 
\cline{2-4}
& $n_Q=n_U=1$ $n_D=0$ & 0.66 & 0.75  \\
\hline
\multirow{2}{*}{VLL} &$n_L=n_E=n_N=1$ & 0.66 & 0.80  \\
\cline{2-4}
& $n_L=n_N=1$ $n_N=0$& 0.77 & 0.94  \\
\hline
\end{tabular}
\caption{Maximal values of VLF Yukawa couplings allowed by the
  conditions \eqref{eq:condition1}-\eqref{eq:condition3} up to the
  scale $\Lambda=100$~TeV for two values of $M_F$.}
\label{tab:SM+VLF}
\end{table}

Results presented in this Section show that in all considered
scenarios the maximal values of VLF Yukawa couplings allowed by the EW
stability condition are always relatively small, $y_{MAX}<1$, which,
as we shall see later on, suggests limited range of phenomenological
consequences of such BSM extensions which could be experimentally
tested in near future.  Many studies considering phenomenology of VLF
models (e.g.~\cite{ Egana-Ugrinovic:2017jib, Fairbairn:2013xaa,
  Angelescu:2018dkk, Chao:2014dpa, Cao:2021yau, Matsedonskyi:2020mlz,
  Baldes:2016rqn, Carena:2004ha}) often require large values of VLF
Yukawa couplings to produce significant observable effects. The
present analysis shows that obtaining those effects may be not
possible in perturbative models with stable EW vacuum.

The discussed above impact of VLF on the EW vacuum stability is in
line with similar study~\cite{Gopalakrishna:2018uxn}, and we were able
to reproduce their results.  However, in our work we studied a wider
set of VLF model scenarios, and focused more on their possible
observable effects on phenomenology (which we discuss in
Section~\ref{sec:pheno}).

Our analysis obviously does not exhaust all possible VLF scenarios,
but shows that the simple EW vacuum stability requirement strongly
constraints allowed parameter space of such models.  Testing VLF
models with more complex pattern of different couplings would require
case by case study, which is beyond the scope of this work.

\subsection{SM extended with real scalar singlet -- case B}
\label{sec:IV:B}

In order to explore the freedom of construction of VLF models and
enlarge their allowed parameter space, we consider also SM extension
containing also the real scalar singlet field, $S$, with a tree-level
potential given in Eq.~\eqref{eq:VSMS}.  We start with reviewing the
theoretical constraints on SM extended by the real scalar singlet
only.

The main effect of adding $S$ comes from the coupling between such
scalar and the SM Higgs scalar, $\lambda_{HS}$, which gives a positive
contribution to $\beta_\lambda$ (see the first equation in
Eq.~\eqref{eq:RGE:S:1}). The singlet self-coupling, $\lambda_S$, also
(indirectly) leads to bigger values of $\lambda$ via its positive
contribution to $\beta_{\lambda_{HS}}$ (second equation in
Eq.~\eqref{eq:RGE:S:1}). Thus, increasing scalar couplings,
$\lambda_{HS}$ and $\lambda_S$, and/or decreasing $M_S$ result in
larger values of $\lambda$ and so has positive impact on the stability
of the EW vacuum.  For example one gets its absolute stability up to
the Planck scale for $M_S=1$~TeV, $\lambda_S=0$ and
$\lambda_{HS}\gtrsim 0.3$.

\begin{figure}[hbt!]
  \centering
\includegraphics[width=1\linewidth]{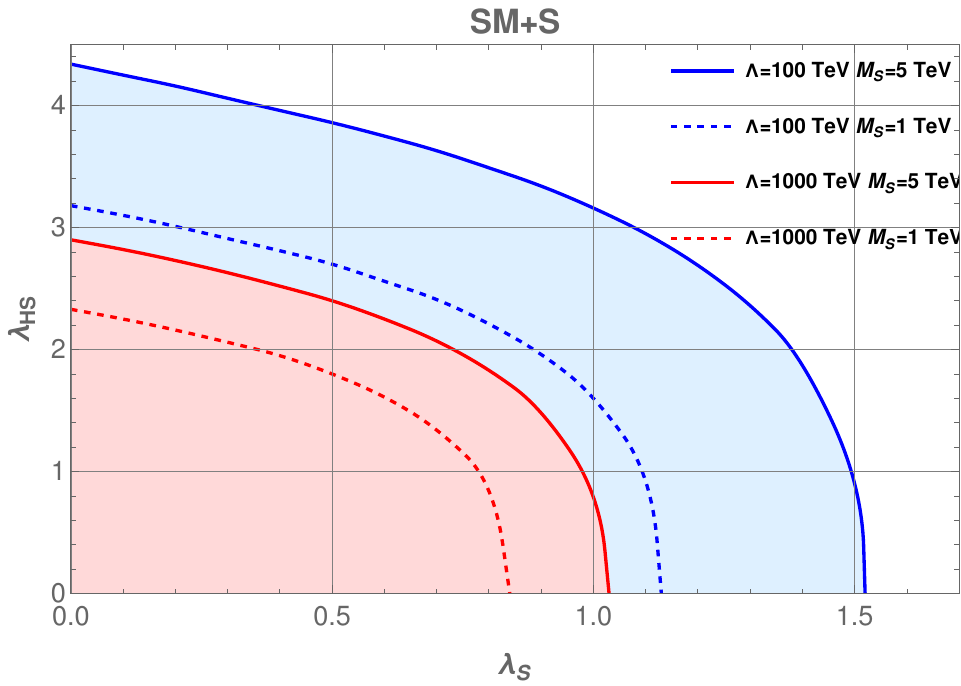}
\caption{Parameter space of the SM extended by real singlet allowed by
  the perturbativity constraints up to $\Lambda=100, \space 1000$ TeV
  with varying $M_S=1, \space 5$ TeV.}
\label{fig:lambdaHS:SMS}
\end{figure}

However, increasing the values of $\lambda_S$ and $\lambda_{HS}$ too
much may lead to the loss of perturbativity, i.e.~condition(s)
\eqref{eq:condition2} and/or \eqref{eq:condition3} may be violated
below the chosen cut-off scale $\Lambda$.  Fig.~\ref{fig:lambdaHS:SMS}
shows regions on the $\lambda_S$-$\lambda_{HS}$ plane compatible with
conditions \eqref{eq:condition2}--\eqref{eq:condition3} for chosen
values of $\Lambda$ and the singlet mass $M_S$.  As expected,
increasing value of $M_S$ weakens the upper bound on $\lambda_{HS}$
while increasing $\lambda_S$ or $\Lambda$ makes this bound stronger.

Our approach provides new and independent way of studying allowed
values of the parameters in the real scalar model and can be seen as
complementary to the previous papers (e.g.~\cite{Curtin:2014jma,
  Profumo:2007wc, Noble:2007kk}) by further constraining allowed
parameter space of this model.  It emphasises the importance of going
beyond the naive study of perturbativity as given by the
condition~\eqref{eq:condition2} by taking into account the differences
between 1- and 2-loop contributions to the RGEs .

\subsection{SM extended with vector-like fermions and real scalar
  singlet -- case C}
\label{sec:IV:C}

Since, as discussed in the previous subsection, adding the scalar
singlet to the model improves stability of the Higgs potential, it
should also help to alleviate constraints on VLF Yukawa couplings
summarised in Table~\ref{tab:SM+VLF}.  However, relaxing the
constraints on $y_{MAX}$ in the presence of extra singlet scalar field
should also take into account potential violation of perturbativity
conditions \eqref{eq:condition2} and \eqref{eq:condition3} when
increasing the values of new scalar couplings.

\begin{figure*}[hbt!]
  \centering
\begin{tabular}{cc}
\includegraphics[width=0.45\linewidth]{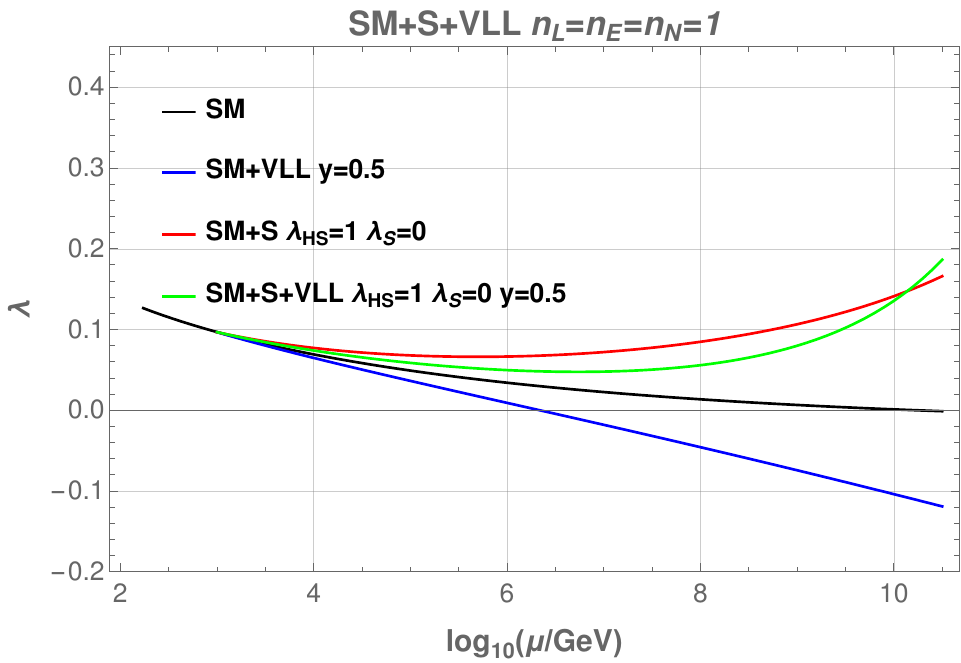}
&
\includegraphics[width=0.45\linewidth]{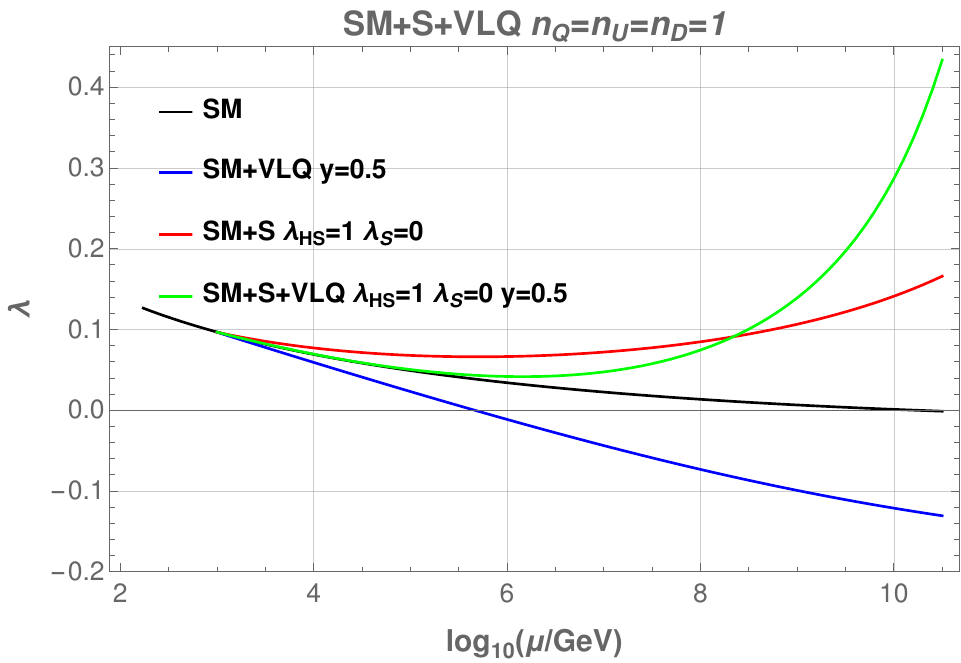}
\end{tabular}
\caption{Running of $\lambda$ in a combined model SM+S+VLF for $M_S=1$
  TeV and $M_F=1$ TeV.}
\label{fig:lambda:SMSVLF:1}
\end{figure*}

In a model with the SM extended by the singlet scalar only (SM+S), all
the conditions \eqref{eq:condition1}-\eqref{eq:condition3} are
fulfilled in the SM+S model for values of $\lambda_{S}$ and
$\lambda_{HS}$ which are not too large in order to avoid problems with
perturbativity.  As it turns out, adding VLF with Yukawa couplings big
enough to have meaningful phenomenological effects further strengthens
the constraints from the conditions
\eqref{eq:condition1}--\eqref{eq:condition3}, as VLF have both
negative impact on the vacuum stability (see Section~\ref{sec:SMVLF})
and lead to stronger problems with perturbativity. The reason of the
latter effect is obvious from Eq.~\eqref{eq:RGE:VLF:S:1}: additional
fermions give positive contribution to $\beta_{\lambda_{HS}}$
proportional to the sum of squares of their Yukawa coupling.

\begin{figure*}[ht!]
\centering
\begin{tabular}{cc}
\includegraphics[width=0.45\linewidth]{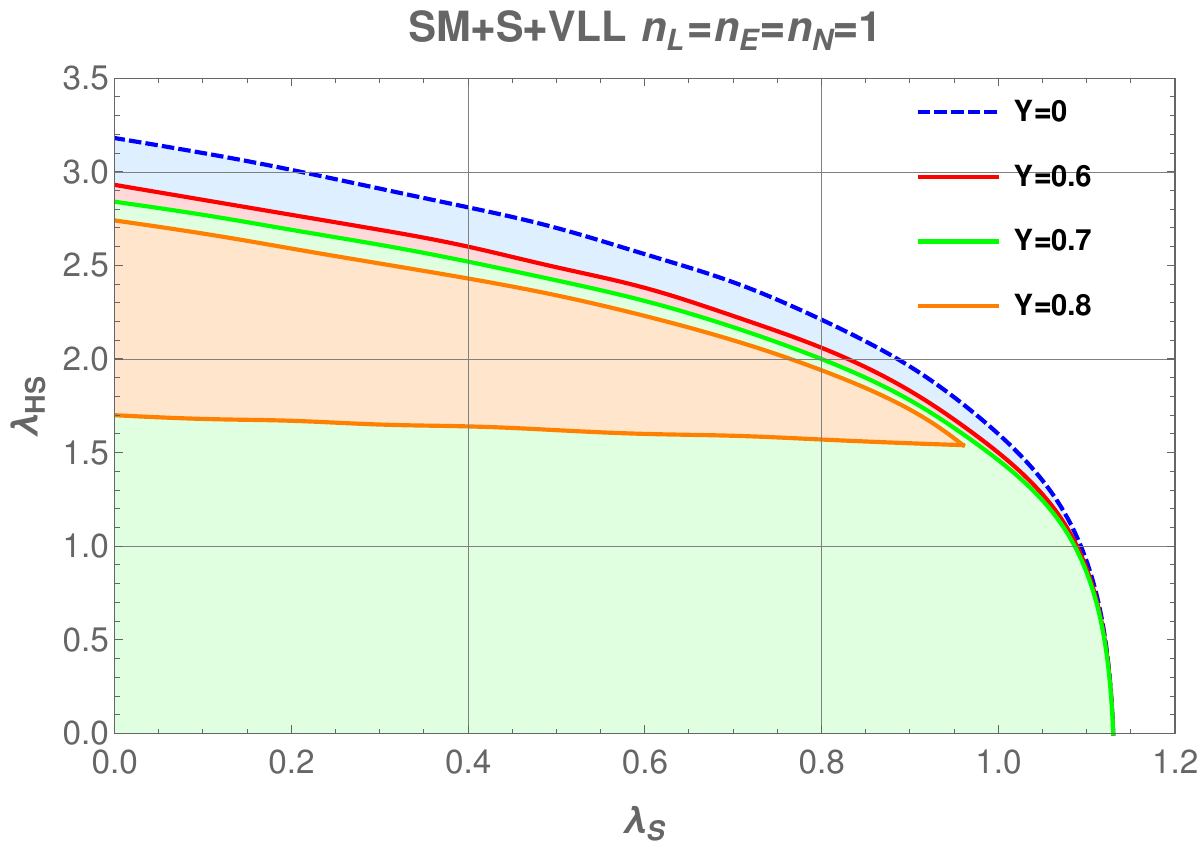}
&
\includegraphics[width=0.45\linewidth]{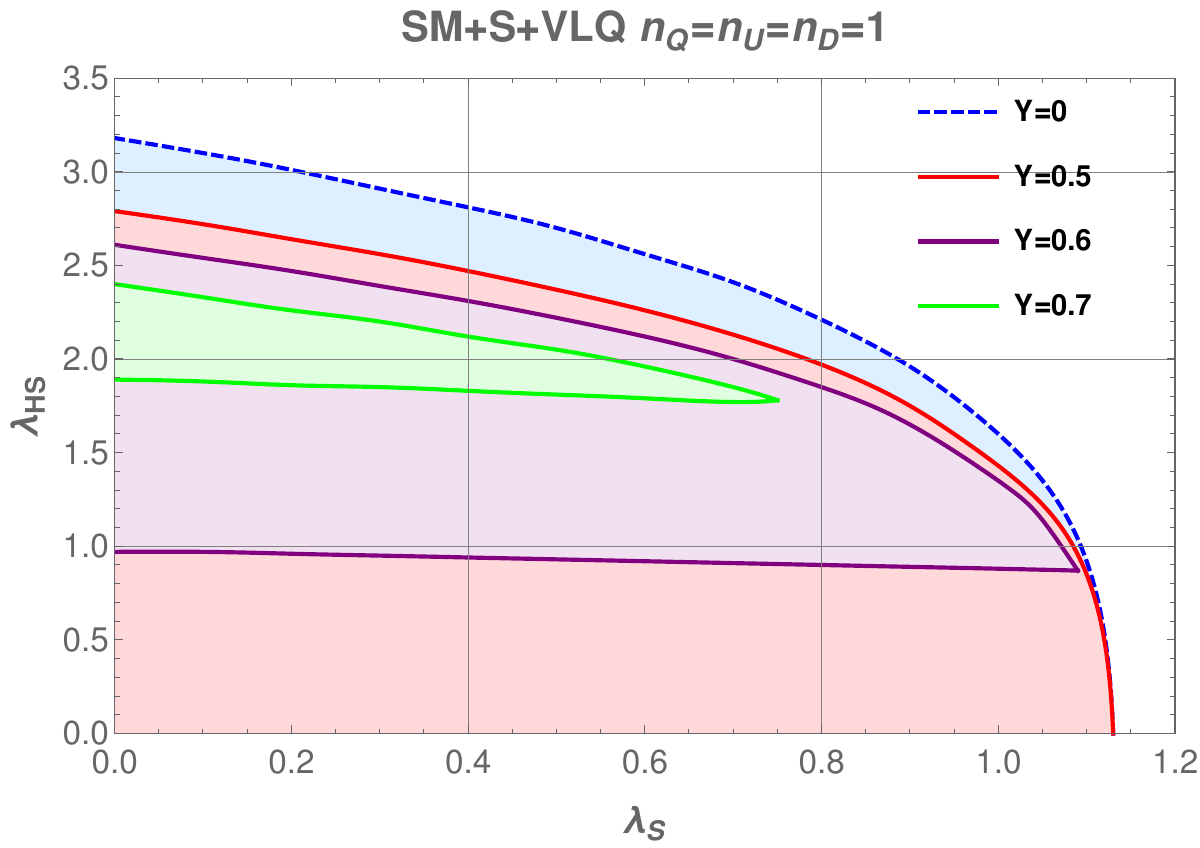}
\end{tabular}
\caption{Allowed regions of the singlet scalar couplings in the SM
  extended by real singlet and VLF for the cut-off scale
  $\Lambda=100$~TeV with $M_F=M_S=1$ TeV and for different values of
  the VLF Yukawa coupling $Y$.  The dashed blue contour is the same as
  in Fig.~\ref{fig:lambdaHS:SMS}.}
\label{fig:paramspace:SMSVLF}
\end{figure*}

Impact of VLF on the RGE running of $\lambda$ is illustrated with
examples in Fig.~\ref{fig:lambda:SMSVLF:1}. The Higgs self-coupling
$\lambda$ in the models with $S$ and VLF (green curves) is lower at
low scales but bigger at high scales as compared to model without VLF
(red curves). Both effects result in stronger constraints on singlet
scalar couplings $\lambda_S$ and especially $\lambda_{HS}$. Stability
condition \eqref{eq:condition1} leads to lower bound on $\lambda_{HS}$
as function of $\lambda_S$, while perturbativity conditions
\eqref{eq:condition2} and \eqref{eq:condition3} lead to corresponding
upper bound.  The impact of VLF on the allowed region for
$\lambda_{HS}$ and $\lambda_S$ couplings is shown, depending on value
of their Yukawa couplings, in Fig.~\ref{fig:paramspace:SMSVLF} (for
comparison, see corresponding plot without VLF shown in
Fig.~\ref{fig:lambdaHS:SMS}) and in Table~\ref{tab:SMSVLF}

The allowed region in the $\lambda_{S}$-$\lambda_{HS}$ parameter space
decreases with increasing VLF Yukawa couplings, eventually shrinking
to the point defining its maximal allowed value $y_{MAX}$.  Adding
singlet $S$ to the theory may allow for maximal Yukawa couplings
bigger by at most up to about 50\% comparing to pure VLF models, as
can be seen by comparing Tables \ref{tab:SM+VLF} and
\ref{tab:SMSVLF}.\footnote{We accept points in the parameter space
which don't satisfy condition~\eqref{eq:condition3} for
$\beta_\lambda$ with $\delta=1$, when it's clear that it's the
consequence of $\beta_\lambda^{(1)}(\mu)\approx 0$ and all other
conditions are satisfied.}

\begin{table*}[ht!]
\centering
\begin{tabular}{|c|c|c|c|c|c|c|}
\hline
&  \multicolumn{2}{|c|}{SM+VLF} &  \multicolumn{4}{|c|}{SM+S+VLF}\\
\cline{2-7}
\multirow{3}{*}{Scenario} & \multicolumn{2}{|c|}{$y_{MAX}$} &
$\lambda_{HS}^{y_{MAX}}$ & $y_{MAX}$ &
$\lambda_{HS}^{y_{MAX}}$ & $y_{MAX}$ \\
\cline{2-7}
& $M_F=1$ TeV &$M_F=10$ TeV & \multicolumn{2}{c|}{$M_F=1$ TeV}
&\multicolumn{2}{c|}{$M_F=10$ TeV}\\
\hline 
$n_Q=n_U=n_D=1$	& 0.55	& 0.63 &2.30 & 0.74 & 2.46 & 0.87 \\ 
\hline
$n_Q=n_U=1$ $n_D=0$ & 0.65 & 0.74& 2.55 & 0.91 & 2.66 & 1.07 \\
\hline
$n_L=n_E=n_N=1$ & 0.65 & 0.79 & 2.59 & 0.93 &2.71& 1.16 \\
\hline
 $n_L=n_N=1$ $n_N=0$& 0.76 & 0.92 & 2.77 & 1.11 &  2.85 & 1.40 \\
\hline
\end{tabular}
\caption{Maximal values of VLF Yukawa couplings $y_{MAX}$ allowed by
  the conditions \eqref{eq:condition1}-\eqref{eq:condition3}, up to
  the scale $\Lambda=100$ TeV with $\Delta=0.4$ and the corresponding
  values of $\lambda_{HS}^{y_{MAX}}$ for which they can be achieved in
  the combined VLF + scalar model for $M_S=1$ TeV, $\lambda_S=0$ and
  two chosen values of $M_F$.}
\label{tab:SMSVLF}
\end{table*}

\begin{figure*}[ht!]
  \centering
\begin{tabular}{cc}
\includegraphics[width=0.45\linewidth]{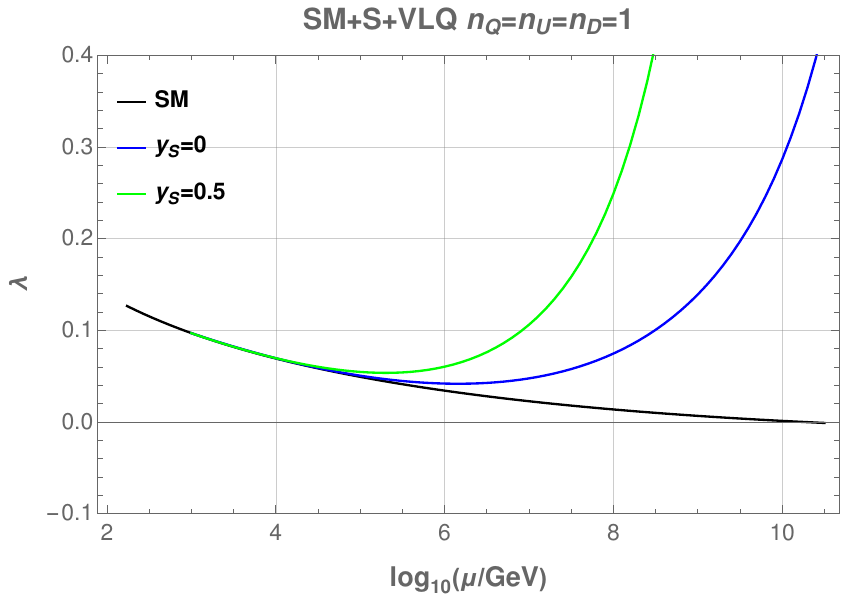}
&
\includegraphics[width=0.45\linewidth]{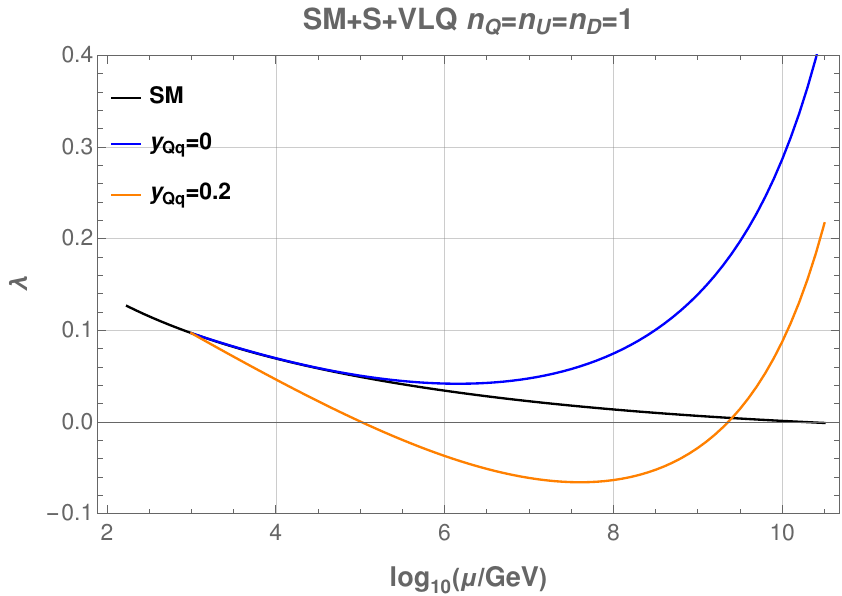}
\end{tabular}
\caption{Running of $\lambda$ in a combined SM+S+VLF model for $M_S=1$
  TeV and $M_F=1$ TeV with $y_F=0.5$, $\lambda_{HS}=1$ and
  $\lambda_S=0$, including VLF-$S$ Yukawa (left) and VLF-SMF (right)
  coupling.}
\label{fig:lambda:SMSVLF:2}
\end{figure*}

\subsection{Models with additional VLF couplings}
\label{sec:IV:extra}

So far we assumed that there are no tree-level interactions between
VLF and the singlet scalar $S$ or the SM fermions. As we discuss
below, adding such couplings in general results in further shrinking
of the allowed parameter space.

\smallskip

{\bf 1. VLF and real scalar couplings.}

\smallskip

The simplest way of including the VLF-$S$ couplings can be realised by
adding the following terms to the Lagrangian:
\begin{equation}
\begin{aligned}
\label{eq:LSMSVLF:Coupled:1}
&- \sum_{i=1}^{n_Q} y_{QS} S\bar{Q}^d_i Q^d_i -
\sum_{j=1}^{n_U} y_{US} S \bar{U}^s_j U^s_j - \sum_{k=1}^{n_D} y_{DS}
S \bar{D}^s_k D^s_k \\
& - \sum_{i=1}^{n_L} y_{LS} S\bar{L}^d_i L^d_i - \sum_{j=1}^{n_N}
y_{NS}S \bar{N}^s_j N^s_j - \sum_{k=1}^{n_E} y_{ES} S \bar{E}^s_k
E^s_k\,.
\end{aligned}
\end{equation}
The presence of non-vanishing VLF-$S$ Yukawa couplings leads to even
stronger bounds from perturbativity condition
\eqref{eq:condition2}. The reason is as follows:
$\beta_{\lambda_{HS}}$ has a positive contributions proportional to
squares of Yukawa couplings present in
Eq.~\eqref{eq:LSMSVLF:Coupled:1}, increasing $\lambda_{HS}$ during
evolution. This indirectly, via term proportional to $\lambda_{HS}^2$
in $\beta_\lambda$~(see eq.~\eqref{eq:RGE:S:1}), leads to a faster
growth of $\lambda$.  Such effect is illustrated in the left plot of
Fig.~\ref{fig:lambda:SMSVLF:2} where for simplicity, we assumed all
new Yukawa couplings to be equal - $y_{XS}\equiv y_S$,
$X=Q,U,D,L,N,E$.  Fig.~\ref{fig:lambda:SMSVLF:2} shows how
non-vanishing $y_S$ pushes $\lambda$ to larger values, implying more
severe problems with perturbativity than in the case of $y_S=0$ and in
consequence leading to smaller allowed maximal values of VLF-Higgs
Yukawa coupling.

\smallskip

\noindent {2. \bf VLF and SM fermion couplings.}

\smallskip

Similarly, we can write down Lagrangian terms corresponding to the
interactions between VLF and SM fermions.  Assuming for simplicity
that the only non-vanishing couplings are between the third family of
SM quarks ($q_L$, $t_R$ and $b_R$) and VLQ in the $n_Q=n_U=n_D=1$
scenario we get the following extra terms in the Lagrangian:
\begin{equation}
\begin{aligned}
\label{eq:LSMSVLF:Coupled:2}
&- y_{Qt} \left(\bar{Q}^d_L \Tilde{\Phi} t_R +
\text{h.c.}\right) - y_{Uq}\left(\bar{q}_L \Tilde{\Phi} U^s_R +
\text{h.c.}\right)
\\ 
&- y_{Qb} \left(\bar{Q}^d_L \Phi b_R +
\text{h.c.}\right) - y_{Dq}\left(\bar{q}_L \Phi D^s_R +
\text{h.c.}\right)\, .
\end{aligned}
\end{equation}
Non-vanishing VLQ-SM quarks Yukawa couplings have similar impact of
the allowed parameter space as ordinary VLF Yukawa couplings discussed
earlier in this Section. They give additional negative contributions
to $\beta_{\lambda}$ resulting in problems with vacuum stability more
severe than in the SM or when VLF and SM fermions are not
interacting. Example of effects of non-vanishing VLQ-SM quarks Yukawa
couplings for a simple case where $y_{Qt}=y_{Uq}= y_{Qb} = y_{Dq} =
y_{Qq}$ is presented in the right panel of
Fig.~\ref{fig:lambda:SMSVLF:2}. One can see that even relatively small
values of these additional Yukawa couplings may destabilise the EW
vacuum.

\subsection{Gauge couplings unification in models with VLF}
\label{sec:IV:gauge}

Fulfilling the vacuum stability and perturbativity conditions
\eqref{eq:condition1}--\eqref{eq:condition3} is obligatory for the
theoretical consistency of any BSM model. However, one can also impose
other, more optional, requirements on the considered model field
content and coupling structure, in order to achieve additional
advantages in its predictions. One such possible extra condition may
concern gauge coupling unification.

Let us define scales $\mu_{ij}$ at which pairs of gauge couplings,
$g_i$ and $g_j$, have equal values, i.e.~when $\Delta g_{ij}\equiv
g_i(\mu_{ij})-g_j(\mu_{ij})=0$. In the SM these three scales are quite
different with $\mu_{23}$ bigger than $\mu_{12}$ by more than 3 orders
of magnitude, thus there is no common point of unification of gauge
couplings.

The idea of grand unification favours models for which all three
scales $\mu_{ij}$ are close to each other. In addition, higher
unification scales, at least of order $10^{15}$~GeV, are preferred in
order to avoid too fast proton decay.  It is interesting to check
whether unification of the gauge couplings may be achieved in models
with extra VLF multiplets modifying their RGE evolution (we do not
consider models with scalar singlet here).

\begin{figure}[t]
\centering
\includegraphics[width=0.95\linewidth]{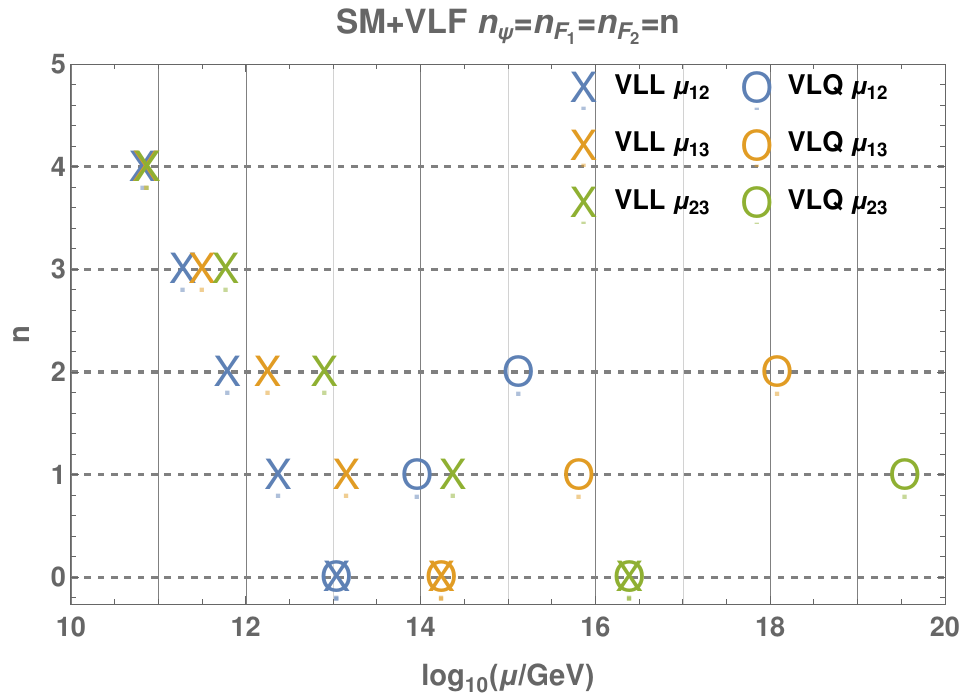} 
\includegraphics[width=0.95\linewidth]{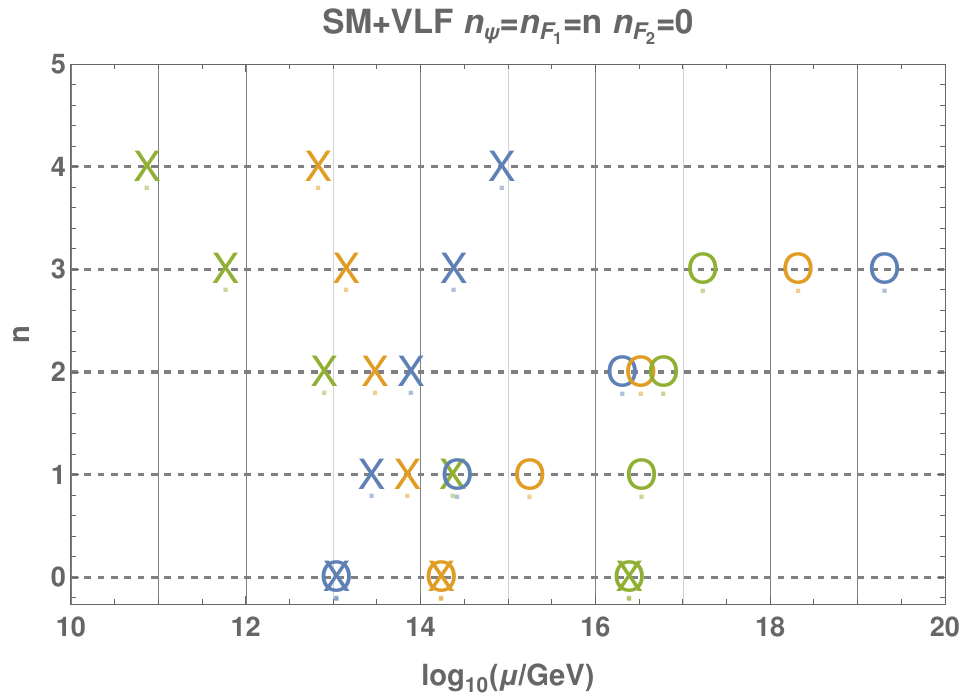} 
\includegraphics[width=0.95\linewidth]{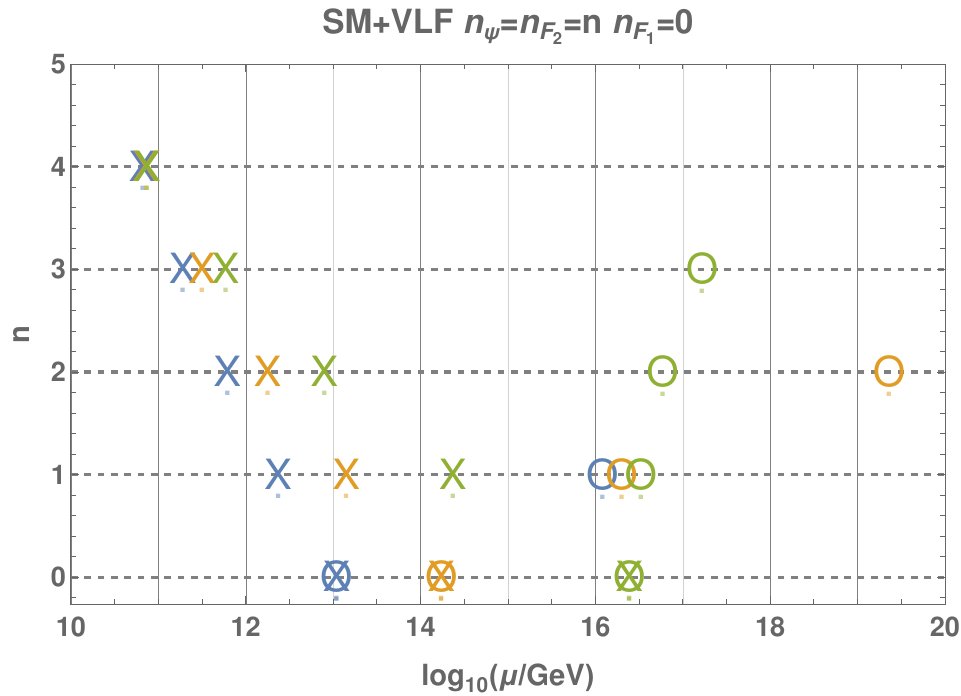}

\caption{Energy scales $\mu_{ij}$ at which pairs of running gauge
  couplings have equal values in models with $n$ families of VLL
  (crosses) of VLQ (circles) with $M_F=3$~TeV. The 2-loop RGEs were
  used. $n_{\psi}\in\{n_Q, n_L\}$, $n_{F_1}\in\{n_U, n_N\}$ and
  $n_{F_2}\in\{n_D, n_E\}$.}
\label{fig:gauge:VLF:1}
\end{figure}

In Fig.~\ref{fig:gauge:VLF:1} we show how the energy scales at which
the gauge couplings may unify depend on the number of VLF multiplet
families for $M_F=3$~TeV (we point out that VLF Yukawa $y$ has no
impact on the running of gauge couplings at 1-loop). Closer look at
the plots on those figures reveals the following (compare
Section~\ref{sec:model} for model classification):

\begin{itemize}
\item SM + VLL models (crosses in Fig.~\ref{fig:gauge:VLF:1}):
  \begin{itemize}
  \item scenario I (top panel) and scenario III (bottom panel): gauge
    couplings convergence points become closer to each other with
    increasing $n$ but the intersection points move to lower energy
    scales;
  \item scenario II (middle panel): gauge couplings converge closer to
    each other for $n=1, \space 2$ but already for $n=3$ they start
    diverging;
  \end{itemize}
\item SM + VLQ models (circles in Fig.~\ref{fig:gauge:VLF:1}):
  \begin{itemize}
  \item scenario I (top panel): the $\Delta g_{23}=0$ point is above
    the Planck scale already for $n=1$;
  \item scenario II (middle panel): gauge couplings converge closer to
    each other for $n=1, \space 2$ but already for $n=3$ they start
    diverging, with all intersection points moving towards higher
    energy scales;
  \item scenario III (bottom panel): for $n=1$ couplings converge
    closer to each other than in the SM, but already for $n=2$ the
    $\Delta g_{12}=0$ point is above the Planck scale.
  \end{itemize}
\end{itemize}
As we checked, varying VLF mass $M_F$ between 1~TeV and 10~TeV does
not make significant difference, general picture and conclusions
remain the same.

Based on the discussion above, one can see that VLQ scenario II with
$n=2$ (to a lesser extent also $n=1, 3$) and scenario III with $n=1$
are the most promising ones for the idea of grand unification of gauge
couplings in VLF models.  All other cases are disfavoured due to
different reasons.  All VLL scenarios I, II and III with not too large
$n$ have a positive impact on unification of couplings when compared
to the SM.  However, the corresponding unification scales are too low.
The VLQ scenario I has a negative impact on unification and already
for $n=2$, $g_2$ and $g_3$ couplings meet above the Planck scale.  The
situation is similar for the VLQ scenario III with $n\geq 2$.

\begin{table}[h!]
\centering
\begin{tabular}{|c|c|c|}
\hline
 & SM+VLL & SM+VLQ \\
\hline
Scenario I & $n_L^{max}=4$ & $n_Q^{max}=2$ \\
Scenario II & $n_L^{max}=12$ & $n_Q^{max}=3$ \\
Scenario III & $n_L^{max}=4$ & $n_Q^{max}=3$ \\
\hline
\end{tabular}
\caption{Maximal number of allowed VLF multiplet families based on the
  perturbativity conditions.}
\label{tab:numbers:1}
\end{table}

Models with large cut-off scales and big number of VLF multiplets are
disfavoured not only from the point of view of possible unification of
gauge couplings but also by perturbativity conditions
\eqref{eq:condition2} and \eqref{eq:condition3}.  New fermions give
positive contributions to the gauge coupling $\beta$-functions. Too
many fermions lead to too fast grow of (at least some of) gauge
couplings.  Demanding the perturbativity up to the Planck scale gives
upper bounds on the number of VLF multiplets presented in
Table~\ref{tab:numbers:1}.

An additional remark concerning models with VLL is in order here. As
we discussed in Sec.~\ref{sec:SMVLF}, EW vacuum in such models becomes
unstable at scales above about $10^{9 \div 10}$~GeV. Therefore, it is
reasonable to consider much bigger cut-off scales only if the singlet
scalar $S$ with suitable couplings is also added to the model,
improving its stability. The singlet does not change the RGEs for the
gauge couplings so the results presented in this Section are valid for
SM+S+VLL models.

\section{Phenomenological consequences}
\label{sec:pheno}

As discussed in Section~\ref{sec:parspace}, the interplay between
stability condition~\eqref{eq:condition1}, and perturbativity
conditions~\eqref{eq:condition2}, \eqref{eq:condition3} leaves
surprisingly small allowed parameter space for the VLF models.
Inclusion of a real scalar weakens somewhat the upper bounds on the
values of VLF Yukawa couplings but as we shall show, these couplings
are still not sufficient to justify expectations of rich
phenomenology.  In fact, the consequences which could be
experimentally tested in near future are quite limited.  We will focus
on three examples of possible phenomena, namely double Higgs boson
production, electroweak precision observables and electroweak phase
transition.

\subsection{Double Higgs boson production}
\label{sec:dihiggs}

\begin{figure}[tb!]
\centering
\includegraphics[width=0.35\linewidth]{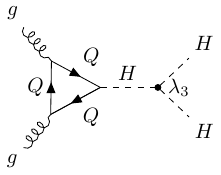}
\qquad
\includegraphics[width=0.32\linewidth]{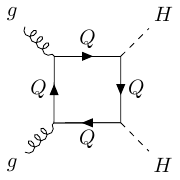}
\caption{Diagrams contributing to the double Higgs production via
  gluon fusion. Internal fermion lines indicated by $Q$ can be either
  SM quarks or VLQ multiplets.  Black dot indicates triple Higgs
  coupling $\lambda_3$.}
\label{fig:DiHiggs:diags}
\end{figure}

The enhancement of the double Higgs production cross section in models
with new VLF multiplets and heavy scalar $S$ can occur by modification
of the triple Higgs coupling or VLQ contributions to the box and
triangle gluon fusion diagrams of
Fig.~\ref{fig:DiHiggs:diags}\footnote{Results in this Section were
obtained using {\tt FeynRules}~\cite{Alloul:2013bka} and {\tt
  MadGraph5\_aMC@NLO}~\cite{Alwall:2011uj} packages.}.
We denote differences between cross sections for the gluon fusion
processes of single and double Higgs boson production in the SM and
its extensions as:
\begin{equation}
K_{\text{final}}^{X}= \frac{\sigma^{SM+X}_{\text{ggF}}(pp\rightarrow
  \text{final})-\sigma^{SM}_{\text{ggF}}(pp\rightarrow
  \text{final})}{\sigma^{SM}_{\text{ggF}}(pp\rightarrow
  \text{final})}\, ,
   \label{eq:KggFinal}
\end{equation}
where $X$ is either $S$ or VLF, and ``final'' state can be $H$ for
single and $HH$ for double Higgs production.  From the experimental
point of view we are interested in scenarios which predict significant
enhancement of $K^{X}_{HH}$ while keeping $K^{X}_{H}$ within present
experimental bounds.

\smallskip

\noindent {\bf 1. Modification of the triple Higgs coupling.}

\smallskip

Interactions of the Higgs scalar with new particles can modify the
triple Higgs coupling $\lambda_3$ which, if sizeable, may directly
impact double Higgs boson production.  This coupling is defined as:
\begin{equation}
\begin{aligned}
\lambda_3 \equiv & \left.\frac{d^3\left(V_0(H,0)+V_0^{C
    W}(H,0)\right)}{d H^3}\right|_{H=v} \\
\equiv & \lambda_3^{SM} + \Delta\lambda_3^{BSM} \equiv
\lambda_{3\text{,tree}}^{SM} +\Delta\lambda_3^{SM} +
\Delta\lambda_3^{BSM}\, ,
\label{eq:3H}
\end{aligned}
\end{equation}
with tree level potential given by Eq.~\eqref{eq:VSMS} and its
Coleman-Weinberg part by Eq.~\eqref{eq:CW}.  The current allowed
$95\%$ confidence level intervals for the trilinear Higgs coupling
modifier $\kappa_{\lambda_3}=\lambda_3/\lambda_3^{SM}$ still leave a
lot of room for the BSM physics and read:
\begin{eqnarray}
\kappa_{\lambda_3} &\in& [-0.4, 6.3],
~\text{ATLAS~\cite{ATLAS:2022jtk}}, \nonumber\\
\kappa_{\lambda_3} &\in&[-1.24, 6.49], ~\text{CMS~\cite{CMS:2022dwd}}.
\end{eqnarray}

Corrections to the leading tree level contribution $\lambda_{3,
  \text{tree}}^{SM} = 3 M_H^2/v$ predicted by the SM can be classified
as follows.
\begin{itemize}
\item Leading SM one-loop top-quark induced contribution:
\begin{equation}
\Delta\lambda_3^{SM} \approx -\frac{3 M_t^4 }{v^3 \pi^2 }.
\label{eq:3H:top}
\end{equation}
\item Leading one-loop real scalar singlet induced contribution:
\begin{equation}
\Delta\lambda_3^{S}\approx\frac{\lambda_{HS}^3
  v^3}{32 \pi^2 M_S^2 }.
\label{eq:3H:S}
\end{equation}
\item Loop contributions from heavy VLF multiplets:
\begin{equation}
\begin{aligned}
\Delta\lambda_3^{VLF}\approx n_{F_1} \frac{N_c^\prime v^3 y_{F_1}^6}{8
  \pi^2 M_{F_1}^2} + n_{F_2} \frac{N_c^\prime v^3 y_{F_2}^6 }{8 \pi^2
  M_{F_2}^2},
\label{eq:3H:VLF}
\end{aligned}
\end{equation}
where $N_c^\prime$ is the number of colours of VLF multiplets and one
should sum over $n_{F_1}\in\{n_U, n_N\}$, $n_{F_2}\in\{n_D, n_E\}$ and
$y_{F_1}\in\{y_U, y_N\}$, $y_{F_2}\in\{y_D, y_E\}$.
\end{itemize}

\begin{figure}[bt!]
\centering
\includegraphics[width=1\linewidth]{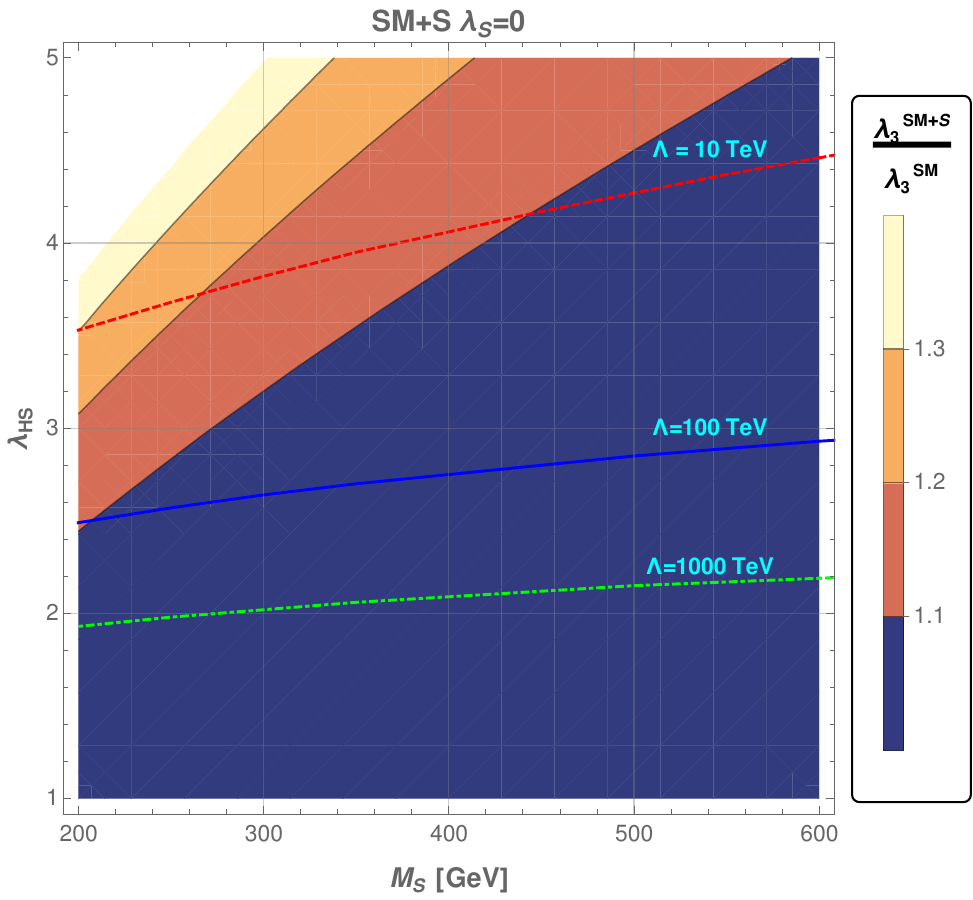}
\caption{Contours of relative triple Higgs coupling enhancement in the
  SM extended by real singlet. Red-dashed, blue-solid and
  green-dot-dashed lines indicate maximal value of $\lambda_{HS}$
  allowed by the perturbativity constraints up to, respectively,
  $\Lambda=10, \space 100, \space 1000$ TeV with $\Delta=0.4$ (compare
  eq.~\eqref{eq:condition3}).}
\label{fig:3H:SMS:Contour}
\end{figure}

Effects of the scalar singlet are presented in
Fig.~\ref{fig:3H:SMS:Contour} which shows the corresponding relative
enhancement of the triple Higgs coupling, $\kappa_{\lambda_3}^{SM+S}
=\lambda_3^{SM+S}/\lambda_3^{SM}$.  The contours of maximal allowed
values of $\lambda_{HS}$ for chosen cut-off scales $\Lambda$ are
obtained using perturbativity conditions~\eqref{eq:condition1}--
\eqref{eq:condition3}.

For the cut-off scale of 100~TeV or lower, the enhancement of
$\lambda_3$ from scalar singlet contributions is at most $\sim10\%$
for $M_S\ge200$~GeV, leading to a moderate, at most
$K^S_{HH}\approx-8\%$, decrease of the double Higgs production cross
section (which is due to destructive interference between the box and
triangle diagrams of Fig.~\ref{fig:DiHiggs:diags}).

Addition of VLF also has some impact on $\lambda_3$, however, this
effect is very small.  We can estimate its maximal possible value by
using numbers from Table~\ref{tab:SMSVLF} and plugging them into the
Eq.~\eqref{eq:3H:VLF}.  As a result, we get negligible contributions
with $|\Delta\lambda_3^{VLF}/\lambda_3^{SM}|\ll 1\%$ for all
considered cases, leaving no chance for significant triple Higgs
coupling modification by VLF fields only.

\smallskip

\noindent{\bf 2. VLQ loop contributions.}

\smallskip

Apart from modifying the triple Higgs boson coupling $\lambda_3$,
heavy VLQ may impact double Higgs boson production through the box and
triangle loop gluon fusion diagrams presented in
Fig.~\ref{fig:DiHiggs:diags}.  Considering such corrections, one
should take into account that loop contributions to the double and
single Higgs boson production are closely related, the Feynman
diagrams for the latter can be obtained from
Fig.~\ref{fig:DiHiggs:diags} by replacing one of the external Higgs
fields by the VEV insertion. As a consequence, all relevant amplitudes
are proportional to the same combination of VLQ parameters:
\begin{equation}
 \mathcal{M}_{gghh}^{VLQ\triangle} \propto
 \mathcal{M}_{gghh}^{VLQ\Box} \propto \mathcal{M}_{ggh}^{VLQ}\propto
 n_{U} \frac{y_{U}^2}{M_U^2} + n_{D} \frac{y_{D}^2}{M_D^2}
\label{eq:ggH:amplitude}
\end{equation}
Therefore, apart from theoretical constraints considered already in
this work, another important source of limits on VLQ model parameters
comes from the current limits on the single Higgs production.
Available experimental data~\cite{CMS:2022dwd} puts strong constraint,
at around $10\%$ at $68\%$ CL and around $18\%$ at $95\%$ CL, on the
deviations of the single Higgs boson production rate via gluon fusion
from its SM value.  This condition may (depending on the choice of
$\Delta$ in~\eqref{eq:condition3}) lead to stronger constraints on the
maximal values of VLF Yukawa couplings then listed in
Table~\ref{tab:SMSVLF}, as illustrated in
Table~\ref{tab:ggHH:box:triangle}.

\begin{table}[bt!]
\centering
{\scriptsize
\begin{tabular}{|c||c|c|c||c|c|c|}
\hline
SM+VLQ & $y^{RGE}_{MAX}$ & $y^{H}_{MAX}$& $K_{H H}^{VLQ}$ &
$y^{RGE}_{MAX}$ & $y^{H}_{MAX}$& $K_{H H}^{VLQ}$ \\ \hline
& \multicolumn{3}{|c||}{$68\%$ CL} &
\multicolumn{3}{|c|}{$95\%$ CL} \\
\hline
$n_Q=n_U=n_D=1$& 0.74 & \textbf{0.65} & $+8.4\%$ & \textbf{0.74} &
0.89 & $+11.3\%$\\
\hline
$n_Q=n_U=n_D=2$& 0.60 &\textbf{0.46}& $+8.2\%$ & \textbf{0.60} & 0.63
& $+15.0\%$\\
\hline
$n_Q=n_U=n_D=3$& 0.50 & \textbf{0.37} & $+8.3\%$ & \textbf{0.50} &
0.51 & $+15.5\%$ \\
\hline
$n_Q=n_U=1$ $n_D=0$ & \textbf{0.91} & 0.92 & $+8.2\%$ & \textbf{0.91}
& 1.26 & $+8.2\%$\\
\hline
$n_Q=n_U=2$ $n_D=0$ & 0.74 & \textbf{0.65} & $+8.4\%$ & \textbf{0.74}
& 0.89 & $+11.3\%$\\
\hline
$n_Q=n_U=3$ $n_D=0$ & 0.65 & \textbf{0.53} & $+8.2\%$ & \textbf{0.66}
& 0.73 & $+13.5\%$ \\
\hline
\end{tabular}
}
\caption{Maximal enhancement of the double Higgs production cross
  section from VLQ loop gluon fusion diagrams for $M_F=1$
  TeV. $y^{RGE}_{MAX}$ and $y^{H}_{MAX}$ indicate maximal values
  allowed respectively by perturbativity and by single Higgs
  production constraints (the latter at CL=68\% or
  95\%~\cite{CMS:2022dwd}). In each case bold font indicates stronger
  constraint that is taken into account. }
\label{tab:ggHH:box:triangle}
\end{table}

Table~\ref{tab:ggHH:box:triangle} shows that the enhancement of the
double Higgs production rate due to the VLQ loop contributions, $K_{H
  H}^{VLQ}$, within the parameter space obtained in this work, is
always small (at most $\sim 15\%$).  Even such minor enhancement
should be considered as optimistic scenario.  As discussed before, in
models with scalar singlet this positive contribution may be in
addition compensated by scalar contribution to the triple Higgs
coupling. Increasing $M_F$ can also only further suppress the VLF
effects.  Finally, effects of increasing number of VLF multiplets can
be deduced from the form of the amplitudes in~\eqref{eq:ggH:amplitude}
- for example, assuming identical masses for all multiplets, case with
$n=1$ and some value of $Y$ is equivalent to $n=N$ with $Y\to
Y/\sqrt{N}$, so that the corresponding bounds on $K_{H H}^{VLQ}$ are
not much affected by varying $n$ (as can be seen in the 4th column of
Table \ref{tab:ggHH:box:triangle}).

Comparing numbers in Table~\ref{tab:ggHH:box:triangle} with still
rather lax current experimental constraints on the Higgs boson pair
production at $95\%$ confidence level:
\begin{eqnarray}
\sigma(pp\rightarrow HH) &<& 2.4\times \sigma^{SM}(pp\rightarrow HH),
~\text{ATLAS~\cite{ATLAS:2022jtk}}\,, \nonumber\\
\sigma(pp\rightarrow HH) &<& 3.4\times \sigma^{SM}(pp\rightarrow HH),
~\text{CMS~\cite{CMS:2022dwd}}\,,
\end{eqnarray}
one can conclude that adding vector-like multiplets to the SM likely
cannot affect the double Higgs production in a way which could be
experimentally observed at present or in the foreseeable future.

\subsection{Electroweak precision observables: corrections to $\mathbb{S}$,
  $\mathbb{T}$ and $\mathbb{U}$ oblique parameters}
\label{sec:STU}

Additional constraints on the parameter space of models considered in
this work may arise from the electroweak precision observables. In our
analysis, we focus on $\mathbb{S}$, $\mathbb{T}$ and $\mathbb{U}$
oblique parameters, which parametrize one-loop contributions to the
electroweak gauge bosons self-energies \cite{Peskin:1990zt,
  Peskin:1991sw}.  In the context of VLF (the singlet scalar does not
contribute to $\mathbb{S}$, $\mathbb{T}$ and $\mathbb{U}$ parameters
due to $\mathbb{Z}_2$ symmetry of its potential), there have been a
number of extensive studies addressing this issue, see
e.g.~\cite{Lavoura:1992np, Kearney:2012zi, Arsenault:2022xty,
  Abouabid:2023mbu}.  Current experimental constraints, obtained with
$\mathbb{U}$ fixed to zero (motivated by the fact that $\mathbb{U}$ is
suppressed by additional factor of $M_i^2/M_Z^2$ comparing with
$\mathbb{S}$, and $\mathbb{T}$, see Ref.~\cite{Grinstein:1991cd}),
read~\cite{ParticleDataGroup:2022pth}:
\begin{equation}
\begin{aligned}
& \mathbb{T}=0.04 \pm 0.6,\\
& \mathbb{S}=-0.01 \pm 0.07.
\end{aligned}
\end{equation}

\begin{figure}[ht!]
\centering
\includegraphics[width=1\linewidth]{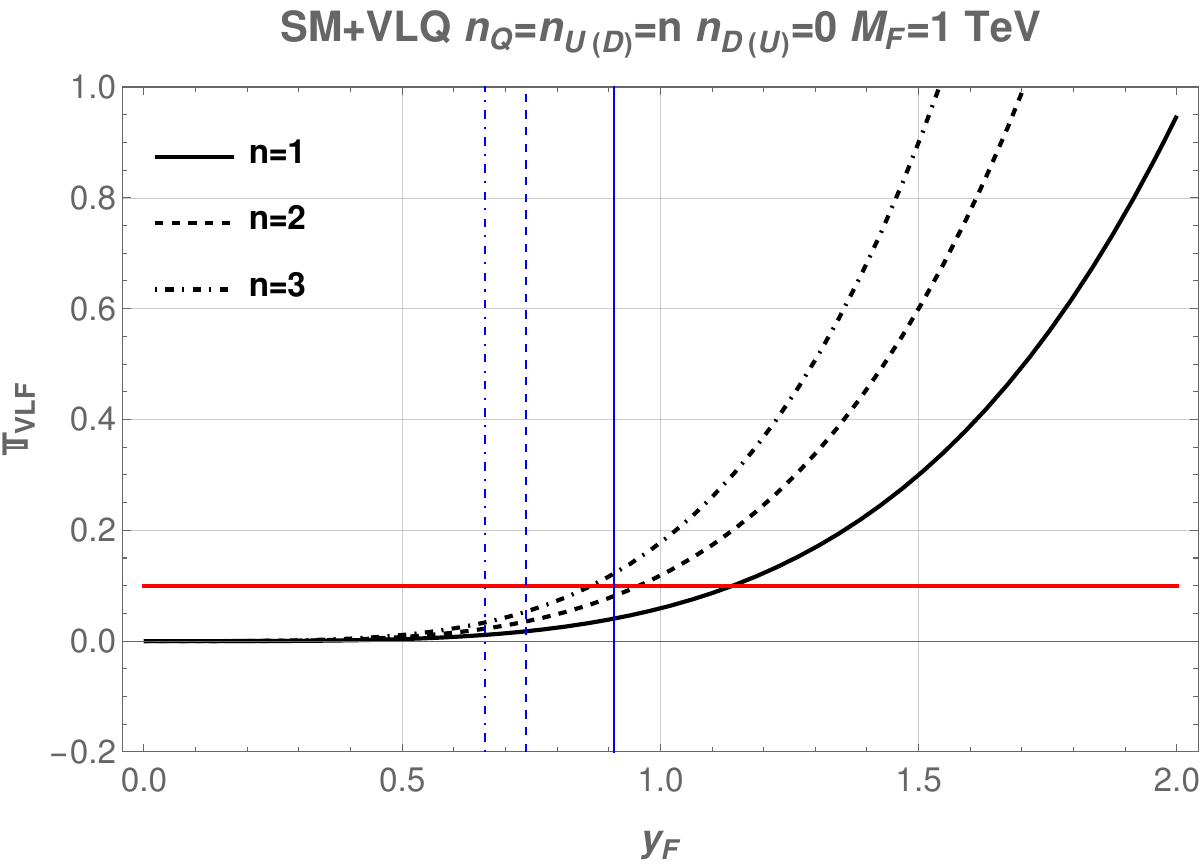}
\caption{Value of $\mathbb{T}$ parameter as function of VLF Yukawa
  coupling $y_F$ (black lines) for two model scenarios and different
  number of VLF families.  Also shown are: theoretical upper bounds on
  $y_F$ (blue lines) and the experimental upper bound on $\mathbb{T}$
  (red line)}.
\label{fig:SMVLF:STU}
\end{figure}

Formulae for the corrections to $\mathbb{T}$ and $\mathbb{S}$
parameters in the presence of VLF have been worked out
in~\cite{Lavoura:1992np} and are collected in
Appendix~\ref{app:STU}. Those parameters by definition come only from
the BSM sector, so that in the absence of mixing between VLF and the
SM fermions one has:
\begin{equation}
\begin{aligned}
   \mathbb{T} &\equiv \mathbb{T}_{VLF},\\
   \mathbb{S}&\equiv \mathbb{S}_{VLF}.
\end{aligned}
\end{equation}
Comparison of constraints on VLF models based on theoretical
considerations of model consistency and/or of single Higgs production
presented in the Section~\ref{sec:dihiggs} (see
Table~\ref{tab:ggHH:box:triangle}), with those based on the precision
$\mathbb{T}$ and $\mathbb{S}$ electroweak observables is illustrated
in Table~\ref{tab:STU} and Figure~\ref{fig:SMVLF:STU}.

Several comments are in place. In the model scenarios considered in
this work, we assume the same number of left- and right-handed VLF
multiplets. This significantly simplifies formulae
from~\cite{Lavoura:1992np}.  Moreover, we assume identical VLF masses
$M_{VLF}$ and Yukawa couplings $y_{VLF}$, which limits the effect of
VLF on $\mathbb{T}$ and $\mathbb{S}$ parameters. More specifically,
for ``Scenario I'' in which we have the same number of VLF doublets
and ``up-type'' and ``down-type'' singlets one has $\mathbb{T}=0$ as
such setup preserves custodial symmetry.  For other scenarios,
contribution to $\mathbb{T}$ is significantly larger than to
$\mathbb{S}$, but still way below the experimental constraints.
Finally, VLL scenarios are far less constrained than VLQ due to the
lack of enhancement by the $N_c$ colour factor (see formulae in
Appendix~\ref{app:STU}).  Since increasing $M_{VLF}$ only leads to
further decrease in the values of $\mathbb{T}$ and $\mathbb{S}$
parameters (even after accounting for the increase of allowed maximal
value of VLF Yukawa couplings), we can conclude that for the model
scenarios studied in this paper, the theoretical stability constraints
are much stronger than those arising from the bounds on the oblique
parameters $\mathbb{T}$ and $\mathbb{S}$.

\begin{table}[bt!]
\centering
\begin{tabular}{|c||c|c|c|c|}
\hline
SM+VLQ & $y^{RGE}_{MAX}$ & $\mathbb{T}_{VLF}$ & $\mathbb{S}_{VLF}$ \\
\hline
$n_Q=n_U=n_D=1$& 0.74 & 0 & 0.006\\
\hline
$n_Q=n_U=n_D=2$& 0.60 & 0 & 0.007 \\
\hline
$n_Q=n_U=n_D=3$& 0.50 & 0 & 0.008 \\
\hline
$n_Q=n_U=1$ $n_D=0$ & 0.91 & 0.041  & 0.012 \\
\hline
$n_Q=n_U=2$ $n_D=0$ & 0.74 & 0.035 & 0.015\\
\hline
$n_Q=n_U=3$ $n_D=0$ & 0.65 & 0.031 & 0.018 \\
\hline
\end{tabular}
\caption{Maximal contributions to $\mathbb{T}$ and $\mathbb{S}$
  oblique parameters for the maximal allowed values of VLQ Yukawa
  couplings for $M_{VLF}=1$ TeV, two different model scenarios and
  varying number of VLF multiplets.}
\label{tab:STU}
\end{table}

\subsection{Electroweak phase transition}
\label{sec:EWPT}

The one-loop effective potential in the SM with additional singlet
scalar field (for review see e.g.~\cite{Quiros:1999jp}) reads:
\begin{equation}
\begin{aligned}
V_{\mathrm{eff}}(H, S, T)=V_0(H, S)+V_{CW}(H, S)+V_{T}(H, S, T)\, .
\end{aligned}
\label{eq:veff}
\end{equation}
with $V_0$ a tree-level potential \eqref{eq:VSMS}, $V_{CW}$ the
one-loop correction known also as the Coleman-Weinberg
potential~\cite{Weinberg:1973am} and $V_T$ the finite temperature
contribution (for further details see Appendix~\ref{sec:effpot}).

Models are considered to be interesting in the context of baryogenesis
if they allow for a strong enough first order phase transition.  In
general, a 1st order electroweak phase transition may occur in the
presence of a barrier in the effective Higgs potential separating two
degenerate minima at $h=0$ and $h=v_C$ (here, $v_C$ is the VEV of the
Higgs field at the critical temperature, $T = T_C$).  The phase
transition is considered strong if the following condition is
satisfied:
\begin{equation}
    \xi = \frac{v_C(T_C)}{T_C}\gtrsim(0.6 \div 1.0).
\end{equation}

\smallskip

\noindent{\bf 1. EWPT with real scalar singlet.}

\smallskip

\begin{figure*}[ht!]
\centering
\includegraphics[width=0.45\linewidth]{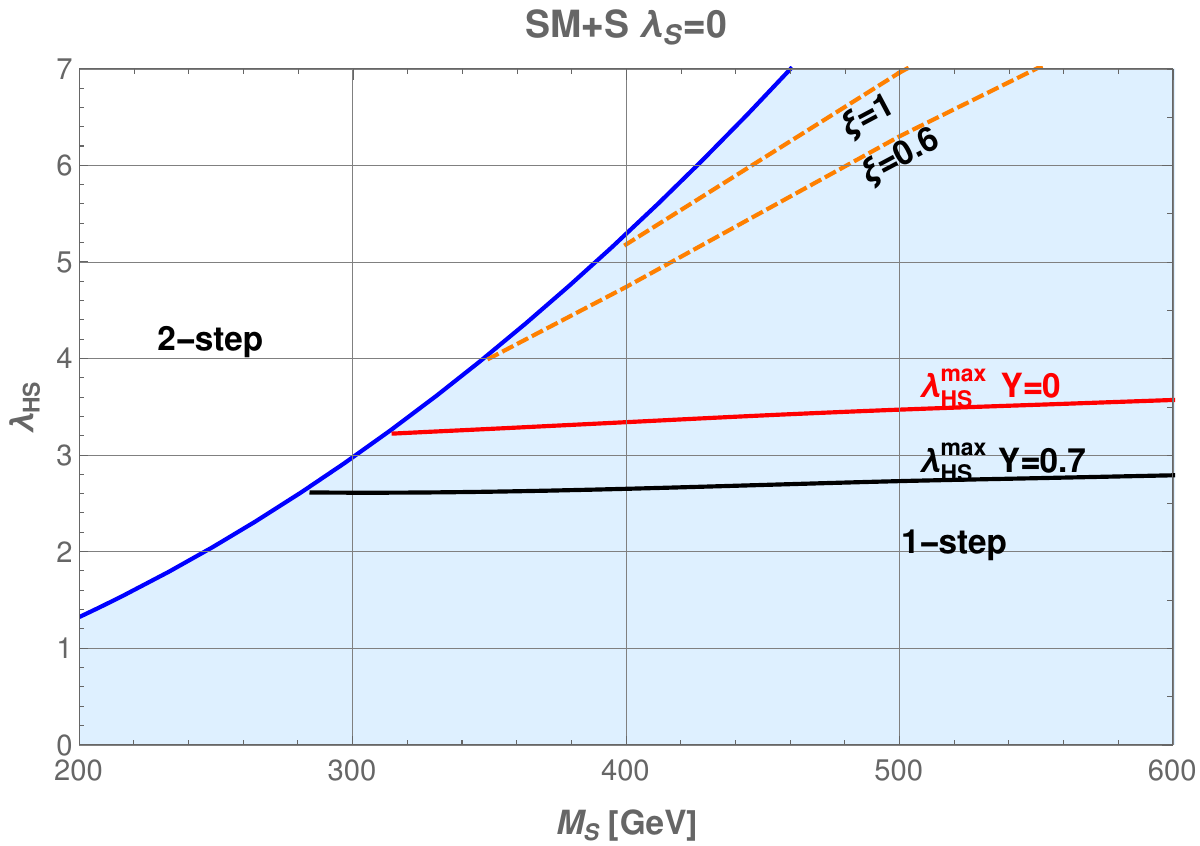}
\includegraphics[width=0.45\linewidth]{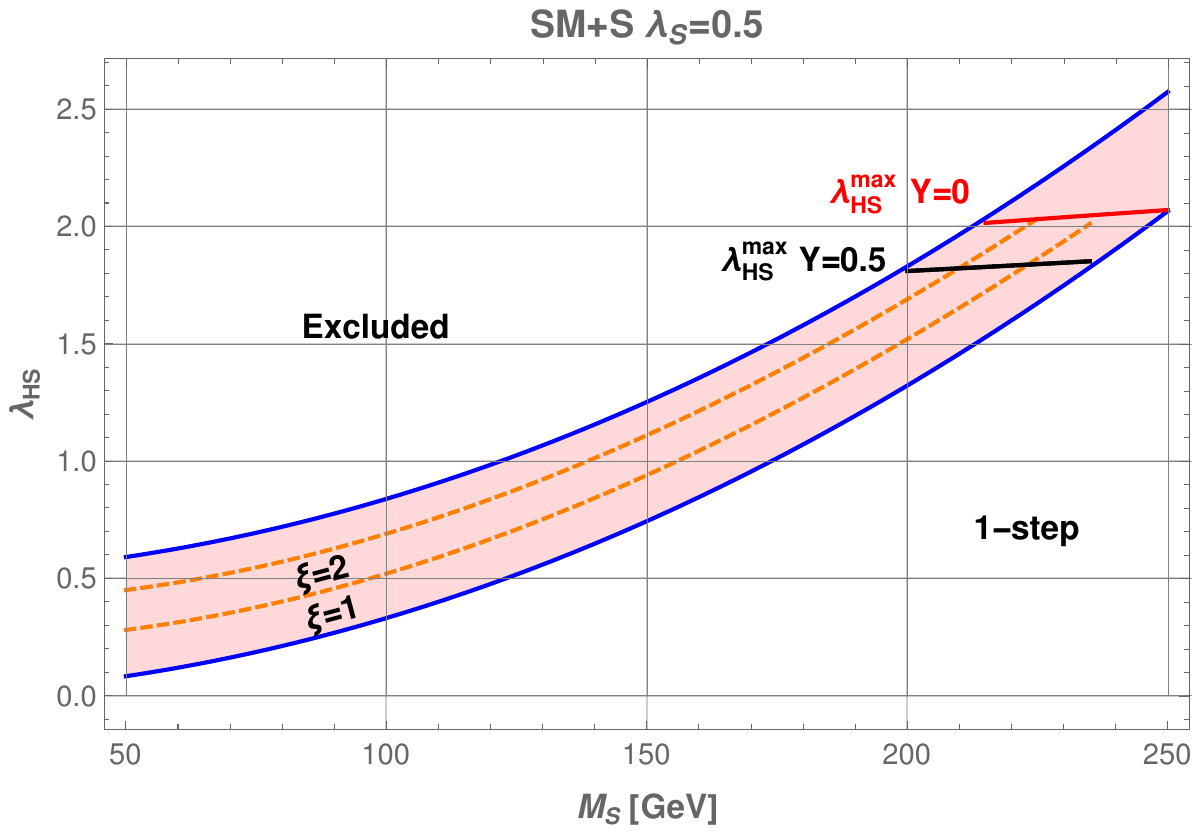}
\caption{Shaded regions show parts of the parameter space allowing for
  the 1-step (left panel) and 2-step (right panel) EWPT in the SM
  extended by a singlet scalar. The orange dashed lines indicate
  strength of the EWPT $\xi=0.6$ or $\xi=1$ (1-step) and $\xi=1$ or
  $\xi=2$ (2-step).  The red and black lines indicate maximal values
  of $\lambda_{HS}$ allowed by the perturbativity constraints up to
  $\Lambda=100$ TeV for two values of the VLF Yukawa couplings. Values
  of $\lambda_S$ were chosen as 0 and 0.5 for the 1- and 2-step
  transition, respectively.}
\label{fig:Xi:SMS:Contour}
\end{figure*}

In the real scalar model with the tree-level potential given by
Eq.~\eqref{eq:VSMS}, depending on the values of the scalar sector
couplings and masses, we can get either 1- or 2-step EWPT. The former
takes place when the system moves from the high-temperature minimum
with $\left<H\right>=\left<S\right>=0$ to the standard EW vacuum with
$\left<H\right>\ne0$, $\left<S\right>=0$. The latter is realised if
for some range of temperatures the minimum of the effective potential
is located at $\left<H\right>=0$, $\left<S\right>\ne0$.
\begin{itemize}
\item 1-step EWPT is realised if $\mu_S^2>0$.  This region is not
  affected by the value of $\lambda_S$, therefore we choose
  $\lambda_S=0$ in order to maximise the allowed value of
  $\lambda_{HS}$ coupling.  The allowed parameter space for the 1-step
  EWPT is presented on the left panel in
  Fig.~\ref{fig:Xi:SMS:Contour}.  The blue line indicates the boundary
  between 1- and 2-step regions.  The red and black lines indicate
  maximal allowed values of $\lambda_{HS}(M_S)$ obtained with
  constraints \eqref{eq:condition2} and \eqref{eq:condition3} assuming
  $\Lambda=100$ TeV for two values of the VLF Yukawa couplings.  The
  orange dashed lines indicate values of $\lambda_{HS}(M_S)$
  corresponding to the strength of the EWPT, $\xi$ equal 0.6 and 1.0.
  As one can note, the region for which $\xi\geq 0.6$ in the singlet
  scalar extension is basically excluded in comparison e.g.~with
  corresponding area presented in~\cite{Curtin:2014jma}.  This follows
  from our more careful treatment of perturbativity conditions.
\item 2-step EWPT may be realised if $ \mu_S^2<0$ with further
  condition guaranteeing that at zero temperature the EW vacuum
  corresponds to the global minimum of the potential:
\begin{equation}
 V_0(v,  0)< V_0(0,  w)\,, 
\end{equation}
which can be translated to the bounds on $\lambda_{HS}$.  A
2-step EWPT is possible if:
\begin{equation}
  2\frac{M_S^2}{v^2}\leq \lambda_{HS} \leq
  2\frac{M_S^2}{v^2} +
  \sqrt{2}\frac{M_H\sqrt{\lambda_S}}{ v}\,.
\label{eq:lamHS:2step}
\end{equation}
The allowed 2-step region is indicated by the red colour on the right
plot in Fig.~\ref{fig:Xi:SMS:Contour}.  The lower and upper limits
from the Eq.~\eqref{eq:lamHS:2step} for $\lambda_S=0.5$ are indicated
by the blue lines. The choice of this particular value of $\lambda_S$
allows for reasonably large $\lambda_{HS}$ while still leaving space
for the 2-step EWPT to occur. The red line, as in the case of 1-step
transition, show the maximal possible values of $\lambda_{HS}$.
\end{itemize}

\smallskip
\noindent{\bf 2. EWPT with vector-like fermions.}
\smallskip

In principle VLF can also affect the mechanism of the EWPT through
thermal and loop corrections to the effective potential (see Appendix
\ref{sec:effpot}). Fermions do not contribute much to inducing
potential barrier for the Higgs field, contrary to the case of bosons
which in general lead to cubic terms in the potential in the high
temperature expansion of thermal effects. However, in the case in
which the critical temperature is much lower than fermion masses, the
\textit{small temperature} expansion is more suitable. In the latter
case, which fits very well the scope of this work in which VLF are
relatively heavy, the leading contribution to the thermal potential is
similar for fermions and bosons. This could potentially lead to
non-trivial impact of VLF on EWPT and thermal history of the Universe
(see e.g. \cite{Egana-Ugrinovic:2017jib, Egana-Ugrinovic:2017jib}).

However, because of strong constraints on the values of the VLF Yukawa
couplings, discussed in Section~\ref{sec:parspace}, the actual impact
of these particles on the EWPT is very limited.  For the maximal
allowed values of $y_F$ in SM+VLF models, presented in
Table~\ref{tab:SMSVLF}, we obtained $\xi_{SM+VLF}$ quite close to
$\xi_{SM}$.  Addition of the real singlet scalar allows for slightly
larger maximal values of $y_F$, see Table~\ref{tab:SMSVLF}, but these
larger values are not sufficient to have any significant impact on the
EWPT, so $\xi_{SM+S+VLF}\approx\xi_{SM+S}$.  We checked that changes
of $\xi$ due to VLF are at most of the order $\mathcal{O}(10\%)$.

\section{Summary and outlook}
\label{sec:concl}    

In this work we studied consequences of extending the Standard Model
by adding vector-like fermions and/or a real scalar singlet for the
stability of the electroweak vacuum, for perturbativity of the model
couplings and for possible unification of the gauge couplings. We
divided our discussion into analysis of three broader model classes
with different sets of extra fields: only VLF; only a singlet scalar;
both VLF and a scalar.  Further, the VLF models were classified into
three types based on numbers of extra multiplets with different
transformation properties with respect to the SM gauge group. In order
to study the most generic features of discussed models, for simplicity
we assumed equal masses and couplings of VLF multiplets, therefore we
did not discuss any effects which could be related to flavor-like
structure of their parameters.

We defined a set of theoretical conditions, summarised by
equations~\eqref{eq:condition1}--\eqref{eq:condition3}, which play a
pivotal role in deriving the results of this work.  These conditions
come solely from the stability and perturbativity requirements of the
considered models.  The primary strategy in our work was to implement
the 2-loop RGEs and derive constraints from the resulting coupling
evolution.  In addition, we incorporated in our discussion also
various experimental bounds, coming from trilinear Higgs coupling,
Higgs production cross-section, bounds on masses of VLF and the real
scalar field.

We obtained strong upper bounds on magnitude of VLF Yukawa couplings.
The precise limits on such couplings, $y_{MAX}$, vary with details of
models and with the assumed cut-off scale $\Lambda$ up to which the
theoretical constraints should be fulfilled. However, in all cases we
get $y_{MAX}<1$ for all considered scenarios if $\Lambda\ge 100$~TeV.

In terms of EW stability, the scalar singlet extension of the SM
offers better prospects. The stability of the EW vacuum improves with
increasing the singlet scalar couplings, $\lambda_{HS}$ and
$\lambda_S$.  However, the model loses perturbativity when scalar
couplings become too large.  Models with both VLF families and the
real scalar have typically less problems with the EW vacuum stability
than models with only VLF.  On the other hand, interplay between VLF
and the scalar results in bigger problems with the perturbativity as
compared to the situation when only the scalar is added to the SM. In
general, our analysis shows that the upper bounds on the VLF Yukawa
couplings can be somewhat relaxed due to the presence of the singlet
scalar, however such enhancement of $y_{MAX}$ is never larger than
about 50\%.

The criteria of perturbativity and electroweak vacuum stability should
be obviously applied on model-by-model basis and the results will vary
accordingly. Or analysis shows that pure VLF models, eventually
extended with one real scalar singlet, have serious limitations
concerning their allowed parameter ranges and in general may not be
sufficient to explain phenomenological effects to which they were
often used.

Adding new fields or more complicated pattern of VLF parameters
modifies RGE evolution and so may affect the position of Landau poles
for various couplings and the stability of the scalar
potential. However, the constraints from the stability and
perturbativity requirements can be also important in such extended
models and should be always checked as a part of realistic
phenomenological analyses of a given SM extension, as we illustrated
in our paper for the set of simplified VLF scenarios.

Finally, we analysed the implications of the constrained parameter
space on the phenomenology of the double Higgs boson production,
electroweak precision observables and the electroweak phase
transition.  Adding scalar singlet to SM can increase somewhat the
triple Higgs coupling $\lambda_3$ enhancing the single Higgs
production but simultaneously suppressing the double Higgs production
rate. Adding VLF fields have negligible effect on $\lambda_3$ but may
affect the Higgs boson pair production via contribution to the
triangle and box loops in the amplitude.  Still, within the allowed
parameter space, the VLF loops may enhance the double Higgs production
rate by at most $15\%$, thus their impact is very limited and
impossible to observe with current or achievable in near future
experimental accuracy.

We also examined the impact of the studied VLF model scenarios on the
electroweak precision observables: $\mathbb{S}$, $\mathbb{T}$ and
$\mathbb{U}$ oblique parameters.  We found that constraints on the
model's parameter space resulting from experimental bounds on these
parameters are weaker than constraints arising from theoretical
stability conditions and perturbativity and from single Higgs
production measurements.

We demonstrated that the parameter space of the SM+S model is heavily
constrained, leaving very little room for a strong first order
EWPT. We showed also that addition of VLF with Yukawa couplings
satisfying the stability and perturbativity conditions has again very
little impact on the EWPT.  Thus, the strong first order phase
transition in the SM with only VLF added (no extra scalars) can not be
realised.

The results presented in this work provide an independent set of
constraints for models that extend the Standard Model with vector-like
fermions and a scalar field.  We believe these constraints are an
important addition to the existing analyses, in the context of the
considered models. Furthermore, the methodology employed here can be
readily applied to study other theoretical frameworks beyond the
specific scenarios explored in the current work.

\section*{Acknowledgements}

Research partially supported by the Norwegian Financial Mechanism for
years 2014-2021, under the grant no 2019/34/H/ST2/00707. A.A. received
support from the French government under the France 2030 investment
plan, as part of the Initiative d'Excellence d'Aix-Marseille
Université - A*MIDEX.  Work of M.O. was partially supported by
National Science Center, Poland, grant DEC-2018/31/B/ST2/02283.  The
work of M.R. and J.R. was supported in part by Polish National Science
Center under research grant DEC-2019/35/B/ST2/02008.  We would like to
thank Stefan Pokorski for numerous discussions and Javier Lizana and
Marco Merchand for sharing their insight into the topic of electroweak
phase transition.

\appendix

\section{RGEs for models with extended scalar and vector-like sectors}
\label{app:rge}

Below we present 1-loop RGEs for all of the relevant couplings in the
model considered in the paper, $\kappa_i = (\lambda, y_t^2, g_1^2,
g_2^2, g_3^2, y_F^2, \lambda_{HS}, \lambda_S)$, separated into the
parts corresponding to various sectors of the theory which (alongside
2-loop contributions which are to lengthy to be displayed), were
obtained using SARAH package~\cite{Staub:2013tta} and validated using
RGBeta~\cite{Thomsen:2021ncy}.
\begin{equation}
\label{eq:RGE}
\begin{aligned}
\beta^{(1)}_{\kappa_i} &= \beta^{SM(1)}_{\kappa_i} +
\beta^{VLF(1)}_{\kappa_i}+\beta^{S(1)}_{\kappa_i}+ \beta^{VLF(1)\times
  S(1)}_{\kappa_i}
\end{aligned}
\end{equation}
with:
\begin{equation}
\frac{d \kappa_i(\mu)}{d \ln \mu} =
\beta_{\kappa_i}\left(\kappa_j(\mu)\right)\equiv\beta_{\kappa_i}
\end{equation}

\vspace{2mm}

\subsection{SM sector}

\begin{equation}
\label{eq:RGE:SM:1}
\begin{aligned}
\beta_\lambda^{SM(1)}&=\frac{1}{16\pi^2}\left[\frac{9}{8}\left(\frac{3}{25}
  g_{1}^{4}+ g_{2}^{4}+\frac{2}{5} g_{1}^{2} g_{2}^{2}\right)\right.\\
  &\left.\qquad\qquad-6
  y_{t}^{4}+24 \lambda^{2}+12 y_{t}^{2} \lambda-\frac{9}{5} g_{1}^{2}
  \lambda-9 g_{2}^{2} \lambda\right] \\
\beta_{y_t^2}^{SM(1)}&=\frac{y_t^2}{16\pi^2}\left[9
  y_{t}^{2}-\frac{17}{10} g_{1}^{2}-\frac{9}{2} g_{2}^{2}-16
  g_{3}^{2}\right] \\
\beta_{g_1^2}^{SM(1)}&=\frac{1}{16\pi^2}\left[\frac{41}{5}
  g_{1}^{4}\right],  \\
\beta_{g_2^2}^{SM(1)} &=\frac{1}{16\pi^2}\left[-\frac{19}{3}
  g_{2}^{4}\right],   \\
\beta_{g_3^2}^{SM(1)} &=\frac{1}{16\pi^2}\left[-14 g_{3}^{4}\right].
\end{aligned}
\end{equation}

\subsection{Vector-like fermion sector}

\begin{equation}
\begin{aligned}
\beta_\lambda^{VLF(1)}& =\frac{1}{16\pi^2}\left[2
  n_{F_1}N_c^{\prime}\left(4 y_{F_1}^2 \lambda-2 y_{F_1}^4\right)\right.\\
  & \left.\qquad\qquad+ 2
  n_{F_2}N_c^{\prime}\left(4 y_{F_2}^2 \lambda-2
  y_{F_2}^4\right)\right], \\
\beta_{y_t^2}^{VLF(1)} & =\frac{y_t^2}{16\pi^2}\left[4 N_c^{\prime}
  \left(n_{F_1} y_{F_1}^2 + n_{F_2} y_{F_2}^2\right)\right], \\
\beta_{g_1^2}^{VLF(1)}&=\frac{g_1^4}{16 \pi^2}\left[\frac{8}{5}
  N_c^{\prime}\left(2 n_\psi Y_{W_\psi}^2+n_{F_1} Y_{W_{F1}}^2+n_{F_2}
  Y_{W_{F2}}^2\right)\right],  \\
\beta_{g_2^2}^{VLF(1)}&=\frac{1}{16\pi^2}\left[\frac{4}{3} N_c^{\prime}
  n_\psi g_2^4\right],   \\
\beta_{g_3^2}^{VLF(1)} &=\frac{1}{16\pi^2}\left[\frac{4}{3} n_3
  g_3^4\right],   
\nonumber
\end{aligned}
\end{equation}
\begin{equation}
\begin{aligned}
\beta_{y_{F_1}^2}^{VLF(1)} &=
\frac{y_{F_1}^2}{16\pi^2}\left[3y_{F_1}^2 +4 n_{F_1}
  N_c^{\prime}y_{F_1}^2+ 6 y_t^2-16 \hat{n}_F^{V L Q} g_3^2\right.\\
  &\qquad\qquad-\frac{9}{2}
  g_2^2-\frac{18}{5} g_1^2\left(Y_{W_H}^2+2 Y_{W_{F1}}
  Y_{W_\psi}\right)\\
&\left. \qquad\qquad  + \Delta_{n_{F_1},  n_{F_2}}\left(5 +
  4n_{F_1}N_c^\prime\right)y_{F_2}^2\right],  \\
\beta_{y_{F_2}^2}^{VLF(1)} &=
\frac{y_{F_2}^2}{16\pi^2}\left[3y_{F_2}^2 +4 n_{F_2}
  N_c^{\prime}y_{F_2}^2+ 6 y_t^2-16 \hat{n}_F^{V L Q} g_3^2\right.\\
  &\qquad\qquad-\frac{9}{2}
  g_2^2-\frac{18}{5} g_1^2\left(Y_{W_H}^2+2 Y_{W_{F2}}
  Y_{W_\psi}\right)\\
&\left.\qquad\qquad+\Delta_{n_{F_1},  n_{F_2}}\left(5 +
  4n_{F_2}N_c^\prime\right)y_{F_1}^2\right]\, ,
\end{aligned}
\label{eq:RGE:VLF:1}
\end{equation}
%
where $N_c^\prime$ is a number of colours of VLF, $n_{F_1}\in\{n_U,
n_N\}$, $n_{F_2}\in\{n_D, n_E\}$, $n_\psi\in\{n_Q, n_L\}$,
$n_3=2n_Q+n_U+n_D$, $y_{F_1}\in\{y_U, y_N\}$, $y_{F_2}\in\{y_D,
y_E\}$, $Y_{W_H}=1/2$, $Y_{W_\psi}\in\{Y_{W_Q}, Y_{W_L}\}$,
$Y_{W_{F1}}\in\{Y_{W_U}, Y_{W_N}\}$, $Y_{W_{F2}}\in\{Y_{W_D},
Y_{W_E}\}$.

\vspace{1mm}

\subsection{Real scalar sector}

\begin{equation}
\label{eq:RGE:S:1}
\begin{aligned}
\beta_\lambda^{S(1)} &=\frac{1}{16\pi^2} \left[\frac{1}{2} \lambda_{H
    S}^{2}\right] \\
\beta_{\lambda_{HS}}^{S(1)} &=\frac{\lambda_{H S}}{16\pi^2}\left[12
  \lambda+6 \lambda_{S}+4 \lambda_{H S}+6 y_{t}^{2}-\frac{3}{2}
  g_{1}^{2}-\frac{9}{2} g_{2}^{2}\right] \\
\beta_{\lambda_S}^{S(1)} &=\frac{1}{16\pi^2}\left[2 \lambda_{H
    S}^{2}+18 \lambda_{S}^{2}\right].
\end{aligned}
\end{equation}

\subsection{Vector-like fermion $\times$ real scalar sector}

\begin{equation}
\label{eq:RGE:VLF:S:1}
\begin{aligned}
 \beta_{\lambda_{HS}}^{VLF(1)\times S(1)} &=\frac{\lambda_{H
     S}}{16\pi^2}\left[ 4 N_c^\prime (n_{F_1} y_{F_1}^{2} + n_{F_2}
   y_{F_2}^{2})\right].
\end{aligned}
\end{equation}

\section{$\mathbb{S}$ and $\mathbb{T}$ oblique parameters in the presence
  of VLF}
\label{app:STU}

General formulae for $\mathbb{S}$ and $\mathbb{T}$ parameters in VLF
models can be found in Ref.~\cite{Lavoura:1992np}.  In the simplified
scenarios considered in this paper, they reduce to:
\begin{equation}
\begin{aligned}
\mathbb{T}_{VLF} = & \frac{N_c}{8 \pi \sin ^2 \theta_W \cos ^2
  \theta_W}\times\\
&\left[\sum_{\alpha, i}\left[|\mathcal{V}_{\alpha
      i}|^2\left(\theta_{+}\left(x_\alpha, x_i\right) +
    \theta_{-}\left(x_\alpha, x_i\right)\right) \right]\right. \\
&\left.-\sum_{\beta<\alpha}\left[|\mathcal{U}_{\alpha
      \beta}|^2\left(\theta_{+}\left(x_\alpha, x_\beta\right) +
    \theta_{-}\left(x_\alpha, x_\beta\right)\right) \right]\right.\\
&\left.-\sum_{j<i}\left[|\mathcal{D}_{i
      j}|^2\left(\theta_{+}\left(x_i, x_j\right) +
    \theta_{-}\left(x_i, x_j\right)\right) \right] \right],
\end{aligned}
\end{equation}

\begin{equation}
\begin{aligned}
\mathbb{S}_{VLF} = & \frac{N_c}{\pi} \times \left[\sum_{\alpha,i}
    \left[|\mathcal{V}_{\alpha i}|^2\left(\psi_{+}\left(x_\alpha, x_i\right) +
      \psi_{-}\left(x_\alpha, x_i\right)\right) \right] \right. \\
&\left.-\sum_{\beta<\alpha}\left[|\mathcal{U}_{\alpha
        \beta}|^2\left(\chi_{+}\left(x_\alpha, x_\beta\right) +
      \chi_{-}\left(x_\alpha, x_\beta\right)\right)\right]\right.\\
&\left.-\sum_{j<i}\left[|\mathcal{D}_{i j}|^2\left(\chi_{+}\left(x_i,
      x_j\right) + \chi_{-}\left(x_i, x_j\right)\right) \right]
    \right],
\end{aligned}
\end{equation}
where:
\begin{equation}
\begin{aligned}
&x_{i(\alpha)} \equiv M_{i(\alpha)}^2 / M_Z^2,\\
& \theta_{+}\left(x_1, x_2\right) \equiv x_1+x_2-\frac{2 x_1
    x_2}{x_1-x_2} \ln \frac{x_1}{x_2}, \\
& \theta_{-}\left(x_1, x_2\right) \equiv 2 \sqrt{x_1
    x_2}\left(\frac{x_1+x_2}{x_1-x_2} \ln \frac{x_1}{x_2}-2\right),
\end{aligned}
\end{equation}

{\footnotesize
\begin{equation}
\begin{aligned}
f\left(x_1, x_2\right) \equiv & \begin{cases}-2
  \sqrt{\Delta}\left(\arctan \frac{x_1-x_2+1}{\sqrt{\Delta}}-\arctan
  \frac{x_1-x_2-1}{\sqrt{\Delta}}\right) & \Delta>0 \\
0 & \Delta=0 \\
\sqrt{-\Delta} \ln \frac{x_1+x_2-1 + \sqrt{-\Delta}}{x_1 +
  x_2-1-\sqrt{-\Delta}} & \Delta<0\end{cases},\\
&\Delta=-1-x_1^2-x_2^2+2 x_1+2 x_2+2 x_1 x_2,
\end{aligned}
\end{equation}
\begin{equation}
\begin{aligned}
\chi_{+}\left(x_1, x_2\right) \equiv &
\frac{x_1+x_2}{2}-\frac{\left(x_1-x_2\right)^2}{3}\\
&+\left[\frac{\left(x_1-x_2\right)^3}{6}-\frac{1}{2}
  \frac{x_1^2+x_2^2}{x_1-x_2}\right] \ln \frac{x_1}{x_2} \\
& +\frac{x_1-1}{6} f\left(x_1, x_1\right)+\frac{x_2-1}{6} f\left(x_2,
x_2\right)\\
&+\left[\frac{1}{3}-\frac{x_1+x_2}{6}-\frac{\left(x_1-x_2\right)^2}{6}\right]
f\left(x_1, x_2\right), \\
\chi_{-}\left(x_1, x_2\right) \equiv & -\sqrt{x_1
  x_2}\left[2+\left(x_1-x_2-\frac{x_1+x_2}{x_1-x_2}\right) \ln
  \frac{x_1}{x_2}\right.\\
&\left.+\frac{f\left(x_1, x_1\right)+f\left(x_2,
    x_2\right)}{2}-f\left(x_1, x_2\right)\right]
\end{aligned}
\end{equation}
\begin{equation}
  \begin{aligned}
\psi_{+}\left(x_\alpha, x_i\right) \equiv & \frac{22 x_\alpha+14
  x_i}{9}-\frac{1}{9} \ln \frac{x_\alpha}{x_i}\\
&+\frac{11 x_\alpha+1}{18} f\left(x_\alpha, x_\alpha\right)+\frac{7
  x_i-1}{18} f\left(x_i, x_i\right) \\
\psi_{-}\left(x_\alpha, x_i\right) \equiv & -\sqrt{x_\alpha
  x_i}\left[4+\frac{f\left(x_\alpha, x_\alpha\right)+f\left(x_i,
    x_i\right)}{2}\right],
\end{aligned}
\end{equation}}
Greek indices indicate summation over ``up-type'' states while Latin
ones over ``down-type'' fields. $\mathcal{V}$, $\mathcal{U}$ and
$\mathcal{D}$ matrices for the scenarios (Section~\ref{sec:model})
considered in this work read:
\begin{itemize}
	\item Scenario I: 
	{\footnotesize	
		\begin{equation}
		\mathcal{V}=\mathcal{U}=\mathcal{D}=
		\begin{pmatrix}
 		\cos\gamma_F^2& \cos\gamma_F \sin\gamma_F \\
 		\cos\gamma_F \sin\gamma_F & \sin\gamma_F^2 \\
		\end{pmatrix},
		\end{equation}}
	\item Scenario II:
	{\footnotesize
		\begin{equation}
		\begin{aligned}
		\mathcal{V}&=
		\begin{pmatrix}
 		\cos\gamma_F \\
 		\sin\gamma_F \\
		\end{pmatrix},\\
		\mathcal{U}&=
		\begin{pmatrix}
 		\cos\gamma_F^2& \cos\gamma_F \sin\gamma_F \\
 		\cos\gamma_F \sin\gamma_F & \sin\gamma_F^2 \\
		\end{pmatrix}
		,\quad
		\mathcal{D}=\mathbb{I},
		\end{aligned}
		\end{equation}}
	\item Scenario III:
	{\footnotesize
		\begin{equation}
		\begin{aligned}
		\mathcal{V}&=
		\left(
 		\cos\gamma_F,\space
 		\sin\gamma_F\right), \quad
		\mathcal{U}=\mathbb{I},\\
		\mathcal{D}&=\begin{pmatrix}
 		\cos\gamma_F^2& \cos\gamma_F \sin\gamma_F \\
 		\cos\gamma_F \sin\gamma_F & \sin\gamma_F^2 \\
		\end{pmatrix}.
		\end{aligned}
		\end{equation}}
\end{itemize}
Assumption of uniform Dirac masses $M_{F^d}=M_{F^s}=M_{F}$ and VLF
Yukawa couplings $y_F=y$ leads to the value of mixing angle
$\gamma_F=\frac{\pi}{4}$.

\section{Effective potential}
\label{sec:effpot}

One-loop effective potential is given schematically in
Eq.~\ref{eq:veff}.  Coleman-Weinberg part of the potential in on-shell
renormalization scheme with cutoff regularisation reads:
%
{\small

\begin{equation}
\begin{aligned}
\label{eq:CW}
V_{C W}(H)=\sum_{k} 
& \frac{r_k N_{k}}{64 \pi^{2}}\left(2
M_{k}^{2}(H) M_{k}^{2}(v)
\right. \\
&\left.+M_{k}^{4}(H)\left(\log
\frac{M_{k}^{2}(H)}{M_{k}^{2}(v)}-\frac{3}{2}\right)\right) \\
-\frac{N_{F}}{64 \pi^{2}}\sum_{i=1,2}&n_{F_i} \left(
M_{F_i}(H)^{4}\left(
\log \frac{M_{F_i}(H)^{2}}{\mu_R^2}-\frac{3}{2}\right)\right.\\
&\left.+C_1\phi^2+C_2\phi^4\right),
\end{aligned}
\end{equation}}
%
%
where we use the following notation:
%
\begin{equation}
\begin{aligned}
&k=(t, W, Z, h, S), \quad  N_k=(12, 6, 3, 1, 1), \\
&M_k(H)^2=M_{0, k}^2+a_k h^2, \quad  M_{0, k}^{2} =\left(0, 0, 0,
  -\mu^{2}, ~ \mu_{S}^{2}\right), \\
&a_{k} =\left(\frac{\lambda_{t}^{2}}{2}, \frac{g^{2}}{4},
  \frac{g^{2}+g^{' 2}}{4}, 3 \lambda,\frac{1}{2} \lambda_{H S}\right).
\end{aligned}
\end{equation}
and
\begin{equation}
\begin{aligned}
&F=(VLQ, VLL),  & \quad N_{F}=(12, 4)\\
&n_{F_1}\in\{n_U, n_N\}, & \quad  n_{F_2}\in\{n_D,
  n_E\}, \\
&M_{F_{1(2)}}=M_F\pm \frac{\sqrt{2}}{2} y_F v,
\end{aligned}
\end{equation}
and particle statistics related sign equals to $r_k=+(-)$ for bosons
(fermions).

We assume renormalization conditions which ensure that Higgs mass and
VEV remain the same as at the tree level, and that there is no
explicit renormalization scale $\mu_R$ dependence in the effective
potential (up to field independent term which can be cancelled by
shifting potential by a constant, such that $V(\phi=0)=0$). They read
as
\begin{equation}
\label{eq:renorm:condition}
\left.\frac{\partial}{\partial \phi} V_{CW}\right|_{\phi=v}=\left.0
\qquad \frac{\partial^2}{\partial \phi^2}V_{CW}\right|_{\phi=v}=0.
\end{equation}
First line of the equation~\eqref{eq:CW} automatically satisfies
conditions~\eqref{eq:renorm:condition}, whereas $C_1$ and $C_2$ in the
second line are chosen to achieve the same outcome.
%
Temperature corrections to the effective potential are given by:
\begin{align}
V_{T}(H, S, T)=&\sum_{k} \frac{N_{k} T^{4}}{2 \pi^{2}}
J_{r_{k}}\left(M_{k}(H,S) / T\right) \nonumber\\
&+N_F\sum_{i=1, 2} \frac{n_{F_i} T^{4}}{2 \pi^{2}}
J_{-}\left(M_{F_i}(H,S) / T\right),
\end{align}
%
where the thermal functions $J_\pm(y)$ with $y=M/T$ are given by:
\begin{equation}
J_{\pm}(y)=\pm \int_{0}^{\infty} \mathrm{d} x x^{2} \log \left[1 \mp
  e^{-\sqrt{x^{2}+y^{2}}}\right].
\end{equation}
At $T\neq 0$, the field dependent scalar and longitudinal gauge boson
masses get modified with thermal loop effects.  They are added as
$\Pi$ to the field dependent masses~\cite{Weinberg:1974hy,
  Curtin:2016urg},
\begin{eqnarray}
\Pi_H(0) &=& \left( \frac{3g^2}{16} + \frac{g'^2}{16} +
\frac{\lambda}{2} + \frac{y_t^2}{4}+\frac12\sum_F y_F^2 +
\frac{\lambda_{HS}}{24} \right) T^2, \nonumber \\
\Pi_s(0) &=&\left( \frac{1}{6}\lambda_{HS} +
\frac{1}{4} \lambda_S \right)T^2, \nonumber \\
\Pi_{G B}^L(0)&=&\frac{11}{6} T^2 \operatorname{diag}\left(g^2, g^2,
g^2, g^{\prime 2}\right).
\end{eqnarray}

\hskip 1cm
\bibliography{DiHiggs.bib}

\begin{thebibliography}{103}%
\makeatletter
\providecommand \@ifxundefined [1]{%
 \@ifx{#1\undefined}
}%
\providecommand \@ifnum [1]{%
 \ifnum #1\expandafter \@firstoftwo
 \else \expandafter \@secondoftwo
 \fi
}%
\providecommand \@ifx [1]{%
 \ifx #1\expandafter \@firstoftwo
 \else \expandafter \@secondoftwo
 \fi
}%
\providecommand \natexlab [1]{#1}%
\providecommand \enquote  [1]{``#1''}%
\providecommand \bibnamefont  [1]{#1}%
\providecommand \bibfnamefont [1]{#1}%
\providecommand \citenamefont [1]{#1}%
\providecommand \href@noop [0]{\@secondoftwo}%
\providecommand \href [0]{\begingroup \@sanitize@url \@href}%
\providecommand \@href[1]{\@@startlink{#1}\@@href}%
\providecommand \@@href[1]{\endgroup#1\@@endlink}%
\providecommand \@sanitize@url [0]{\catcode `\\12\catcode `\$12\catcode
  `\&12\catcode `\#12\catcode `\^12\catcode `\_12\catcode `\%12\relax}%
\providecommand \@@startlink[1]{}%
\providecommand \@@endlink[0]{}%
\providecommand \url  [0]{\begingroup\@sanitize@url \@url }%
\providecommand \@url [1]{\endgroup\@href {#1}{\urlprefix }}%
\providecommand \urlprefix  [0]{URL }%
\providecommand \Eprint [0]{\href }%
\providecommand \doibase [0]{https://doi.org/}%
\providecommand \selectlanguage [0]{\@gobble}%
\providecommand \bibinfo  [0]{\@secondoftwo}%
\providecommand \bibfield  [0]{\@secondoftwo}%
\providecommand \translation [1]{[#1]}%
\providecommand \BibitemOpen [0]{}%
\providecommand \bibitemStop [0]{}%
\providecommand \bibitemNoStop [0]{.\EOS\space}%
\providecommand \EOS [0]{\spacefactor3000\relax}%
\providecommand \BibitemShut  [1]{\csname bibitem#1\endcsname}%
\let\auto@bib@innerbib\@empty
\bibitem [{\citenamefont {Chatrchyan}\ \emph {et~al.}(2012)\citenamefont
  {Chatrchyan} \emph {et~al.}}]{CMS:2012qbp}%
  \BibitemOpen
  \bibfield  {author} {\bibinfo {author} {\bibfnamefont {S.}~\bibnamefont
  {Chatrchyan}} \emph {et~al.} (\bibinfo {collaboration} {CMS}),\ }\bibfield
  {title} {\bibinfo {title} {{Observation of a New Boson at a Mass of 125 GeV
  with the CMS Experiment at the LHC}},\ }\href
  {https://doi.org/10.1016/j.physletb.2012.08.021} {\bibfield  {journal}
  {\bibinfo  {journal} {Phys. Lett. B}\ }\textbf {\bibinfo {volume} {716}},\
  \bibinfo {pages} {30} (\bibinfo {year} {2012})},\ \Eprint
  {https://arxiv.org/abs/1207.7235} {arXiv:1207.7235 [hep-ex]} \BibitemShut
  {NoStop}%
\bibitem [{\citenamefont {Aad}\ \emph {et~al.}(2012)\citenamefont {Aad} \emph
  {et~al.}}]{ATLAS:2012yve}%
  \BibitemOpen
  \bibfield  {author} {\bibinfo {author} {\bibfnamefont {G.}~\bibnamefont
  {Aad}} \emph {et~al.} (\bibinfo {collaboration} {ATLAS}),\ }\bibfield
  {title} {\bibinfo {title} {{Observation of a new particle in the search for
  the Standard Model Higgs boson with the ATLAS detector at the LHC}},\ }\href
  {https://doi.org/10.1016/j.physletb.2012.08.020} {\bibfield  {journal}
  {\bibinfo  {journal} {Phys. Lett. B}\ }\textbf {\bibinfo {volume} {716}},\
  \bibinfo {pages} {1} (\bibinfo {year} {2012})},\ \Eprint
  {https://arxiv.org/abs/1207.7214} {arXiv:1207.7214 [hep-ex]} \BibitemShut
  {NoStop}%
\bibitem [{\citenamefont {Cheng}\ \emph {et~al.}(2000)\citenamefont {Cheng},
  \citenamefont {Dobrescu},\ and\ \citenamefont {Hill}}]{Cheng:1999bg}%
  \BibitemOpen
  \bibfield  {author} {\bibinfo {author} {\bibfnamefont {H.-C.}\ \bibnamefont
  {Cheng}}, \bibinfo {author} {\bibfnamefont {B.~A.}\ \bibnamefont
  {Dobrescu}},\ and\ \bibinfo {author} {\bibfnamefont {C.~T.}\ \bibnamefont
  {Hill}},\ }\bibfield  {title} {\bibinfo {title} {Electroweak symmetry
  breaking and extra dimensions},\ }\href
  {https://doi.org/10.1016/S0550-3213(00)00401-6} {\bibfield  {journal}
  {\bibinfo  {journal} {Nucl. Phys. B}\ }\textbf {\bibinfo {volume} {589}},\
  \bibinfo {pages} {249} (\bibinfo {year} {2000})},\ \Eprint
  {https://arxiv.org/abs/hep-ph/9912343} {arXiv:hep-ph/9912343} \BibitemShut
  {NoStop}%
\bibitem [{\citenamefont {Arkani-Hamed}\ \emph {et~al.}(2002)\citenamefont
  {Arkani-Hamed}, \citenamefont {Cohen}, \citenamefont {Katz},\ and\
  \citenamefont {Nelson}}]{Arkani-Hamed:2002ikv}%
  \BibitemOpen
  \bibfield  {author} {\bibinfo {author} {\bibfnamefont {N.}~\bibnamefont
  {Arkani-Hamed}}, \bibinfo {author} {\bibfnamefont {A.~G.}\ \bibnamefont
  {Cohen}}, \bibinfo {author} {\bibfnamefont {E.}~\bibnamefont {Katz}},\ and\
  \bibinfo {author} {\bibfnamefont {A.~E.}\ \bibnamefont {Nelson}},\ }\bibfield
   {title} {\bibinfo {title} {{The Littlest Higgs}},\ }\href
  {https://doi.org/10.1088/1126-6708/2002/07/034} {\bibfield  {journal}
  {\bibinfo  {journal} {JHEP}\ }\textbf {\bibinfo {volume} {07}},\ \bibinfo
  {pages} {034}},\ \Eprint {https://arxiv.org/abs/hep-ph/0206021}
  {arXiv:hep-ph/0206021} \BibitemShut {NoStop}%
\bibitem [{\citenamefont {Han}\ \emph {et~al.}(2003)\citenamefont {Han},
  \citenamefont {Logan}, \citenamefont {McElrath},\ and\ \citenamefont
  {Wang}}]{Han:2003gf}%
  \BibitemOpen
  \bibfield  {author} {\bibinfo {author} {\bibfnamefont {T.}~\bibnamefont
  {Han}}, \bibinfo {author} {\bibfnamefont {H.~E.}\ \bibnamefont {Logan}},
  \bibinfo {author} {\bibfnamefont {B.}~\bibnamefont {McElrath}},\ and\
  \bibinfo {author} {\bibfnamefont {L.-T.}\ \bibnamefont {Wang}},\ }\bibfield
  {title} {\bibinfo {title} {{Loop induced decays of the little Higgs: H
  ---\ensuremath{>} gg, gamma gamma}},\ }\href
  {https://doi.org/10.1016/j.physletb.2004.10.021} {\bibfield  {journal}
  {\bibinfo  {journal} {Phys. Lett. B}\ }\textbf {\bibinfo {volume} {563}},\
  \bibinfo {pages} {191} (\bibinfo {year} {2003})},\ \bibinfo {note} {[Erratum:
  Phys.Lett.B 603, 257--259 (2004)]},\ \Eprint
  {https://arxiv.org/abs/hep-ph/0302188} {arXiv:hep-ph/0302188} \BibitemShut
  {NoStop}%
\bibitem [{\citenamefont {Cheng}\ \emph {et~al.}(2006)\citenamefont {Cheng},
  \citenamefont {Low},\ and\ \citenamefont {Wang}}]{Cheng:2005as}%
  \BibitemOpen
  \bibfield  {author} {\bibinfo {author} {\bibfnamefont {H.-C.}\ \bibnamefont
  {Cheng}}, \bibinfo {author} {\bibfnamefont {I.}~\bibnamefont {Low}},\ and\
  \bibinfo {author} {\bibfnamefont {L.-T.}\ \bibnamefont {Wang}},\ }\bibfield
  {title} {\bibinfo {title} {{Top partners in little Higgs theories with
  T-parity}},\ }\href {https://doi.org/10.1103/PhysRevD.74.055001} {\bibfield
  {journal} {\bibinfo  {journal} {Phys. Rev. D}\ }\textbf {\bibinfo {volume}
  {74}},\ \bibinfo {pages} {055001} (\bibinfo {year} {2006})},\ \Eprint
  {https://arxiv.org/abs/hep-ph/0510225} {arXiv:hep-ph/0510225} \BibitemShut
  {NoStop}%
\bibitem [{\citenamefont {Kang}\ \emph {et~al.}(2008)\citenamefont {Kang},
  \citenamefont {Langacker},\ and\ \citenamefont {Nelson}}]{Kang:2007ib}%
  \BibitemOpen
  \bibfield  {author} {\bibinfo {author} {\bibfnamefont {J.}~\bibnamefont
  {Kang}}, \bibinfo {author} {\bibfnamefont {P.}~\bibnamefont {Langacker}},\
  and\ \bibinfo {author} {\bibfnamefont {B.~D.}\ \bibnamefont {Nelson}},\
  }\bibfield  {title} {\bibinfo {title} {{Theory and Phenomenology of Exotic
  Isosinglet Quarks and Squarks}},\ }\href
  {https://doi.org/10.1103/PhysRevD.77.035003} {\bibfield  {journal} {\bibinfo
  {journal} {Phys. Rev. D}\ }\textbf {\bibinfo {volume} {77}},\ \bibinfo
  {pages} {035003} (\bibinfo {year} {2008})},\ \Eprint
  {https://arxiv.org/abs/0708.2701} {arXiv:0708.2701 [hep-ph]} \BibitemShut
  {NoStop}%
\bibitem [{\citenamefont {Cacciapaglia}\ \emph {et~al.}(2019)\citenamefont
  {Cacciapaglia}, \citenamefont {Carvalho}, \citenamefont {Deandrea},
  \citenamefont {Flacke}, \citenamefont {Fuks}, \citenamefont {Majumder},
  \citenamefont {Panizzi},\ and\ \citenamefont {Shao}}]{Cacciapaglia:2018qep}%
  \BibitemOpen
  \bibfield  {author} {\bibinfo {author} {\bibfnamefont {G.}~\bibnamefont
  {Cacciapaglia}}, \bibinfo {author} {\bibfnamefont {A.}~\bibnamefont
  {Carvalho}}, \bibinfo {author} {\bibfnamefont {A.}~\bibnamefont {Deandrea}},
  \bibinfo {author} {\bibfnamefont {T.}~\bibnamefont {Flacke}}, \bibinfo
  {author} {\bibfnamefont {B.}~\bibnamefont {Fuks}}, \bibinfo {author}
  {\bibfnamefont {D.}~\bibnamefont {Majumder}}, \bibinfo {author}
  {\bibfnamefont {L.}~\bibnamefont {Panizzi}},\ and\ \bibinfo {author}
  {\bibfnamefont {H.-S.}\ \bibnamefont {Shao}},\ }\bibfield  {title} {\bibinfo
  {title} {{Next-to-leading-order predictions for single vector-like quark
  production at the LHC}},\ }\href
  {https://doi.org/10.1016/j.physletb.2019.04.056} {\bibfield  {journal}
  {\bibinfo  {journal} {Phys. Lett. B}\ }\textbf {\bibinfo {volume} {793}},\
  \bibinfo {pages} {206} (\bibinfo {year} {2019})},\ \Eprint
  {https://arxiv.org/abs/1811.05055} {arXiv:1811.05055 [hep-ph]} \BibitemShut
  {NoStop}%
\bibitem [{\citenamefont {Cacciapaglia}\ \emph {et~al.}(2012)\citenamefont
  {Cacciapaglia}, \citenamefont {Deandrea}, \citenamefont {Panizzi},
  \citenamefont {Gaur}, \citenamefont {Harada},\ and\ \citenamefont
  {Okada}}]{Cacciapaglia:2011fx}%
  \BibitemOpen
  \bibfield  {author} {\bibinfo {author} {\bibfnamefont {G.}~\bibnamefont
  {Cacciapaglia}}, \bibinfo {author} {\bibfnamefont {A.}~\bibnamefont
  {Deandrea}}, \bibinfo {author} {\bibfnamefont {L.}~\bibnamefont {Panizzi}},
  \bibinfo {author} {\bibfnamefont {N.}~\bibnamefont {Gaur}}, \bibinfo {author}
  {\bibfnamefont {D.}~\bibnamefont {Harada}},\ and\ \bibinfo {author}
  {\bibfnamefont {Y.}~\bibnamefont {Okada}},\ }\bibfield  {title} {\bibinfo
  {title} {{Heavy Vector-like Top Partners at the LHC and flavour
  constraints}},\ }\href {https://doi.org/10.1007/JHEP03(2012)070} {\bibfield
  {journal} {\bibinfo  {journal} {JHEP}\ }\textbf {\bibinfo {volume} {03}},\
  \bibinfo {pages} {070}},\ \Eprint {https://arxiv.org/abs/1108.6329}
  {arXiv:1108.6329 [hep-ph]} \BibitemShut {NoStop}%
\bibitem [{\citenamefont {Aguilar-Saavedra}\ \emph {et~al.}(2013)\citenamefont
  {Aguilar-Saavedra}, \citenamefont {Benbrik}, \citenamefont {Heinemeyer},\
  and\ \citenamefont {P\'erez-Victoria}}]{Aguilar-Saavedra:2013qpa}%
  \BibitemOpen
  \bibfield  {author} {\bibinfo {author} {\bibfnamefont {J.~A.}\ \bibnamefont
  {Aguilar-Saavedra}}, \bibinfo {author} {\bibfnamefont {R.}~\bibnamefont
  {Benbrik}}, \bibinfo {author} {\bibfnamefont {S.}~\bibnamefont
  {Heinemeyer}},\ and\ \bibinfo {author} {\bibfnamefont {M.}~\bibnamefont
  {P\'erez-Victoria}},\ }\bibfield  {title} {\bibinfo {title} {{Handbook of
  vectorlike quarks: Mixing and single production}},\ }\href
  {https://doi.org/10.1103/PhysRevD.88.094010} {\bibfield  {journal} {\bibinfo
  {journal} {Phys. Rev. D}\ }\textbf {\bibinfo {volume} {88}},\ \bibinfo
  {pages} {094010} (\bibinfo {year} {2013})},\ \Eprint
  {https://arxiv.org/abs/1306.0572} {arXiv:1306.0572 [hep-ph]} \BibitemShut
  {NoStop}%
\bibitem [{\citenamefont {Ellis}\ \emph {et~al.}(2014)\citenamefont {Ellis},
  \citenamefont {Godbole}, \citenamefont {Gopalakrishna},\ and\ \citenamefont
  {Wells}}]{Ellis:2014dza}%
  \BibitemOpen
  \bibfield  {author} {\bibinfo {author} {\bibfnamefont {S.~A.~R.}\
  \bibnamefont {Ellis}}, \bibinfo {author} {\bibfnamefont {R.~M.}\ \bibnamefont
  {Godbole}}, \bibinfo {author} {\bibfnamefont {S.}~\bibnamefont
  {Gopalakrishna}},\ and\ \bibinfo {author} {\bibfnamefont {J.~D.}\
  \bibnamefont {Wells}},\ }\bibfield  {title} {\bibinfo {title} {{Survey of
  vector-like fermion extensions of the Standard Model and their
  phenomenological implications}},\ }\href
  {https://doi.org/10.1007/JHEP09(2014)130} {\bibfield  {journal} {\bibinfo
  {journal} {JHEP}\ }\textbf {\bibinfo {volume} {09}},\ \bibinfo {pages}
  {130}},\ \Eprint {https://arxiv.org/abs/1404.4398} {arXiv:1404.4398 [hep-ph]}
  \BibitemShut {NoStop}%
\bibitem [{\citenamefont {Angelescu}\ \emph {et~al.}(2016)\citenamefont
  {Angelescu}, \citenamefont {Djouadi},\ and\ \citenamefont
  {Moreau}}]{Angelescu:2015uiz}%
  \BibitemOpen
  \bibfield  {author} {\bibinfo {author} {\bibfnamefont {A.}~\bibnamefont
  {Angelescu}}, \bibinfo {author} {\bibfnamefont {A.}~\bibnamefont {Djouadi}},\
  and\ \bibinfo {author} {\bibfnamefont {G.}~\bibnamefont {Moreau}},\
  }\bibfield  {title} {\bibinfo {title} {{Scenarii for interpretations of the
  LHC diphoton excess: two Higgs doublets and vector-like quarks and
  leptons}},\ }\href {https://doi.org/10.1016/j.physletb.2016.02.064}
  {\bibfield  {journal} {\bibinfo  {journal} {Phys. Lett. B}\ }\textbf
  {\bibinfo {volume} {756}},\ \bibinfo {pages} {126} (\bibinfo {year}
  {2016})},\ \Eprint {https://arxiv.org/abs/1512.04921} {arXiv:1512.04921
  [hep-ph]} \BibitemShut {NoStop}%
\bibitem [{\citenamefont {Arhrib}\ \emph {et~al.}(2018)\citenamefont {Arhrib},
  \citenamefont {Benbrik}, \citenamefont {King}, \citenamefont {Manaut},
  \citenamefont {Moretti},\ and\ \citenamefont {Un}}]{Arhrib:2016rlj}%
  \BibitemOpen
  \bibfield  {author} {\bibinfo {author} {\bibfnamefont {A.}~\bibnamefont
  {Arhrib}}, \bibinfo {author} {\bibfnamefont {R.}~\bibnamefont {Benbrik}},
  \bibinfo {author} {\bibfnamefont {S.~J.~D.}\ \bibnamefont {King}}, \bibinfo
  {author} {\bibfnamefont {B.}~\bibnamefont {Manaut}}, \bibinfo {author}
  {\bibfnamefont {S.}~\bibnamefont {Moretti}},\ and\ \bibinfo {author}
  {\bibfnamefont {C.~S.}\ \bibnamefont {Un}},\ }\bibfield  {title} {\bibinfo
  {title} {{Phenomenology of 2HDM with vectorlike quarks}},\ }\href
  {https://doi.org/10.1103/PhysRevD.97.095015} {\bibfield  {journal} {\bibinfo
  {journal} {Phys. Rev. D}\ }\textbf {\bibinfo {volume} {97}},\ \bibinfo
  {pages} {095015} (\bibinfo {year} {2018})},\ \Eprint
  {https://arxiv.org/abs/1607.08517} {arXiv:1607.08517 [hep-ph]} \BibitemShut
  {NoStop}%
\bibitem [{\citenamefont {Barducci}\ and\ \citenamefont
  {Panizzi}(2017)}]{Barducci:2017xtw}%
  \BibitemOpen
  \bibfield  {author} {\bibinfo {author} {\bibfnamefont {D.}~\bibnamefont
  {Barducci}}\ and\ \bibinfo {author} {\bibfnamefont {L.}~\bibnamefont
  {Panizzi}},\ }\bibfield  {title} {\bibinfo {title} {{Vector-like quarks
  coupling discrimination at the LHC and future hadron colliders}},\ }\href
  {https://doi.org/10.1007/JHEP12(2017)057} {\bibfield  {journal} {\bibinfo
  {journal} {JHEP}\ }\textbf {\bibinfo {volume} {12}},\ \bibinfo {pages}
  {057}},\ \Eprint {https://arxiv.org/abs/1710.02325} {arXiv:1710.02325
  [hep-ph]} \BibitemShut {NoStop}%
\bibitem [{\citenamefont {Arhrib}\ \emph {et~al.}(2019)\citenamefont {Arhrib},
  \citenamefont {Benbrik}, \citenamefont {El~Falaki}, \citenamefont {Sampaio},\
  and\ \citenamefont {Santos}}]{Arhrib:2018pdi}%
  \BibitemOpen
  \bibfield  {author} {\bibinfo {author} {\bibfnamefont {A.}~\bibnamefont
  {Arhrib}}, \bibinfo {author} {\bibfnamefont {R.}~\bibnamefont {Benbrik}},
  \bibinfo {author} {\bibfnamefont {J.}~\bibnamefont {El~Falaki}}, \bibinfo
  {author} {\bibfnamefont {M.}~\bibnamefont {Sampaio}},\ and\ \bibinfo {author}
  {\bibfnamefont {R.}~\bibnamefont {Santos}},\ }\bibfield  {title} {\bibinfo
  {title} {{Pseudoscalar decays to gauge bosons at the LHC and at a future 100
  TeV collider}},\ }\href {https://doi.org/10.1103/PhysRevD.99.035043}
  {\bibfield  {journal} {\bibinfo  {journal} {Phys. Rev. D}\ }\textbf {\bibinfo
  {volume} {99}},\ \bibinfo {pages} {035043} (\bibinfo {year} {2019})},\
  \Eprint {https://arxiv.org/abs/1809.04805} {arXiv:1809.04805 [hep-ph]}
  \BibitemShut {NoStop}%
\bibitem [{\citenamefont {Song}\ and\ \citenamefont
  {Yoon}(2019)}]{Song:2019aav}%
  \BibitemOpen
  \bibfield  {author} {\bibinfo {author} {\bibfnamefont {J.}~\bibnamefont
  {Song}}\ and\ \bibinfo {author} {\bibfnamefont {Y.~W.}\ \bibnamefont
  {Yoon}},\ }\bibfield  {title} {\bibinfo {title} {{$W\gamma$ decay of the
  elusive charged Higgs boson in the two-Higgs-doublet model with vectorlike
  fermions}},\ }\href {https://doi.org/10.1103/PhysRevD.100.055006} {\bibfield
  {journal} {\bibinfo  {journal} {Phys. Rev. D}\ }\textbf {\bibinfo {volume}
  {100}},\ \bibinfo {pages} {055006} (\bibinfo {year} {2019})},\ \Eprint
  {https://arxiv.org/abs/1904.06521} {arXiv:1904.06521 [hep-ph]} \BibitemShut
  {NoStop}%
\bibitem [{\citenamefont {del Aguila}\ \emph {et~al.}(1990)\citenamefont {del
  Aguila}, \citenamefont {Ametller}, \citenamefont {Kane},\ and\ \citenamefont
  {Vidal}}]{delAguila:1989rq}%
  \BibitemOpen
  \bibfield  {author} {\bibinfo {author} {\bibfnamefont {F.}~\bibnamefont {del
  Aguila}}, \bibinfo {author} {\bibfnamefont {L.}~\bibnamefont {Ametller}},
  \bibinfo {author} {\bibfnamefont {G.~L.}\ \bibnamefont {Kane}},\ and\
  \bibinfo {author} {\bibfnamefont {J.}~\bibnamefont {Vidal}},\ }\bibfield
  {title} {\bibinfo {title} {{Vector Like Fermion and Standard Higgs Production
  at Hadron Colliders}},\ }\href {https://doi.org/10.1016/0550-3213(90)90655-W}
  {\bibfield  {journal} {\bibinfo  {journal} {Nucl. Phys. B}\ }\textbf
  {\bibinfo {volume} {334}},\ \bibinfo {pages} {1} (\bibinfo {year}
  {1990})}\BibitemShut {NoStop}%
\bibitem [{\citenamefont {Bhattacharya}\ \emph {et~al.}(2022)\citenamefont
  {Bhattacharya}, \citenamefont {Jahedi},\ and\ \citenamefont
  {Wudka}}]{Bhattacharya:2021ltd}%
  \BibitemOpen
  \bibfield  {author} {\bibinfo {author} {\bibfnamefont {S.}~\bibnamefont
  {Bhattacharya}}, \bibinfo {author} {\bibfnamefont {S.}~\bibnamefont
  {Jahedi}},\ and\ \bibinfo {author} {\bibfnamefont {J.}~\bibnamefont
  {Wudka}},\ }\bibfield  {title} {\bibinfo {title} {{Probing heavy charged
  fermions at $e^{+}e^{-}$ collider using the optimal observable technique}},\
  }\href {https://doi.org/10.1007/JHEP05(2022)009} {\bibfield  {journal}
  {\bibinfo  {journal} {JHEP}\ }\textbf {\bibinfo {volume} {05}},\ \bibinfo
  {pages} {009}},\ \Eprint {https://arxiv.org/abs/2106.02846} {arXiv:2106.02846
  [hep-ph]} \BibitemShut {NoStop}%
\bibitem [{\citenamefont {C\'arcamo~Hern\'andez}\ \emph
  {et~al.}(2023)\citenamefont {C\'arcamo~Hern\'andez}, \citenamefont
  {Kowalska}, \citenamefont {Lee},\ and\ \citenamefont
  {Rizzo}}]{CarcamoHernandez:2023wzf}%
  \BibitemOpen
  \bibfield  {author} {\bibinfo {author} {\bibfnamefont {A.~E.}\ \bibnamefont
  {C\'arcamo~Hern\'andez}}, \bibinfo {author} {\bibfnamefont {K.}~\bibnamefont
  {Kowalska}}, \bibinfo {author} {\bibfnamefont {H.}~\bibnamefont {Lee}},\ and\
  \bibinfo {author} {\bibfnamefont {D.}~\bibnamefont {Rizzo}},\ }\href@noop {}
  {\bibinfo {title} {{Global analysis and LHC study of a vector-like extension
  of the Standard Model with extra scalars}}} (\bibinfo {year} {2023}),\
  \Eprint {https://arxiv.org/abs/2309.13968} {arXiv:2309.13968 [hep-ph]}
  \BibitemShut {NoStop}%
\bibitem [{\citenamefont {Blum}\ \emph {et~al.}(2015)\citenamefont {Blum},
  \citenamefont {D'Agnolo},\ and\ \citenamefont {Fan}}]{Blum:2015rpa}%
  \BibitemOpen
  \bibfield  {author} {\bibinfo {author} {\bibfnamefont {K.}~\bibnamefont
  {Blum}}, \bibinfo {author} {\bibfnamefont {R.~T.}\ \bibnamefont {D'Agnolo}},\
  and\ \bibinfo {author} {\bibfnamefont {J.}~\bibnamefont {Fan}},\ }\bibfield
  {title} {\bibinfo {title} {{Vacuum stability bounds on Higgs coupling
  deviations in the absence of new bosons}},\ }\href
  {https://doi.org/10.1007/JHEP03(2015)166} {\bibfield  {journal} {\bibinfo
  {journal} {JHEP}\ }\textbf {\bibinfo {volume} {03}},\ \bibinfo {pages}
  {166}},\ \Eprint {https://arxiv.org/abs/1502.01045} {arXiv:1502.01045
  [hep-ph]} \BibitemShut {NoStop}%
\bibitem [{\citenamefont {Gopalakrishna}\ and\ \citenamefont
  {Velusamy}(2019)}]{Gopalakrishna:2018uxn}%
  \BibitemOpen
  \bibfield  {author} {\bibinfo {author} {\bibfnamefont {S.}~\bibnamefont
  {Gopalakrishna}}\ and\ \bibinfo {author} {\bibfnamefont {A.}~\bibnamefont
  {Velusamy}},\ }\bibfield  {title} {\bibinfo {title} {{Higgs vacuum stability
  with vectorlike fermions}},\ }\href
  {https://doi.org/10.1103/PhysRevD.99.115020} {\bibfield  {journal} {\bibinfo
  {journal} {Phys. Rev. D}\ }\textbf {\bibinfo {volume} {99}},\ \bibinfo
  {pages} {115020} (\bibinfo {year} {2019})},\ \Eprint
  {https://arxiv.org/abs/1812.11303} {arXiv:1812.11303 [hep-ph]} \BibitemShut
  {NoStop}%
\bibitem [{\citenamefont {Arsenault}\ \emph {et~al.}(2023)\citenamefont
  {Arsenault}, \citenamefont {Cingiloglu},\ and\ \citenamefont
  {Frank}}]{Arsenault:2022xty}%
  \BibitemOpen
  \bibfield  {author} {\bibinfo {author} {\bibfnamefont {A.}~\bibnamefont
  {Arsenault}}, \bibinfo {author} {\bibfnamefont {K.~Y.}\ \bibnamefont
  {Cingiloglu}},\ and\ \bibinfo {author} {\bibfnamefont {M.}~\bibnamefont
  {Frank}},\ }\bibfield  {title} {\bibinfo {title} {{Vacuum stability in the
  Standard Model with vectorlike fermions}},\ }\href
  {https://doi.org/10.1103/PhysRevD.107.036018} {\bibfield  {journal} {\bibinfo
   {journal} {Phys. Rev. D}\ }\textbf {\bibinfo {volume} {107}},\ \bibinfo
  {pages} {036018} (\bibinfo {year} {2023})},\ \Eprint
  {https://arxiv.org/abs/2207.10332} {arXiv:2207.10332 [hep-ph]} \BibitemShut
  {NoStop}%
\bibitem [{\citenamefont {Hiller}\ \emph {et~al.}(2022)\citenamefont {Hiller},
  \citenamefont {H\"ohne}, \citenamefont {Litim},\ and\ \citenamefont
  {Steudtner}}]{Hiller:2022rla}%
  \BibitemOpen
  \bibfield  {author} {\bibinfo {author} {\bibfnamefont {G.}~\bibnamefont
  {Hiller}}, \bibinfo {author} {\bibfnamefont {T.}~\bibnamefont {H\"ohne}},
  \bibinfo {author} {\bibfnamefont {D.~F.}\ \bibnamefont {Litim}},\ and\
  \bibinfo {author} {\bibfnamefont {T.}~\bibnamefont {Steudtner}},\ }\bibfield
  {title} {\bibinfo {title} {{Portals into Higgs vacuum stability}},\ }\href
  {https://doi.org/10.1103/PhysRevD.106.115004} {\bibfield  {journal} {\bibinfo
   {journal} {Phys. Rev. D}\ }\textbf {\bibinfo {volume} {106}},\ \bibinfo
  {pages} {115004} (\bibinfo {year} {2022})},\ \Eprint
  {https://arxiv.org/abs/2207.07737} {arXiv:2207.07737 [hep-ph]} \BibitemShut
  {NoStop}%
\bibitem [{\citenamefont {Hiller}\ \emph {et~al.}(2023)\citenamefont {Hiller},
  \citenamefont {H\"ohne}, \citenamefont {Litim},\ and\ \citenamefont
  {Steudtner}}]{Hiller:2023bdb}%
  \BibitemOpen
  \bibfield  {author} {\bibinfo {author} {\bibfnamefont {G.}~\bibnamefont
  {Hiller}}, \bibinfo {author} {\bibfnamefont {T.}~\bibnamefont {H\"ohne}},
  \bibinfo {author} {\bibfnamefont {D.~F.}\ \bibnamefont {Litim}},\ and\
  \bibinfo {author} {\bibfnamefont {T.}~\bibnamefont {Steudtner}},\ }\bibfield
  {title} {\bibinfo {title} {{Vacuum Stability as a Guide for Model Bulding}}\
  }(\bibinfo  {publisher} {{57th Rencontres de Moriond on Electroweak
  Interactions and Unified Theories}},\ \bibinfo {year} {2023})\ \Eprint
  {https://arxiv.org/abs/2305.18520} {arXiv:2305.18520 [hep-ph]} \BibitemShut
  {NoStop}%
\bibitem [{\citenamefont {Cacciapaglia}\ \emph {et~al.}(2017)\citenamefont
  {Cacciapaglia}, \citenamefont {Cai}, \citenamefont {Carvalho}, \citenamefont
  {Deandrea}, \citenamefont {Flacke}, \citenamefont {Fuks}, \citenamefont
  {Majumder},\ and\ \citenamefont {Shao}}]{Cacciapaglia:2017gzh}%
  \BibitemOpen
  \bibfield  {author} {\bibinfo {author} {\bibfnamefont {G.}~\bibnamefont
  {Cacciapaglia}}, \bibinfo {author} {\bibfnamefont {H.}~\bibnamefont {Cai}},
  \bibinfo {author} {\bibfnamefont {A.}~\bibnamefont {Carvalho}}, \bibinfo
  {author} {\bibfnamefont {A.}~\bibnamefont {Deandrea}}, \bibinfo {author}
  {\bibfnamefont {T.}~\bibnamefont {Flacke}}, \bibinfo {author} {\bibfnamefont
  {B.}~\bibnamefont {Fuks}}, \bibinfo {author} {\bibfnamefont {D.}~\bibnamefont
  {Majumder}},\ and\ \bibinfo {author} {\bibfnamefont {H.-S.}\ \bibnamefont
  {Shao}},\ }\bibfield  {title} {\bibinfo {title} {{Probing vector-like quark
  models with Higgs-boson pair production}},\ }\href
  {https://doi.org/10.1007/JHEP07(2017)005} {\bibfield  {journal} {\bibinfo
  {journal} {JHEP}\ }\textbf {\bibinfo {volume} {07}},\ \bibinfo {pages}
  {005}},\ \Eprint {https://arxiv.org/abs/1703.10614} {arXiv:1703.10614
  [hep-ph]} \BibitemShut {NoStop}%
\bibitem [{\citenamefont {Cheung}\ \emph {et~al.}(2021)\citenamefont {Cheung},
  \citenamefont {Jueid}, \citenamefont {Lu}, \citenamefont {Song},\ and\
  \citenamefont {Yoon}}]{Cheung:2020xij}%
  \BibitemOpen
  \bibfield  {author} {\bibinfo {author} {\bibfnamefont {K.}~\bibnamefont
  {Cheung}}, \bibinfo {author} {\bibfnamefont {A.}~\bibnamefont {Jueid}},
  \bibinfo {author} {\bibfnamefont {C.-T.}\ \bibnamefont {Lu}}, \bibinfo
  {author} {\bibfnamefont {J.}~\bibnamefont {Song}},\ and\ \bibinfo {author}
  {\bibfnamefont {Y.~W.}\ \bibnamefont {Yoon}},\ }\bibfield  {title} {\bibinfo
  {title} {{Disentangling new physics effects on nonresonant Higgs boson pair
  production from gluon fusion}},\ }\href
  {https://doi.org/10.1103/PhysRevD.103.015019} {\bibfield  {journal} {\bibinfo
   {journal} {Phys. Rev. D}\ }\textbf {\bibinfo {volume} {103}},\ \bibinfo
  {pages} {015019} (\bibinfo {year} {2021})},\ \Eprint
  {https://arxiv.org/abs/2003.11043} {arXiv:2003.11043 [hep-ph]} \BibitemShut
  {NoStop}%
\bibitem [{\citenamefont {Egana-Ugrinovic}(2017)}]{Egana-Ugrinovic:2017jib}%
  \BibitemOpen
  \bibfield  {author} {\bibinfo {author} {\bibfnamefont {D.}~\bibnamefont
  {Egana-Ugrinovic}},\ }\bibfield  {title} {\bibinfo {title} {{The minimal
  fermionic model of electroweak baryogenesis}},\ }\href
  {https://doi.org/10.1007/JHEP12(2017)064} {\bibfield  {journal} {\bibinfo
  {journal} {JHEP}\ }\textbf {\bibinfo {volume} {12}},\ \bibinfo {pages}
  {064}},\ \Eprint {https://arxiv.org/abs/1707.02306} {arXiv:1707.02306
  [hep-ph]} \BibitemShut {NoStop}%
\bibitem [{\citenamefont {Bell}\ \emph {et~al.}(2019)\citenamefont {Bell},
  \citenamefont {Dolan}, \citenamefont {Friedrich}, \citenamefont
  {Ramsey-Musolf},\ and\ \citenamefont {Volkas}}]{Bell:2019mbn}%
  \BibitemOpen
  \bibfield  {author} {\bibinfo {author} {\bibfnamefont {N.~F.}\ \bibnamefont
  {Bell}}, \bibinfo {author} {\bibfnamefont {M.~J.}\ \bibnamefont {Dolan}},
  \bibinfo {author} {\bibfnamefont {L.~S.}\ \bibnamefont {Friedrich}}, \bibinfo
  {author} {\bibfnamefont {M.~J.}\ \bibnamefont {Ramsey-Musolf}},\ and\
  \bibinfo {author} {\bibfnamefont {R.~R.}\ \bibnamefont {Volkas}},\ }\bibfield
   {title} {\bibinfo {title} {{Electroweak Baryogenesis with Vector-like
  Leptons and Scalar Singlets}},\ }\href
  {https://doi.org/10.1007/JHEP09(2019)012} {\bibfield  {journal} {\bibinfo
  {journal} {JHEP}\ }\textbf {\bibinfo {volume} {09}},\ \bibinfo {pages}
  {012}},\ \Eprint {https://arxiv.org/abs/1903.11255} {arXiv:1903.11255
  [hep-ph]} \BibitemShut {NoStop}%
\bibitem [{\citenamefont {Davoudiasl}\ \emph {et~al.}(2013)\citenamefont
  {Davoudiasl}, \citenamefont {Lewis},\ and\ \citenamefont
  {Ponton}}]{Davoudiasl:2012tu}%
  \BibitemOpen
  \bibfield  {author} {\bibinfo {author} {\bibfnamefont {H.}~\bibnamefont
  {Davoudiasl}}, \bibinfo {author} {\bibfnamefont {I.}~\bibnamefont {Lewis}},\
  and\ \bibinfo {author} {\bibfnamefont {E.}~\bibnamefont {Ponton}},\
  }\bibfield  {title} {\bibinfo {title} {{Electroweak Phase Transition, Higgs
  Diphoton Rate, and New Heavy Fermions}},\ }\href
  {https://doi.org/10.1103/PhysRevD.87.093001} {\bibfield  {journal} {\bibinfo
  {journal} {Phys. Rev. D}\ }\textbf {\bibinfo {volume} {87}},\ \bibinfo
  {pages} {093001} (\bibinfo {year} {2013})},\ \Eprint
  {https://arxiv.org/abs/1211.3449} {arXiv:1211.3449 [hep-ph]} \BibitemShut
  {NoStop}%
\bibitem [{\citenamefont {Fairbairn}\ and\ \citenamefont
  {Grothaus}(2013)}]{Fairbairn:2013xaa}%
  \BibitemOpen
  \bibfield  {author} {\bibinfo {author} {\bibfnamefont {M.}~\bibnamefont
  {Fairbairn}}\ and\ \bibinfo {author} {\bibfnamefont {P.}~\bibnamefont
  {Grothaus}},\ }\bibfield  {title} {\bibinfo {title} {{Baryogenesis and Dark
  Matter with Vector-like Fermions}},\ }\href
  {https://doi.org/10.1007/JHEP10(2013)176} {\bibfield  {journal} {\bibinfo
  {journal} {JHEP}\ }\textbf {\bibinfo {volume} {10}},\ \bibinfo {pages}
  {176}},\ \Eprint {https://arxiv.org/abs/1307.8011} {arXiv:1307.8011 [hep-ph]}
  \BibitemShut {NoStop}%
\bibitem [{\citenamefont {Angelescu}\ and\ \citenamefont
  {Huang}(2019)}]{Angelescu:2018dkk}%
  \BibitemOpen
  \bibfield  {author} {\bibinfo {author} {\bibfnamefont {A.}~\bibnamefont
  {Angelescu}}\ and\ \bibinfo {author} {\bibfnamefont {P.}~\bibnamefont
  {Huang}},\ }\bibfield  {title} {\bibinfo {title} {{Multistep Strongly First
  Order Phase Transitions from New Fermions at the TeV Scale}},\ }\href
  {https://doi.org/10.1103/PhysRevD.99.055023} {\bibfield  {journal} {\bibinfo
  {journal} {Phys. Rev. D}\ }\textbf {\bibinfo {volume} {99}},\ \bibinfo
  {pages} {055023} (\bibinfo {year} {2019})},\ \Eprint
  {https://arxiv.org/abs/1812.08293} {arXiv:1812.08293 [hep-ph]} \BibitemShut
  {NoStop}%
\bibitem [{\citenamefont {Chao}\ and\ \citenamefont
  {Ramsey-Musolf}(2014)}]{Chao:2014dpa}%
  \BibitemOpen
  \bibfield  {author} {\bibinfo {author} {\bibfnamefont {W.}~\bibnamefont
  {Chao}}\ and\ \bibinfo {author} {\bibfnamefont {M.~J.}\ \bibnamefont
  {Ramsey-Musolf}},\ }\bibfield  {title} {\bibinfo {title} {{Electroweak
  Baryogenesis, Electric Dipole Moments, and Higgs Diphoton Decays}},\ }\href
  {https://doi.org/10.1007/JHEP10(2014)180} {\bibfield  {journal} {\bibinfo
  {journal} {JHEP}\ }\textbf {\bibinfo {volume} {10}},\ \bibinfo {pages}
  {180}},\ \Eprint {https://arxiv.org/abs/1406.0517} {arXiv:1406.0517 [hep-ph]}
  \BibitemShut {NoStop}%
\bibitem [{\citenamefont {Cao}\ \emph {et~al.}(2022)\citenamefont {Cao},
  \citenamefont {Hashino}, \citenamefont {Li}, \citenamefont {Ren},\ and\
  \citenamefont {Yu}}]{Cao:2021yau}%
  \BibitemOpen
  \bibfield  {author} {\bibinfo {author} {\bibfnamefont {Q.-H.}\ \bibnamefont
  {Cao}}, \bibinfo {author} {\bibfnamefont {K.}~\bibnamefont {Hashino}},
  \bibinfo {author} {\bibfnamefont {X.-X.}\ \bibnamefont {Li}}, \bibinfo
  {author} {\bibfnamefont {Z.}~\bibnamefont {Ren}},\ and\ \bibinfo {author}
  {\bibfnamefont {J.-H.}\ \bibnamefont {Yu}},\ }\bibfield  {title} {\bibinfo
  {title} {{Electroweak phase transition triggered by fermion sector}},\ }\href
  {https://doi.org/10.1007/JHEP01(2022)001} {\bibfield  {journal} {\bibinfo
  {journal} {JHEP}\ }\textbf {\bibinfo {volume} {01}},\ \bibinfo {pages}
  {001}},\ \Eprint {https://arxiv.org/abs/2103.05688} {arXiv:2103.05688
  [hep-ph]} \BibitemShut {NoStop}%
\bibitem [{\citenamefont {Matsedonskyi}\ and\ \citenamefont
  {Servant}(2020)}]{Matsedonskyi:2020mlz}%
  \BibitemOpen
  \bibfield  {author} {\bibinfo {author} {\bibfnamefont {O.}~\bibnamefont
  {Matsedonskyi}}\ and\ \bibinfo {author} {\bibfnamefont {G.}~\bibnamefont
  {Servant}},\ }\bibfield  {title} {\bibinfo {title} {{High-Temperature
  Electroweak Symmetry Non-Restoration from New Fermions and Implications for
  Baryogenesis}},\ }\href {https://doi.org/10.1007/JHEP09(2020)012} {\bibfield
  {journal} {\bibinfo  {journal} {JHEP}\ }\textbf {\bibinfo {volume} {09}},\
  \bibinfo {pages} {012}},\ \Eprint {https://arxiv.org/abs/2002.05174}
  {arXiv:2002.05174 [hep-ph]} \BibitemShut {NoStop}%
\bibitem [{\citenamefont {Baldes}\ \emph {et~al.}(2018)\citenamefont {Baldes},
  \citenamefont {Konstandin},\ and\ \citenamefont {Servant}}]{Baldes:2016rqn}%
  \BibitemOpen
  \bibfield  {author} {\bibinfo {author} {\bibfnamefont {I.}~\bibnamefont
  {Baldes}}, \bibinfo {author} {\bibfnamefont {T.}~\bibnamefont {Konstandin}},\
  and\ \bibinfo {author} {\bibfnamefont {G.}~\bibnamefont {Servant}},\
  }\bibfield  {title} {\bibinfo {title} {{A first-order electroweak phase
  transition from varying Yukawas}},\ }\href
  {https://doi.org/10.1016/j.physletb.2018.10.015} {\bibfield  {journal}
  {\bibinfo  {journal} {Phys. Lett. B}\ }\textbf {\bibinfo {volume} {786}},\
  \bibinfo {pages} {373} (\bibinfo {year} {2018})},\ \Eprint
  {https://arxiv.org/abs/1604.04526} {arXiv:1604.04526 [hep-ph]} \BibitemShut
  {NoStop}%
\bibitem [{\citenamefont {Carena}\ \emph {et~al.}(2005)\citenamefont {Carena},
  \citenamefont {Megevand}, \citenamefont {Quiros},\ and\ \citenamefont
  {Wagner}}]{Carena:2004ha}%
  \BibitemOpen
  \bibfield  {author} {\bibinfo {author} {\bibfnamefont {M.}~\bibnamefont
  {Carena}}, \bibinfo {author} {\bibfnamefont {A.}~\bibnamefont {Megevand}},
  \bibinfo {author} {\bibfnamefont {M.}~\bibnamefont {Quiros}},\ and\ \bibinfo
  {author} {\bibfnamefont {C.~E.~M.}\ \bibnamefont {Wagner}},\ }\bibfield
  {title} {\bibinfo {title} {{Electroweak baryogenesis and new TeV fermions}},\
  }\href {https://doi.org/10.1016/j.nuclphysb.2005.03.025} {\bibfield
  {journal} {\bibinfo  {journal} {Nucl. Phys. B}\ }\textbf {\bibinfo {volume}
  {716}},\ \bibinfo {pages} {319} (\bibinfo {year} {2005})},\ \Eprint
  {https://arxiv.org/abs/hep-ph/0410352} {arXiv:hep-ph/0410352} \BibitemShut
  {NoStop}%
\bibitem [{\citenamefont {Poh}\ and\ \citenamefont {Raby}(2017)}]{Poh:2017tfo}%
  \BibitemOpen
  \bibfield  {author} {\bibinfo {author} {\bibfnamefont {Z.}~\bibnamefont
  {Poh}}\ and\ \bibinfo {author} {\bibfnamefont {S.}~\bibnamefont {Raby}},\
  }\bibfield  {title} {\bibinfo {title} {{Vectorlike leptons: Muon g-2 anomaly,
  lepton flavor violation, Higgs boson decays, and lepton nonuniversality}},\
  }\href {https://doi.org/10.1103/PhysRevD.96.015032} {\bibfield  {journal}
  {\bibinfo  {journal} {Phys. Rev. D}\ }\textbf {\bibinfo {volume} {96}},\
  \bibinfo {pages} {015032} (\bibinfo {year} {2017})},\ \Eprint
  {https://arxiv.org/abs/1705.07007} {arXiv:1705.07007 [hep-ph]} \BibitemShut
  {NoStop}%
\bibitem [{\citenamefont {Crivellin}\ \emph {et~al.}(2018)\citenamefont
  {Crivellin}, \citenamefont {Hoferichter},\ and\ \citenamefont
  {Schmidt-Wellenburg}}]{Crivellin:2018qmi}%
  \BibitemOpen
  \bibfield  {author} {\bibinfo {author} {\bibfnamefont {A.}~\bibnamefont
  {Crivellin}}, \bibinfo {author} {\bibfnamefont {M.}~\bibnamefont
  {Hoferichter}},\ and\ \bibinfo {author} {\bibfnamefont {P.}~\bibnamefont
  {Schmidt-Wellenburg}},\ }\bibfield  {title} {\bibinfo {title} {{Combined
  explanations of $(g-2)_{\mu,e}$ and implications for a large muon EDM}},\
  }\href {https://doi.org/10.1103/PhysRevD.98.113002} {\bibfield  {journal}
  {\bibinfo  {journal} {Phys. Rev. D}\ }\textbf {\bibinfo {volume} {98}},\
  \bibinfo {pages} {113002} (\bibinfo {year} {2018})},\ \Eprint
  {https://arxiv.org/abs/1807.11484} {arXiv:1807.11484 [hep-ph]} \BibitemShut
  {NoStop}%
\bibitem [{\citenamefont {Athron}\ \emph {et~al.}(2021)\citenamefont {Athron},
  \citenamefont {Bal\'azs}, \citenamefont {Jacob}, \citenamefont {Kotlarski},
  \citenamefont {St\"ockinger},\ and\ \citenamefont
  {St\"ockinger-Kim}}]{Athron:2021iuf}%
  \BibitemOpen
  \bibfield  {author} {\bibinfo {author} {\bibfnamefont {P.}~\bibnamefont
  {Athron}}, \bibinfo {author} {\bibfnamefont {C.}~\bibnamefont {Bal\'azs}},
  \bibinfo {author} {\bibfnamefont {D.~H.}\ \bibnamefont {Jacob}}, \bibinfo
  {author} {\bibfnamefont {W.}~\bibnamefont {Kotlarski}}, \bibinfo {author}
  {\bibfnamefont {D.}~\bibnamefont {St\"ockinger}},\ and\ \bibinfo {author}
  {\bibfnamefont {H.}~\bibnamefont {St\"ockinger-Kim}},\ }\bibfield  {title}
  {\bibinfo {title} {{New physics explanations of $a_\mu$ in light of the FNAL
  muon $g-2$ measurement}},\ }\href {https://doi.org/10.1007/JHEP09(2021)080}
  {\bibfield  {journal} {\bibinfo  {journal} {JHEP}\ }\textbf {\bibinfo
  {volume} {09}},\ \bibinfo {pages} {080}},\ \Eprint
  {https://arxiv.org/abs/2104.03691} {arXiv:2104.03691 [hep-ph]} \BibitemShut
  {NoStop}%
\bibitem [{\citenamefont {Abi}\ \emph {et~al.}(2021)\citenamefont {Abi} \emph
  {et~al.}}]{Muong-2:2021ojo}%
  \BibitemOpen
  \bibfield  {author} {\bibinfo {author} {\bibfnamefont {B.}~\bibnamefont
  {Abi}} \emph {et~al.} (\bibinfo {collaboration} {Muon g-2}),\ }\bibfield
  {title} {\bibinfo {title} {{Measurement of the Positive Muon Anomalous
  Magnetic Moment to 0.46 ppm}},\ }\href
  {https://doi.org/10.1103/PhysRevLett.126.141801} {\bibfield  {journal}
  {\bibinfo  {journal} {Phys. Rev. Lett.}\ }\textbf {\bibinfo {volume} {126}},\
  \bibinfo {pages} {141801} (\bibinfo {year} {2021})},\ \Eprint
  {https://arxiv.org/abs/2104.03281} {arXiv:2104.03281 [hep-ex]} \BibitemShut
  {NoStop}%
\bibitem [{\citenamefont {Hiller}\ \emph
  {et~al.}(2020{\natexlab{a}})\citenamefont {Hiller}, \citenamefont
  {Hormigos-Feliu}, \citenamefont {Litim},\ and\ \citenamefont
  {Steudtner}}]{Hiller:2019mou}%
  \BibitemOpen
  \bibfield  {author} {\bibinfo {author} {\bibfnamefont {G.}~\bibnamefont
  {Hiller}}, \bibinfo {author} {\bibfnamefont {C.}~\bibnamefont
  {Hormigos-Feliu}}, \bibinfo {author} {\bibfnamefont {D.~F.}\ \bibnamefont
  {Litim}},\ and\ \bibinfo {author} {\bibfnamefont {T.}~\bibnamefont
  {Steudtner}},\ }\bibfield  {title} {\bibinfo {title} {{Anomalous magnetic
  moments from asymptotic safety}},\ }\href
  {https://doi.org/10.1103/PhysRevD.102.071901} {\bibfield  {journal} {\bibinfo
   {journal} {Phys. Rev. D}\ }\textbf {\bibinfo {volume} {102}},\ \bibinfo
  {pages} {071901} (\bibinfo {year} {2020}{\natexlab{a}})},\ \Eprint
  {https://arxiv.org/abs/1910.14062} {arXiv:1910.14062 [hep-ph]} \BibitemShut
  {NoStop}%
\bibitem [{\citenamefont {Hiller}\ \emph
  {et~al.}(2020{\natexlab{b}})\citenamefont {Hiller}, \citenamefont
  {Hormigos-Feliu}, \citenamefont {Litim},\ and\ \citenamefont
  {Steudtner}}]{Hiller:2020fbu}%
  \BibitemOpen
  \bibfield  {author} {\bibinfo {author} {\bibfnamefont {G.}~\bibnamefont
  {Hiller}}, \bibinfo {author} {\bibfnamefont {C.}~\bibnamefont
  {Hormigos-Feliu}}, \bibinfo {author} {\bibfnamefont {D.~F.}\ \bibnamefont
  {Litim}},\ and\ \bibinfo {author} {\bibfnamefont {T.}~\bibnamefont
  {Steudtner}},\ }\bibfield  {title} {\bibinfo {title} {{Model Building from
  Asymptotic Safety with Higgs and Flavor Portals}},\ }\href
  {https://doi.org/10.1103/PhysRevD.102.095023} {\bibfield  {journal} {\bibinfo
   {journal} {Phys. Rev. D}\ }\textbf {\bibinfo {volume} {102}},\ \bibinfo
  {pages} {095023} (\bibinfo {year} {2020}{\natexlab{b}})},\ \Eprint
  {https://arxiv.org/abs/2008.08606} {arXiv:2008.08606 [hep-ph]} \BibitemShut
  {NoStop}%
\bibitem [{\citenamefont {Hiller}\ \emph {et~al.}(2019)\citenamefont {Hiller},
  \citenamefont {Hormigos-Feliu}, \citenamefont {Litim},\ and\ \citenamefont
  {Steudtner}}]{Hiller:2019tvg}%
  \BibitemOpen
  \bibfield  {author} {\bibinfo {author} {\bibfnamefont {G.}~\bibnamefont
  {Hiller}}, \bibinfo {author} {\bibfnamefont {C.}~\bibnamefont
  {Hormigos-Feliu}}, \bibinfo {author} {\bibfnamefont {D.~F.}\ \bibnamefont
  {Litim}},\ and\ \bibinfo {author} {\bibfnamefont {T.}~\bibnamefont
  {Steudtner}},\ }\bibfield  {title} {\bibinfo {title} {{Asymptotically safe
  extensions of the Standard Model with flavour phenomenology}}\ }(\bibinfo
  {publisher} {{54th Rencontres de Moriond on Electroweak Interactions and
  Unified Theories}},\ \bibinfo {year} {2019})\ pp.\ \bibinfo {pages}
  {415--418},\ \Eprint {https://arxiv.org/abs/1905.11020} {arXiv:1905.11020
  [hep-ph]} \BibitemShut {NoStop}%
\bibitem [{\citenamefont {Dermisek}(2013)}]{Dermisek:2012ke}%
  \BibitemOpen
  \bibfield  {author} {\bibinfo {author} {\bibfnamefont {R.}~\bibnamefont
  {Dermisek}},\ }\bibfield  {title} {\bibinfo {title} {{Unification of gauge
  couplings in the standard model with extra vectorlike families}},\ }\href
  {https://doi.org/10.1103/PhysRevD.87.055008} {\bibfield  {journal} {\bibinfo
  {journal} {Phys. Rev. D}\ }\textbf {\bibinfo {volume} {87}},\ \bibinfo
  {pages} {055008} (\bibinfo {year} {2013})},\ \Eprint
  {https://arxiv.org/abs/1212.3035} {arXiv:1212.3035 [hep-ph]} \BibitemShut
  {NoStop}%
\bibitem [{\citenamefont {Bhattacherjee}\ \emph {et~al.}(2018)\citenamefont
  {Bhattacherjee}, \citenamefont {Byakti}, \citenamefont {Kushwaha},\ and\
  \citenamefont {Vempati}}]{Bhattacherjee:2017cxh}%
  \BibitemOpen
  \bibfield  {author} {\bibinfo {author} {\bibfnamefont {B.}~\bibnamefont
  {Bhattacherjee}}, \bibinfo {author} {\bibfnamefont {P.}~\bibnamefont
  {Byakti}}, \bibinfo {author} {\bibfnamefont {A.}~\bibnamefont {Kushwaha}},\
  and\ \bibinfo {author} {\bibfnamefont {S.~K.}\ \bibnamefont {Vempati}},\
  }\bibfield  {title} {\bibinfo {title} {{Unification with Vector-like fermions
  and signals at LHC}},\ }\href {https://doi.org/10.1007/JHEP05(2018)090}
  {\bibfield  {journal} {\bibinfo  {journal} {JHEP}\ }\textbf {\bibinfo
  {volume} {05}},\ \bibinfo {pages} {090}},\ \Eprint
  {https://arxiv.org/abs/1702.06417} {arXiv:1702.06417 [hep-ph]} \BibitemShut
  {NoStop}%
\bibitem [{\citenamefont {Emmanuel-Costa}\ and\ \citenamefont
  {Gonzalez~Felipe}(2005)}]{Emmanuel-Costa:2005qsv}%
  \BibitemOpen
  \bibfield  {author} {\bibinfo {author} {\bibfnamefont {D.}~\bibnamefont
  {Emmanuel-Costa}}\ and\ \bibinfo {author} {\bibfnamefont {R.}~\bibnamefont
  {Gonzalez~Felipe}},\ }\bibfield  {title} {\bibinfo {title} {{Minimal
  string-scale unification of gauge couplings}},\ }\href
  {https://doi.org/10.1016/j.physletb.2005.07.038} {\bibfield  {journal}
  {\bibinfo  {journal} {Phys. Lett. B}\ }\textbf {\bibinfo {volume} {623}},\
  \bibinfo {pages} {111} (\bibinfo {year} {2005})},\ \Eprint
  {https://arxiv.org/abs/hep-ph/0505257} {arXiv:hep-ph/0505257} \BibitemShut
  {NoStop}%
\bibitem [{\citenamefont {Barger}\ \emph {et~al.}(2007)\citenamefont {Barger},
  \citenamefont {Jiang}, \citenamefont {Langacker},\ and\ \citenamefont
  {Li}}]{Barger:2006fm}%
  \BibitemOpen
  \bibfield  {author} {\bibinfo {author} {\bibfnamefont {V.}~\bibnamefont
  {Barger}}, \bibinfo {author} {\bibfnamefont {J.}~\bibnamefont {Jiang}},
  \bibinfo {author} {\bibfnamefont {P.}~\bibnamefont {Langacker}},\ and\
  \bibinfo {author} {\bibfnamefont {T.}~\bibnamefont {Li}},\ }\bibfield
  {title} {\bibinfo {title} {{String scale gauge coupling unification with
  vector-like exotics and non-canonical U(1)(Y) normalization}},\ }\href
  {https://doi.org/10.1142/S0217751X07038128} {\bibfield  {journal} {\bibinfo
  {journal} {Int. J. Mod. Phys. A}\ }\textbf {\bibinfo {volume} {22}},\
  \bibinfo {pages} {6203} (\bibinfo {year} {2007})},\ \Eprint
  {https://arxiv.org/abs/hep-ph/0612206} {arXiv:hep-ph/0612206} \BibitemShut
  {NoStop}%
\bibitem [{\citenamefont {Dorsner}\ \emph {et~al.}(2014)\citenamefont
  {Dorsner}, \citenamefont {Fajfer},\ and\ \citenamefont
  {Mustac}}]{Dorsner:2014wva}%
  \BibitemOpen
  \bibfield  {author} {\bibinfo {author} {\bibfnamefont {I.}~\bibnamefont
  {Dorsner}}, \bibinfo {author} {\bibfnamefont {S.}~\bibnamefont {Fajfer}},\
  and\ \bibinfo {author} {\bibfnamefont {I.}~\bibnamefont {Mustac}},\
  }\bibfield  {title} {\bibinfo {title} {{Light vector-like fermions in a
  minimal SU(5) setup}},\ }\href {https://doi.org/10.1103/PhysRevD.89.115004}
  {\bibfield  {journal} {\bibinfo  {journal} {Phys. Rev. D}\ }\textbf {\bibinfo
  {volume} {89}},\ \bibinfo {pages} {115004} (\bibinfo {year} {2014})},\
  \Eprint {https://arxiv.org/abs/1401.6870} {arXiv:1401.6870 [hep-ph]}
  \BibitemShut {NoStop}%
\bibitem [{\citenamefont {Kowalska}\ and\ \citenamefont
  {Kumar}(2019)}]{Kowalska:2019qxm}%
  \BibitemOpen
  \bibfield  {author} {\bibinfo {author} {\bibfnamefont {K.}~\bibnamefont
  {Kowalska}}\ and\ \bibinfo {author} {\bibfnamefont {D.}~\bibnamefont
  {Kumar}},\ }\bibfield  {title} {\bibinfo {title} {{Road map through the
  desert: unification with vector-like fermions}},\ }\href
  {https://doi.org/10.1007/JHEP12(2019)094} {\bibfield  {journal} {\bibinfo
  {journal} {JHEP}\ }\textbf {\bibinfo {volume} {12}},\ \bibinfo {pages}
  {094}},\ \Eprint {https://arxiv.org/abs/1910.00847} {arXiv:1910.00847
  [hep-ph]} \BibitemShut {NoStop}%
\bibitem [{\citenamefont {Olivas}\ \emph {et~al.}(2022)\citenamefont {Olivas},
  \citenamefont {Kowalska},\ and\ \citenamefont {Kumar}}]{Olivas:2021nft}%
  \BibitemOpen
  \bibfield  {author} {\bibinfo {author} {\bibfnamefont {U.~C.}\ \bibnamefont
  {Olivas}}, \bibinfo {author} {\bibfnamefont {K.}~\bibnamefont {Kowalska}},\
  and\ \bibinfo {author} {\bibfnamefont {D.}~\bibnamefont {Kumar}},\ }\bibfield
   {title} {\bibinfo {title} {{Road map through the desert with scalars}},\
  }\href {https://doi.org/10.1007/JHEP03(2022)132} {\bibfield  {journal}
  {\bibinfo  {journal} {JHEP}\ }\textbf {\bibinfo {volume} {03}},\ \bibinfo
  {pages} {132}},\ \Eprint {https://arxiv.org/abs/2112.11742} {arXiv:2112.11742
  [hep-ph]} \BibitemShut {NoStop}%
\bibitem [{\citenamefont {del Aguila}\ \emph {et~al.}(2000)\citenamefont {del
  Aguila}, \citenamefont {Perez-Victoria},\ and\ \citenamefont
  {Santiago}}]{delAguila:2000rc}%
  \BibitemOpen
  \bibfield  {author} {\bibinfo {author} {\bibfnamefont {F.}~\bibnamefont {del
  Aguila}}, \bibinfo {author} {\bibfnamefont {M.}~\bibnamefont
  {Perez-Victoria}},\ and\ \bibinfo {author} {\bibfnamefont {J.}~\bibnamefont
  {Santiago}},\ }\bibfield  {title} {\bibinfo {title} {{Observable
  contributions of new exotic quarks to quark mixing}},\ }\href
  {https://doi.org/10.1088/1126-6708/2000/09/011} {\bibfield  {journal}
  {\bibinfo  {journal} {JHEP}\ }\textbf {\bibinfo {volume} {09}},\ \bibinfo
  {pages} {011}},\ \Eprint {https://arxiv.org/abs/hep-ph/0007316}
  {arXiv:hep-ph/0007316} \BibitemShut {NoStop}%
\bibitem [{\citenamefont {Bobeth}\ \emph {et~al.}(2017)\citenamefont {Bobeth},
  \citenamefont {Buras}, \citenamefont {Celis},\ and\ \citenamefont
  {Jung}}]{Bobeth:2016llm}%
  \BibitemOpen
  \bibfield  {author} {\bibinfo {author} {\bibfnamefont {C.}~\bibnamefont
  {Bobeth}}, \bibinfo {author} {\bibfnamefont {A.~J.}\ \bibnamefont {Buras}},
  \bibinfo {author} {\bibfnamefont {A.}~\bibnamefont {Celis}},\ and\ \bibinfo
  {author} {\bibfnamefont {M.}~\bibnamefont {Jung}},\ }\bibfield  {title}
  {\bibinfo {title} {{Patterns of Flavour Violation in Models with Vector-Like
  Quarks}},\ }\href {https://doi.org/10.1007/JHEP04(2017)079} {\bibfield
  {journal} {\bibinfo  {journal} {JHEP}\ }\textbf {\bibinfo {volume} {04}},\
  \bibinfo {pages} {079}},\ \Eprint {https://arxiv.org/abs/1609.04783}
  {arXiv:1609.04783 [hep-ph]} \BibitemShut {NoStop}%
\bibitem [{\citenamefont {Crivellin}\ \emph {et~al.}(2021)\citenamefont
  {Crivellin}, \citenamefont {Hoferichter}, \citenamefont {Kirk}, \citenamefont
  {Manzari},\ and\ \citenamefont {Schnell}}]{Crivellin:2021bkd}%
  \BibitemOpen
  \bibfield  {author} {\bibinfo {author} {\bibfnamefont {A.}~\bibnamefont
  {Crivellin}}, \bibinfo {author} {\bibfnamefont {M.}~\bibnamefont
  {Hoferichter}}, \bibinfo {author} {\bibfnamefont {M.}~\bibnamefont {Kirk}},
  \bibinfo {author} {\bibfnamefont {C.~A.}\ \bibnamefont {Manzari}},\ and\
  \bibinfo {author} {\bibfnamefont {L.}~\bibnamefont {Schnell}},\ }\bibfield
  {title} {\bibinfo {title} {{First-generation new physics in simplified
  models: from low-energy parity violation to the LHC}},\ }\href
  {https://doi.org/10.1007/JHEP10(2021)221} {\bibfield  {journal} {\bibinfo
  {journal} {JHEP}\ }\textbf {\bibinfo {volume} {10}},\ \bibinfo {pages}
  {221}},\ \Eprint {https://arxiv.org/abs/2107.13569} {arXiv:2107.13569
  [hep-ph]} \BibitemShut {NoStop}%
\bibitem [{\citenamefont {Alves}\ \emph {et~al.}(2024)\citenamefont {Alves},
  \citenamefont {Branco}, \citenamefont {Cherchiglia}, \citenamefont {Nishi},
  \citenamefont {Penedo}, \citenamefont {Pereira}, \citenamefont {Rebelo},\
  and\ \citenamefont {Silva-Marcos}}]{Alves:2023ufm}%
  \BibitemOpen
  \bibfield  {author} {\bibinfo {author} {\bibfnamefont {J.~a.~M.}\
  \bibnamefont {Alves}}, \bibinfo {author} {\bibfnamefont {G.~C.}\ \bibnamefont
  {Branco}}, \bibinfo {author} {\bibfnamefont {A.~L.}\ \bibnamefont
  {Cherchiglia}}, \bibinfo {author} {\bibfnamefont {C.~C.}\ \bibnamefont
  {Nishi}}, \bibinfo {author} {\bibfnamefont {J.~T.}\ \bibnamefont {Penedo}},
  \bibinfo {author} {\bibfnamefont {P.~M.~F.}\ \bibnamefont {Pereira}},
  \bibinfo {author} {\bibfnamefont {M.~N.}\ \bibnamefont {Rebelo}},\ and\
  \bibinfo {author} {\bibfnamefont {J.~I.}\ \bibnamefont {Silva-Marcos}},\
  }\bibfield  {title} {\bibinfo {title} {{Vector-like singlet quarks: A
  roadmap}},\ }\href {https://doi.org/10.1016/j.physrep.2023.12.004} {\bibfield
   {journal} {\bibinfo  {journal} {Phys. Rept.}\ }\textbf {\bibinfo {volume}
  {1057}},\ \bibinfo {pages} {1} (\bibinfo {year} {2024})},\ \Eprint
  {https://arxiv.org/abs/2304.10561} {arXiv:2304.10561 [hep-ph]} \BibitemShut
  {NoStop}%
\bibitem [{\citenamefont {Lavoura}\ and\ \citenamefont
  {Silva}(1993)}]{Lavoura:1992np}%
  \BibitemOpen
  \bibfield  {author} {\bibinfo {author} {\bibfnamefont {L.}~\bibnamefont
  {Lavoura}}\ and\ \bibinfo {author} {\bibfnamefont {J.~P.}\ \bibnamefont
  {Silva}},\ }\bibfield  {title} {\bibinfo {title} {{The Oblique corrections
  from vector - like singlet and doublet quarks}},\ }\href
  {https://doi.org/10.1103/PhysRevD.47.2046} {\bibfield  {journal} {\bibinfo
  {journal} {Phys. Rev. D}\ }\textbf {\bibinfo {volume} {47}},\ \bibinfo
  {pages} {2046} (\bibinfo {year} {1993})}\BibitemShut {NoStop}%
\bibitem [{\citenamefont {Kearney}\ \emph {et~al.}(2012)\citenamefont
  {Kearney}, \citenamefont {Pierce},\ and\ \citenamefont
  {Weiner}}]{Kearney:2012zi}%
  \BibitemOpen
  \bibfield  {author} {\bibinfo {author} {\bibfnamefont {J.}~\bibnamefont
  {Kearney}}, \bibinfo {author} {\bibfnamefont {A.}~\bibnamefont {Pierce}},\
  and\ \bibinfo {author} {\bibfnamefont {N.}~\bibnamefont {Weiner}},\
  }\bibfield  {title} {\bibinfo {title} {{Vectorlike Fermions and Higgs
  Couplings}},\ }\href {https://doi.org/10.1103/PhysRevD.86.113005} {\bibfield
  {journal} {\bibinfo  {journal} {Phys. Rev. D}\ }\textbf {\bibinfo {volume}
  {86}},\ \bibinfo {pages} {113005} (\bibinfo {year} {2012})},\ \Eprint
  {https://arxiv.org/abs/1207.7062} {arXiv:1207.7062 [hep-ph]} \BibitemShut
  {NoStop}%
\bibitem [{\citenamefont {Abouabid}\ \emph {et~al.}(2024)\citenamefont
  {Abouabid}, \citenamefont {Arhrib}, \citenamefont {Benbrik}, \citenamefont
  {Boukidi},\ and\ \citenamefont {Falaki}}]{Abouabid:2023mbu}%
  \BibitemOpen
  \bibfield  {author} {\bibinfo {author} {\bibfnamefont {H.}~\bibnamefont
  {Abouabid}}, \bibinfo {author} {\bibfnamefont {A.}~\bibnamefont {Arhrib}},
  \bibinfo {author} {\bibfnamefont {R.}~\bibnamefont {Benbrik}}, \bibinfo
  {author} {\bibfnamefont {M.}~\bibnamefont {Boukidi}},\ and\ \bibinfo {author}
  {\bibfnamefont {J.~E.}\ \bibnamefont {Falaki}},\ }\bibfield  {title}
  {\bibinfo {title} {{The oblique parameters in the 2HDM with vector-like
  quarks: confronting M $_{W}$ CDF-II anomaly}},\ }\href
  {https://doi.org/10.1088/1361-6471/ad3f34} {\bibfield  {journal} {\bibinfo
  {journal} {J. Phys. G}\ }\textbf {\bibinfo {volume} {51}},\ \bibinfo {pages}
  {075001} (\bibinfo {year} {2024})},\ \Eprint
  {https://arxiv.org/abs/2302.07149} {arXiv:2302.07149 [hep-ph]} \BibitemShut
  {NoStop}%
\bibitem [{\citenamefont {Fan}\ \emph {et~al.}(2016)\citenamefont {Fan},
  \citenamefont {Koushiappas},\ and\ \citenamefont {Landsberg}}]{Fan:2015sza}%
  \BibitemOpen
  \bibfield  {author} {\bibinfo {author} {\bibfnamefont {J.}~\bibnamefont
  {Fan}}, \bibinfo {author} {\bibfnamefont {S.~M.}\ \bibnamefont
  {Koushiappas}},\ and\ \bibinfo {author} {\bibfnamefont {G.}~\bibnamefont
  {Landsberg}},\ }\bibfield  {title} {\bibinfo {title} {{Pseudoscalar Portal
  Dark Matter and New Signatures of Vector-like Fermions}},\ }\href
  {https://doi.org/10.1007/JHEP01(2016)111} {\bibfield  {journal} {\bibinfo
  {journal} {JHEP}\ }\textbf {\bibinfo {volume} {01}},\ \bibinfo {pages}
  {111}},\ \Eprint {https://arxiv.org/abs/1507.06993} {arXiv:1507.06993
  [hep-ph]} \BibitemShut {NoStop}%
\bibitem [{\citenamefont {Belyaev}\ \emph {et~al.}(2023)\citenamefont
  {Belyaev}, \citenamefont {Deandrea}, \citenamefont {Moretti}, \citenamefont
  {Panizzi}, \citenamefont {Ross},\ and\ \citenamefont
  {Thongyoi}}]{Belyaev:2022shr}%
  \BibitemOpen
  \bibfield  {author} {\bibinfo {author} {\bibfnamefont {A.}~\bibnamefont
  {Belyaev}}, \bibinfo {author} {\bibfnamefont {A.}~\bibnamefont {Deandrea}},
  \bibinfo {author} {\bibfnamefont {S.}~\bibnamefont {Moretti}}, \bibinfo
  {author} {\bibfnamefont {L.}~\bibnamefont {Panizzi}}, \bibinfo {author}
  {\bibfnamefont {D.~A.}\ \bibnamefont {Ross}},\ and\ \bibinfo {author}
  {\bibfnamefont {N.}~\bibnamefont {Thongyoi}},\ }\bibfield  {title} {\bibinfo
  {title} {{Fermionic portal to vector dark matter from a new gauge sector}},\
  }\href {https://doi.org/10.1103/PhysRevD.108.095001} {\bibfield  {journal}
  {\bibinfo  {journal} {Phys. Rev. D}\ }\textbf {\bibinfo {volume} {108}},\
  \bibinfo {pages} {095001} (\bibinfo {year} {2023})},\ \Eprint
  {https://arxiv.org/abs/2204.03510} {arXiv:2204.03510 [hep-ph]} \BibitemShut
  {NoStop}%
\bibitem [{\citenamefont {Barman}\ \emph {et~al.}(2019)\citenamefont {Barman},
  \citenamefont {Bhattacharya}, \citenamefont {Ghosh}, \citenamefont {Kadam},\
  and\ \citenamefont {Sahu}}]{Barman:2019tuo}%
  \BibitemOpen
  \bibfield  {author} {\bibinfo {author} {\bibfnamefont {B.}~\bibnamefont
  {Barman}}, \bibinfo {author} {\bibfnamefont {S.}~\bibnamefont
  {Bhattacharya}}, \bibinfo {author} {\bibfnamefont {P.}~\bibnamefont {Ghosh}},
  \bibinfo {author} {\bibfnamefont {S.}~\bibnamefont {Kadam}},\ and\ \bibinfo
  {author} {\bibfnamefont {N.}~\bibnamefont {Sahu}},\ }\bibfield  {title}
  {\bibinfo {title} {{Fermion Dark Matter with Scalar Triplet at Direct and
  Collider Searches}},\ }\href {https://doi.org/10.1103/PhysRevD.100.015027}
  {\bibfield  {journal} {\bibinfo  {journal} {Phys. Rev. D}\ }\textbf {\bibinfo
  {volume} {100}},\ \bibinfo {pages} {015027} (\bibinfo {year} {2019})},\
  \Eprint {https://arxiv.org/abs/1902.01217} {arXiv:1902.01217 [hep-ph]}
  \BibitemShut {NoStop}%
\bibitem [{\citenamefont {Aad}\ \emph {et~al.}(2022)\citenamefont {Aad} \emph
  {et~al.}}]{ATLAS:2022ozf}%
  \BibitemOpen
  \bibfield  {author} {\bibinfo {author} {\bibfnamefont {G.}~\bibnamefont
  {Aad}} \emph {et~al.} (\bibinfo {collaboration} {ATLAS}),\ }\bibfield
  {title} {\bibinfo {title} {{Search for single production of a vectorlike $T$
  quark decaying into a Higgs boson and top quark with fully hadronic final
  states using the ATLAS detector}},\ }\href
  {https://doi.org/10.1103/PhysRevD.105.092012} {\bibfield  {journal} {\bibinfo
   {journal} {Phys. Rev. D}\ }\textbf {\bibinfo {volume} {105}},\ \bibinfo
  {pages} {092012} (\bibinfo {year} {2022})},\ \Eprint
  {https://arxiv.org/abs/2201.07045} {arXiv:2201.07045 [hep-ex]} \BibitemShut
  {NoStop}%
\bibitem [{\citenamefont {Aad}\ \emph {et~al.}(2023{\natexlab{a}})\citenamefont
  {Aad} \emph {et~al.}}]{ATLAS:2022hnn}%
  \BibitemOpen
  \bibfield  {author} {\bibinfo {author} {\bibfnamefont {G.}~\bibnamefont
  {Aad}} \emph {et~al.} (\bibinfo {collaboration} {ATLAS}),\ }\bibfield
  {title} {\bibinfo {title} {{Search for pair-production of vector-like quarks
  in pp collision events at s=13 TeV with at least one leptonically decaying Z
  boson and a third-generation quark with the ATLAS detector}},\ }\href
  {https://doi.org/10.1016/j.physletb.2023.138019} {\bibfield  {journal}
  {\bibinfo  {journal} {Phys. Lett. B}\ }\textbf {\bibinfo {volume} {843}},\
  \bibinfo {pages} {138019} (\bibinfo {year} {2023}{\natexlab{a}})},\ \Eprint
  {https://arxiv.org/abs/2210.15413} {arXiv:2210.15413 [hep-ex]} \BibitemShut
  {NoStop}%
\bibitem [{\citenamefont {Aad}\ \emph {et~al.}(2023{\natexlab{b}})\citenamefont
  {Aad} \emph {et~al.}}]{ATLAS:2022tla}%
  \BibitemOpen
  \bibfield  {author} {\bibinfo {author} {\bibfnamefont {G.}~\bibnamefont
  {Aad}} \emph {et~al.} (\bibinfo {collaboration} {ATLAS}),\ }\bibfield
  {title} {\bibinfo {title} {{Search for pair-produced vector-like top and
  bottom partners in events with large missing transverse momentum in pp
  collisions with the ATLAS detector}},\ }\href
  {https://doi.org/10.1140/epjc/s10052-023-11790-7} {\bibfield  {journal}
  {\bibinfo  {journal} {Eur. Phys. J. C}\ }\textbf {\bibinfo {volume} {83}},\
  \bibinfo {pages} {719} (\bibinfo {year} {2023}{\natexlab{b}})},\ \Eprint
  {https://arxiv.org/abs/2212.05263} {arXiv:2212.05263 [hep-ex]} \BibitemShut
  {NoStop}%
\bibitem [{\citenamefont {Aad}\ \emph {et~al.}(2023{\natexlab{c}})\citenamefont
  {Aad} \emph {et~al.}}]{ATLAS:2023sbu}%
  \BibitemOpen
  \bibfield  {author} {\bibinfo {author} {\bibfnamefont {G.}~\bibnamefont
  {Aad}} \emph {et~al.} (\bibinfo {collaboration} {ATLAS}),\ }\bibfield
  {title} {\bibinfo {title} {{Search for third-generation vector-like leptons
  in $pp$ collisions at $\sqrt{s} = 13\,\text{TeV}$ with the ATLAS detector}},\
  }\href {https://doi.org/10.1007/JHEP07(2023)118} {\bibfield  {journal}
  {\bibinfo  {journal} {JHEP}\ }\textbf {\bibinfo {volume} {07}},\ \bibinfo
  {pages} {118}},\ \Eprint {https://arxiv.org/abs/2303.05441} {arXiv:2303.05441
  [hep-ex]} \BibitemShut {NoStop}%
\bibitem [{\citenamefont {Aad}\ \emph {et~al.}(2023{\natexlab{d}})\citenamefont
  {Aad} \emph {et~al.}}]{ATLAS:2023pja}%
  \BibitemOpen
  \bibfield  {author} {\bibinfo {author} {\bibfnamefont {G.}~\bibnamefont
  {Aad}} \emph {et~al.} (\bibinfo {collaboration} {ATLAS}),\ }\bibfield
  {title} {\bibinfo {title} {{Search for single production of vector-like T
  quarks decaying into Ht or Zt in pp collisions at $ \sqrt{s} $ = 13 TeV with
  the ATLAS detector}},\ }\href {https://doi.org/10.1007/JHEP08(2023)153}
  {\bibfield  {journal} {\bibinfo  {journal} {JHEP}\ }\textbf {\bibinfo
  {volume} {08}},\ \bibinfo {pages} {153}},\ \Eprint
  {https://arxiv.org/abs/2305.03401} {arXiv:2305.03401 [hep-ex]} \BibitemShut
  {NoStop}%
\bibitem [{\citenamefont {Aad}\ \emph {et~al.}(2023{\natexlab{e}})\citenamefont
  {Aad} \emph {et~al.}}]{ATLAS:2023bfh}%
  \BibitemOpen
  \bibfield  {author} {\bibinfo {author} {\bibfnamefont {G.}~\bibnamefont
  {Aad}} \emph {et~al.} (\bibinfo {collaboration} {ATLAS}),\ }\href@noop {}
  {\bibinfo {title} {{Search for singly produced vector-like top partners in
  multilepton final states with 139 $\mathrm{fb}^{-1}$ of $pp$ collision data
  at $\sqrt{s} = 13$ TeV with the ATLAS detector}}} (\bibinfo {year}
  {2023}{\natexlab{e}}),\ \Eprint {https://arxiv.org/abs/2307.07584}
  {arXiv:2307.07584 [hep-ex]} \BibitemShut {NoStop}%
\bibitem [{\citenamefont {Tumasyan}\ \emph
  {et~al.}(2022{\natexlab{a}})\citenamefont {Tumasyan} \emph
  {et~al.}}]{CMS:2022yxp}%
  \BibitemOpen
  \bibfield  {author} {\bibinfo {author} {\bibfnamefont {A.}~\bibnamefont
  {Tumasyan}} \emph {et~al.} (\bibinfo {collaboration} {CMS}),\ }\bibfield
  {title} {\bibinfo {title} {{Search for single production of a vector-like T
  quark decaying to a top quark and a Z boson in the final state with jets and
  missing transverse momentum at $ \sqrt{s} $ = 13 TeV}},\ }\href
  {https://doi.org/10.1007/JHEP05(2022)093} {\bibfield  {journal} {\bibinfo
  {journal} {JHEP}\ }\textbf {\bibinfo {volume} {05}},\ \bibinfo {pages}
  {093}},\ \Eprint {https://arxiv.org/abs/2201.02227} {arXiv:2201.02227
  [hep-ex]} \BibitemShut {NoStop}%
\bibitem [{\citenamefont {Tumasyan}\ \emph
  {et~al.}(2022{\natexlab{b}})\citenamefont {Tumasyan} \emph
  {et~al.}}]{CMS:2022tdo}%
  \BibitemOpen
  \bibfield  {author} {\bibinfo {author} {\bibfnamefont {A.}~\bibnamefont
  {Tumasyan}} \emph {et~al.} (\bibinfo {collaboration} {CMS}),\ }\bibfield
  {title} {\bibinfo {title} {{Search for a W' boson decaying to a vector-like
  quark and a top or bottom quark in the all-jets final state at $
  \sqrt{\mathrm{s}} $ = 13 TeV}},\ }\href
  {https://doi.org/10.1007/JHEP09(2022)088} {\bibfield  {journal} {\bibinfo
  {journal} {JHEP}\ }\textbf {\bibinfo {volume} {09}},\ \bibinfo {pages}
  {088}},\ \Eprint {https://arxiv.org/abs/2202.12988} {arXiv:2202.12988
  [hep-ex]} \BibitemShut {NoStop}%
\bibitem [{\citenamefont {Tumasyan}\ \emph
  {et~al.}(2023{\natexlab{a}})\citenamefont {Tumasyan} \emph
  {et~al.}}]{CMS:2022cpe}%
  \BibitemOpen
  \bibfield  {author} {\bibinfo {author} {\bibfnamefont {A.}~\bibnamefont
  {Tumasyan}} \emph {et~al.} (\bibinfo {collaboration} {CMS}),\ }\bibfield
  {title} {\bibinfo {title} {{Search for pair-produced vector-like leptons in
  final states with third-generation leptons and at least three b quark jets in
  proton-proton collisions at s=13TeV}},\ }\href
  {https://doi.org/10.1016/j.physletb.2023.137713} {\bibfield  {journal}
  {\bibinfo  {journal} {Phys. Lett. B}\ }\textbf {\bibinfo {volume} {846}},\
  \bibinfo {pages} {137713} (\bibinfo {year} {2023}{\natexlab{a}})},\ \Eprint
  {https://arxiv.org/abs/2208.09700} {arXiv:2208.09700 [hep-ex]} \BibitemShut
  {NoStop}%
\bibitem [{\citenamefont {Tumasyan}\ \emph
  {et~al.}(2023{\natexlab{b}})\citenamefont {Tumasyan} \emph
  {et~al.}}]{CMS:2022fck}%
  \BibitemOpen
  \bibfield  {author} {\bibinfo {author} {\bibfnamefont {A.}~\bibnamefont
  {Tumasyan}} \emph {et~al.} (\bibinfo {collaboration} {CMS}),\ }\bibfield
  {title} {\bibinfo {title} {{Search for pair production of vector-like quarks
  in leptonic final states in proton-proton collisions at $ \sqrt{s} $ = 13
  TeV}},\ }\href {https://doi.org/10.1007/JHEP07(2023)020} {\bibfield
  {journal} {\bibinfo  {journal} {JHEP}\ }\textbf {\bibinfo {volume} {07}},\
  \bibinfo {pages} {020}},\ \Eprint {https://arxiv.org/abs/2209.07327}
  {arXiv:2209.07327 [hep-ex]} \BibitemShut {NoStop}%
\bibitem [{\citenamefont {Tumasyan}\ \emph
  {et~al.}(2023{\natexlab{c}})\citenamefont {Tumasyan} \emph
  {et~al.}}]{CMS:2023agg}%
  \BibitemOpen
  \bibfield  {author} {\bibinfo {author} {\bibfnamefont {A.}~\bibnamefont
  {Tumasyan}} \emph {et~al.} (\bibinfo {collaboration} {CMS}),\ }\bibfield
  {title} {\bibinfo {title} {{Search for a vector-like quark T$'$$\to$ tH via
  the diphoton decay mode of the Higgs boson in proton-proton collisions at
  $\sqrt{s}$ = 13 TeV}},\ }\href {https://doi.org/10.1007/JHEP09(2023)057}
  {\bibfield  {journal} {\bibinfo  {journal} {JHEP}\ }\textbf {\bibinfo
  {volume} {09}},\ \bibinfo {pages} {057}},\ \Eprint
  {https://arxiv.org/abs/2302.12802} {arXiv:2302.12802 [hep-ex]} \BibitemShut
  {NoStop}%
\bibitem [{\citenamefont {Curtin}\ \emph {et~al.}(2014)\citenamefont {Curtin},
  \citenamefont {Meade},\ and\ \citenamefont {Yu}}]{Curtin:2014jma}%
  \BibitemOpen
  \bibfield  {author} {\bibinfo {author} {\bibfnamefont {D.}~\bibnamefont
  {Curtin}}, \bibinfo {author} {\bibfnamefont {P.}~\bibnamefont {Meade}},\ and\
  \bibinfo {author} {\bibfnamefont {C.-T.}\ \bibnamefont {Yu}},\ }\bibfield
  {title} {\bibinfo {title} {{Testing Electroweak Baryogenesis with Future
  Colliders}},\ }\href {https://doi.org/10.1007/JHEP11(2014)127} {\bibfield
  {journal} {\bibinfo  {journal} {JHEP}\ }\textbf {\bibinfo {volume} {11}},\
  \bibinfo {pages} {127}},\ \Eprint {https://arxiv.org/abs/1409.0005}
  {arXiv:1409.0005 [hep-ph]} \BibitemShut {NoStop}%
\bibitem [{\citenamefont {Profumo}\ \emph {et~al.}(2007)\citenamefont
  {Profumo}, \citenamefont {Ramsey-Musolf},\ and\ \citenamefont
  {Shaughnessy}}]{Profumo:2007wc}%
  \BibitemOpen
  \bibfield  {author} {\bibinfo {author} {\bibfnamefont {S.}~\bibnamefont
  {Profumo}}, \bibinfo {author} {\bibfnamefont {M.~J.}\ \bibnamefont
  {Ramsey-Musolf}},\ and\ \bibinfo {author} {\bibfnamefont {G.}~\bibnamefont
  {Shaughnessy}},\ }\bibfield  {title} {\bibinfo {title} {{Singlet Higgs
  phenomenology and the electroweak phase transition}},\ }\href
  {https://doi.org/10.1088/1126-6708/2007/08/010} {\bibfield  {journal}
  {\bibinfo  {journal} {JHEP}\ }\textbf {\bibinfo {volume} {08}},\ \bibinfo
  {pages} {010}},\ \Eprint {https://arxiv.org/abs/0705.2425} {arXiv:0705.2425
  [hep-ph]} \BibitemShut {NoStop}%
\bibitem [{\citenamefont {Noble}\ and\ \citenamefont
  {Perelstein}(2008)}]{Noble:2007kk}%
  \BibitemOpen
  \bibfield  {author} {\bibinfo {author} {\bibfnamefont {A.}~\bibnamefont
  {Noble}}\ and\ \bibinfo {author} {\bibfnamefont {M.}~\bibnamefont
  {Perelstein}},\ }\bibfield  {title} {\bibinfo {title} {{Higgs self-coupling
  as a probe of electroweak phase transition}},\ }\href
  {https://doi.org/10.1103/PhysRevD.78.063518} {\bibfield  {journal} {\bibinfo
  {journal} {Phys. Rev. D}\ }\textbf {\bibinfo {volume} {78}},\ \bibinfo
  {pages} {063518} (\bibinfo {year} {2008})},\ \Eprint
  {https://arxiv.org/abs/0711.3018} {arXiv:0711.3018 [hep-ph]} \BibitemShut
  {NoStop}%
\bibitem [{\citenamefont {Espinosa}\ \emph {et~al.}(2012)\citenamefont
  {Espinosa}, \citenamefont {Konstandin},\ and\ \citenamefont
  {Riva}}]{Espinosa:2011ax}%
  \BibitemOpen
  \bibfield  {author} {\bibinfo {author} {\bibfnamefont {J.~R.}\ \bibnamefont
  {Espinosa}}, \bibinfo {author} {\bibfnamefont {T.}~\bibnamefont
  {Konstandin}},\ and\ \bibinfo {author} {\bibfnamefont {F.}~\bibnamefont
  {Riva}},\ }\bibfield  {title} {\bibinfo {title} {{Strong Electroweak Phase
  Transitions in the Standard Model with a Singlet}},\ }\href
  {https://doi.org/10.1016/j.nuclphysb.2011.09.010} {\bibfield  {journal}
  {\bibinfo  {journal} {Nucl. Phys. B}\ }\textbf {\bibinfo {volume} {854}},\
  \bibinfo {pages} {592} (\bibinfo {year} {2012})},\ \Eprint
  {https://arxiv.org/abs/1107.5441} {arXiv:1107.5441 [hep-ph]} \BibitemShut
  {NoStop}%
\bibitem [{\citenamefont {Espinosa}\ and\ \citenamefont
  {Quiros}(1993)}]{Espinosa:1993bs}%
  \BibitemOpen
  \bibfield  {author} {\bibinfo {author} {\bibfnamefont {J.~R.}\ \bibnamefont
  {Espinosa}}\ and\ \bibinfo {author} {\bibfnamefont {M.}~\bibnamefont
  {Quiros}},\ }\bibfield  {title} {\bibinfo {title} {{The Electroweak phase
  transition with a singlet}},\ }\href
  {https://doi.org/10.1016/0370-2693(93)91111-Y} {\bibfield  {journal}
  {\bibinfo  {journal} {Phys. Lett. B}\ }\textbf {\bibinfo {volume} {305}},\
  \bibinfo {pages} {98} (\bibinfo {year} {1993})},\ \Eprint
  {https://arxiv.org/abs/hep-ph/9301285} {arXiv:hep-ph/9301285} \BibitemShut
  {NoStop}%
\bibitem [{\citenamefont {Fern\'andez-Mart\'\i{}nez}\ \emph
  {et~al.}(2023)\citenamefont {Fern\'andez-Mart\'\i{}nez}, \citenamefont
  {L\'opez-Pav\'on}, \citenamefont {No}, \citenamefont {Ota},\ and\
  \citenamefont {Rosauro-Alcaraz}}]{Fernandez-Martinez:2022stj}%
  \BibitemOpen
  \bibfield  {author} {\bibinfo {author} {\bibfnamefont {E.}~\bibnamefont
  {Fern\'andez-Mart\'\i{}nez}}, \bibinfo {author} {\bibfnamefont
  {J.}~\bibnamefont {L\'opez-Pav\'on}}, \bibinfo {author} {\bibfnamefont
  {J.~M.}\ \bibnamefont {No}}, \bibinfo {author} {\bibfnamefont
  {T.}~\bibnamefont {Ota}},\ and\ \bibinfo {author} {\bibfnamefont
  {S.}~\bibnamefont {Rosauro-Alcaraz}},\ }\bibfield  {title} {\bibinfo {title}
  {{$\nu $ Electroweak baryogenesis: the scalar singlet strikes back}},\ }\href
  {https://doi.org/10.1140/epjc/s10052-023-11887-z} {\bibfield  {journal}
  {\bibinfo  {journal} {Eur. Phys. J. C}\ }\textbf {\bibinfo {volume} {83}},\
  \bibinfo {pages} {715} (\bibinfo {year} {2023})},\ \Eprint
  {https://arxiv.org/abs/2210.16279} {arXiv:2210.16279 [hep-ph]} \BibitemShut
  {NoStop}%
\bibitem [{\citenamefont {No}\ and\ \citenamefont
  {Ramsey-Musolf}(2014)}]{No:2013wsa}%
  \BibitemOpen
  \bibfield  {author} {\bibinfo {author} {\bibfnamefont {J.~M.}\ \bibnamefont
  {No}}\ and\ \bibinfo {author} {\bibfnamefont {M.}~\bibnamefont
  {Ramsey-Musolf}},\ }\bibfield  {title} {\bibinfo {title} {{Probing the Higgs
  Portal at the LHC Through Resonant di-Higgs Production}},\ }\href
  {https://doi.org/10.1103/PhysRevD.89.095031} {\bibfield  {journal} {\bibinfo
  {journal} {Phys. Rev. D}\ }\textbf {\bibinfo {volume} {89}},\ \bibinfo
  {pages} {095031} (\bibinfo {year} {2014})},\ \Eprint
  {https://arxiv.org/abs/1310.6035} {arXiv:1310.6035 [hep-ph]} \BibitemShut
  {NoStop}%
\bibitem [{\citenamefont {Barger}\ \emph {et~al.}(2008)\citenamefont {Barger},
  \citenamefont {Langacker}, \citenamefont {McCaskey}, \citenamefont
  {Ramsey-Musolf},\ and\ \citenamefont {Shaughnessy}}]{Barger:2007im}%
  \BibitemOpen
  \bibfield  {author} {\bibinfo {author} {\bibfnamefont {V.}~\bibnamefont
  {Barger}}, \bibinfo {author} {\bibfnamefont {P.}~\bibnamefont {Langacker}},
  \bibinfo {author} {\bibfnamefont {M.}~\bibnamefont {McCaskey}}, \bibinfo
  {author} {\bibfnamefont {M.~J.}\ \bibnamefont {Ramsey-Musolf}},\ and\
  \bibinfo {author} {\bibfnamefont {G.}~\bibnamefont {Shaughnessy}},\
  }\bibfield  {title} {\bibinfo {title} {{LHC Phenomenology of an Extended
  Standard Model with a Real Scalar Singlet}},\ }\href
  {https://doi.org/10.1103/PhysRevD.77.035005} {\bibfield  {journal} {\bibinfo
  {journal} {Phys. Rev. D}\ }\textbf {\bibinfo {volume} {77}},\ \bibinfo
  {pages} {035005} (\bibinfo {year} {2008})},\ \Eprint
  {https://arxiv.org/abs/0706.4311} {arXiv:0706.4311 [hep-ph]} \BibitemShut
  {NoStop}%
\bibitem [{\citenamefont {Huang}\ \emph {et~al.}(2017)\citenamefont {Huang},
  \citenamefont {No}, \citenamefont {Perni\'e}, \citenamefont {Ramsey-Musolf},
  \citenamefont {Safonov}, \citenamefont {Spannowsky},\ and\ \citenamefont
  {Winslow}}]{Huang:2017jws}%
  \BibitemOpen
  \bibfield  {author} {\bibinfo {author} {\bibfnamefont {T.}~\bibnamefont
  {Huang}}, \bibinfo {author} {\bibfnamefont {J.~M.}\ \bibnamefont {No}},
  \bibinfo {author} {\bibfnamefont {L.}~\bibnamefont {Perni\'e}}, \bibinfo
  {author} {\bibfnamefont {M.}~\bibnamefont {Ramsey-Musolf}}, \bibinfo {author}
  {\bibfnamefont {A.}~\bibnamefont {Safonov}}, \bibinfo {author} {\bibfnamefont
  {M.}~\bibnamefont {Spannowsky}},\ and\ \bibinfo {author} {\bibfnamefont
  {P.}~\bibnamefont {Winslow}},\ }\bibfield  {title} {\bibinfo {title}
  {{Resonant di-Higgs boson production in the $b{\bar b}WW$ channel: Probing
  the electroweak phase transition at the LHC}},\ }\href
  {https://doi.org/10.1103/PhysRevD.96.035007} {\bibfield  {journal} {\bibinfo
  {journal} {Phys. Rev. D}\ }\textbf {\bibinfo {volume} {96}},\ \bibinfo
  {pages} {035007} (\bibinfo {year} {2017})},\ \Eprint
  {https://arxiv.org/abs/1701.04442} {arXiv:1701.04442 [hep-ph]} \BibitemShut
  {NoStop}%
\bibitem [{\citenamefont {Profumo}\ \emph {et~al.}(2015)\citenamefont
  {Profumo}, \citenamefont {Ramsey-Musolf}, \citenamefont {Wainwright},\ and\
  \citenamefont {Winslow}}]{Profumo:2014opa}%
  \BibitemOpen
  \bibfield  {author} {\bibinfo {author} {\bibfnamefont {S.}~\bibnamefont
  {Profumo}}, \bibinfo {author} {\bibfnamefont {M.~J.}\ \bibnamefont
  {Ramsey-Musolf}}, \bibinfo {author} {\bibfnamefont {C.~L.}\ \bibnamefont
  {Wainwright}},\ and\ \bibinfo {author} {\bibfnamefont {P.}~\bibnamefont
  {Winslow}},\ }\bibfield  {title} {\bibinfo {title} {{Singlet-catalyzed
  electroweak phase transitions and precision Higgs boson studies}},\ }\href
  {https://doi.org/10.1103/PhysRevD.91.035018} {\bibfield  {journal} {\bibinfo
  {journal} {Phys. Rev. D}\ }\textbf {\bibinfo {volume} {91}},\ \bibinfo
  {pages} {035018} (\bibinfo {year} {2015})},\ \Eprint
  {https://arxiv.org/abs/1407.5342} {arXiv:1407.5342 [hep-ph]} \BibitemShut
  {NoStop}%
\bibitem [{\citenamefont {Chen}\ \emph {et~al.}(2017)\citenamefont {Chen},
  \citenamefont {Kozaczuk},\ and\ \citenamefont {Lewis}}]{Chen:2017qcz}%
  \BibitemOpen
  \bibfield  {author} {\bibinfo {author} {\bibfnamefont {C.-Y.}\ \bibnamefont
  {Chen}}, \bibinfo {author} {\bibfnamefont {J.}~\bibnamefont {Kozaczuk}},\
  and\ \bibinfo {author} {\bibfnamefont {I.~M.}\ \bibnamefont {Lewis}},\
  }\bibfield  {title} {\bibinfo {title} {{Non-resonant Collider Signatures of a
  Singlet-Driven Electroweak Phase Transition}},\ }\href
  {https://doi.org/10.1007/JHEP08(2017)096} {\bibfield  {journal} {\bibinfo
  {journal} {JHEP}\ }\textbf {\bibinfo {volume} {08}},\ \bibinfo {pages}
  {096}},\ \Eprint {https://arxiv.org/abs/1704.05844} {arXiv:1704.05844
  [hep-ph]} \BibitemShut {NoStop}%
\bibitem [{\citenamefont {Hashino}\ \emph {et~al.}(2017)\citenamefont
  {Hashino}, \citenamefont {Kakizaki}, \citenamefont {Kanemura}, \citenamefont
  {Ko},\ and\ \citenamefont {Matsui}}]{Hashino:2016xoj}%
  \BibitemOpen
  \bibfield  {author} {\bibinfo {author} {\bibfnamefont {K.}~\bibnamefont
  {Hashino}}, \bibinfo {author} {\bibfnamefont {M.}~\bibnamefont {Kakizaki}},
  \bibinfo {author} {\bibfnamefont {S.}~\bibnamefont {Kanemura}}, \bibinfo
  {author} {\bibfnamefont {P.}~\bibnamefont {Ko}},\ and\ \bibinfo {author}
  {\bibfnamefont {T.}~\bibnamefont {Matsui}},\ }\bibfield  {title} {\bibinfo
  {title} {{Gravitational waves and Higgs boson couplings for exploring first
  order phase transition in the model with a singlet scalar field}},\ }\href
  {https://doi.org/10.1016/j.physletb.2016.12.052} {\bibfield  {journal}
  {\bibinfo  {journal} {Phys. Lett. B}\ }\textbf {\bibinfo {volume} {766}},\
  \bibinfo {pages} {49} (\bibinfo {year} {2017})},\ \Eprint
  {https://arxiv.org/abs/1609.00297} {arXiv:1609.00297 [hep-ph]} \BibitemShut
  {NoStop}%
\bibitem [{\citenamefont {Ellis}\ \emph {et~al.}(2023)\citenamefont {Ellis},
  \citenamefont {Lewicki}, \citenamefont {Merchand}, \citenamefont {No},\ and\
  \citenamefont {Zych}}]{Ellis:2022lft}%
  \BibitemOpen
  \bibfield  {author} {\bibinfo {author} {\bibfnamefont {J.}~\bibnamefont
  {Ellis}}, \bibinfo {author} {\bibfnamefont {M.}~\bibnamefont {Lewicki}},
  \bibinfo {author} {\bibfnamefont {M.}~\bibnamefont {Merchand}}, \bibinfo
  {author} {\bibfnamefont {J.~M.}\ \bibnamefont {No}},\ and\ \bibinfo {author}
  {\bibfnamefont {M.}~\bibnamefont {Zych}},\ }\bibfield  {title} {\bibinfo
  {title} {{The scalar singlet extension of the Standard Model: gravitational
  waves versus baryogenesis}},\ }\href
  {https://doi.org/10.1007/JHEP01(2023)093} {\bibfield  {journal} {\bibinfo
  {journal} {JHEP}\ }\textbf {\bibinfo {volume} {01}},\ \bibinfo {pages}
  {093}},\ \Eprint {https://arxiv.org/abs/2210.16305} {arXiv:2210.16305
  [hep-ph]} \BibitemShut {NoStop}%
\bibitem [{\citenamefont {Gonderinger}\ \emph {et~al.}(2010)\citenamefont
  {Gonderinger}, \citenamefont {Li}, \citenamefont {Patel},\ and\ \citenamefont
  {Ramsey-Musolf}}]{Gonderinger:2009jp}%
  \BibitemOpen
  \bibfield  {author} {\bibinfo {author} {\bibfnamefont {M.}~\bibnamefont
  {Gonderinger}}, \bibinfo {author} {\bibfnamefont {Y.}~\bibnamefont {Li}},
  \bibinfo {author} {\bibfnamefont {H.}~\bibnamefont {Patel}},\ and\ \bibinfo
  {author} {\bibfnamefont {M.~J.}\ \bibnamefont {Ramsey-Musolf}},\ }\bibfield
  {title} {\bibinfo {title} {{Vacuum Stability, Perturbativity, and Scalar
  Singlet Dark Matter}},\ }\href {https://doi.org/10.1007/JHEP01(2010)053}
  {\bibfield  {journal} {\bibinfo  {journal} {JHEP}\ }\textbf {\bibinfo
  {volume} {01}},\ \bibinfo {pages} {053}},\ \Eprint
  {https://arxiv.org/abs/0910.3167} {arXiv:0910.3167 [hep-ph]} \BibitemShut
  {NoStop}%
\bibitem [{\citenamefont {Cline}\ \emph {et~al.}(2013)\citenamefont {Cline},
  \citenamefont {Kainulainen}, \citenamefont {Scott},\ and\ \citenamefont
  {Weniger}}]{Cline:2013gha}%
  \BibitemOpen
  \bibfield  {author} {\bibinfo {author} {\bibfnamefont {J.~M.}\ \bibnamefont
  {Cline}}, \bibinfo {author} {\bibfnamefont {K.}~\bibnamefont {Kainulainen}},
  \bibinfo {author} {\bibfnamefont {P.}~\bibnamefont {Scott}},\ and\ \bibinfo
  {author} {\bibfnamefont {C.}~\bibnamefont {Weniger}},\ }\bibfield  {title}
  {\bibinfo {title} {{Update on scalar singlet dark matter}},\ }\href
  {https://doi.org/10.1103/PhysRevD.88.055025} {\bibfield  {journal} {\bibinfo
  {journal} {Phys. Rev. D}\ }\textbf {\bibinfo {volume} {88}},\ \bibinfo
  {pages} {055025} (\bibinfo {year} {2013})},\ \bibinfo {note} {[Erratum:
  Phys.Rev.D 92, 039906 (2015)]},\ \Eprint {https://arxiv.org/abs/1306.4710}
  {arXiv:1306.4710 [hep-ph]} \BibitemShut {NoStop}%
\bibitem [{\citenamefont {He}\ \emph {et~al.}(2010)\citenamefont {He},
  \citenamefont {Li}, \citenamefont {Li}, \citenamefont {Tandean},\ and\
  \citenamefont {Tsai}}]{He:2009yd}%
  \BibitemOpen
  \bibfield  {author} {\bibinfo {author} {\bibfnamefont {X.-G.}\ \bibnamefont
  {He}}, \bibinfo {author} {\bibfnamefont {T.}~\bibnamefont {Li}}, \bibinfo
  {author} {\bibfnamefont {X.-Q.}\ \bibnamefont {Li}}, \bibinfo {author}
  {\bibfnamefont {J.}~\bibnamefont {Tandean}},\ and\ \bibinfo {author}
  {\bibfnamefont {H.-C.}\ \bibnamefont {Tsai}},\ }\bibfield  {title} {\bibinfo
  {title} {{The Simplest Dark-Matter Model, CDMS II Results, and Higgs
  Detection at LHC}},\ }\href {https://doi.org/10.1016/j.physletb.2010.04.026}
  {\bibfield  {journal} {\bibinfo  {journal} {Phys. Lett. B}\ }\textbf
  {\bibinfo {volume} {688}},\ \bibinfo {pages} {332} (\bibinfo {year}
  {2010})},\ \Eprint {https://arxiv.org/abs/0912.4722} {arXiv:0912.4722
  [hep-ph]} \BibitemShut {NoStop}%
\bibitem [{\citenamefont {Aad}\ \emph {et~al.}(2023{\natexlab{f}})\citenamefont
  {Aad} \emph {et~al.}}]{ATLAS:2022jtk}%
  \BibitemOpen
  \bibfield  {author} {\bibinfo {author} {\bibfnamefont {G.}~\bibnamefont
  {Aad}} \emph {et~al.} (\bibinfo {collaboration} {ATLAS}),\ }\bibfield
  {title} {\bibinfo {title} {{Constraints on the Higgs boson self-coupling from
  single- and double-Higgs production with the ATLAS detector using pp
  collisions at s=13 TeV}},\ }\href
  {https://doi.org/10.1016/j.physletb.2023.137745} {\bibfield  {journal}
  {\bibinfo  {journal} {Phys. Lett. B}\ }\textbf {\bibinfo {volume} {843}},\
  \bibinfo {pages} {137745} (\bibinfo {year} {2023}{\natexlab{f}})},\ \Eprint
  {https://arxiv.org/abs/2211.01216} {arXiv:2211.01216 [hep-ex]} \BibitemShut
  {NoStop}%
\bibitem [{\citenamefont {Tumasyan}\ \emph
  {et~al.}(2022{\natexlab{c}})\citenamefont {Tumasyan} \emph
  {et~al.}}]{CMS:2022dwd}%
  \BibitemOpen
  \bibfield  {author} {\bibinfo {author} {\bibfnamefont {A.}~\bibnamefont
  {Tumasyan}} \emph {et~al.} (\bibinfo {collaboration} {CMS}),\ }\bibfield
  {title} {\bibinfo {title} {{A portrait of the Higgs boson by the CMS
  experiment ten years after the discovery.}},\ }\href
  {https://doi.org/10.1038/s41586-022-04892-x} {\bibfield  {journal} {\bibinfo
  {journal} {Nature}\ }\textbf {\bibinfo {volume} {607}},\ \bibinfo {pages}
  {60} (\bibinfo {year} {2022}{\natexlab{c}})},\ \Eprint
  {https://arxiv.org/abs/2207.00043} {arXiv:2207.00043 [hep-ex]} \BibitemShut
  {NoStop}%
\bibitem [{\citenamefont {Hieu}\ \emph {et~al.}(2020)\citenamefont {Hieu},
  \citenamefont {Sang},\ and\ \citenamefont {Trang}}]{Hieu:2020hti}%
  \BibitemOpen
  \bibfield  {author} {\bibinfo {author} {\bibfnamefont {T.~M.}\ \bibnamefont
  {Hieu}}, \bibinfo {author} {\bibfnamefont {Q.~S.}\ \bibnamefont {Sang}},\
  and\ \bibinfo {author} {\bibfnamefont {T.~Q.}\ \bibnamefont {Trang}},\
  }\bibfield  {title} {\bibinfo {title} {{On a standard model extension with
  vector-like fermions and Abelian symmetry}},\ }\href
  {https://doi.org/10.15625/0868-3166/30/3/15071} {\bibfield  {journal}
  {\bibinfo  {journal} {Commun. in Phys.}\ }\textbf {\bibinfo {volume} {30}},\
  \bibinfo {pages} {231} (\bibinfo {year} {2020})}\BibitemShut {NoStop}%
\bibitem [{\citenamefont {Buttazzo}\ \emph {et~al.}(2013)\citenamefont
  {Buttazzo}, \citenamefont {Degrassi}, \citenamefont {Giardino}, \citenamefont
  {Giudice}, \citenamefont {Sala}, \citenamefont {Salvio},\ and\ \citenamefont
  {Strumia}}]{Buttazzo:2013uya}%
  \BibitemOpen
  \bibfield  {author} {\bibinfo {author} {\bibfnamefont {D.}~\bibnamefont
  {Buttazzo}}, \bibinfo {author} {\bibfnamefont {G.}~\bibnamefont {Degrassi}},
  \bibinfo {author} {\bibfnamefont {P.~P.}\ \bibnamefont {Giardino}}, \bibinfo
  {author} {\bibfnamefont {G.~F.}\ \bibnamefont {Giudice}}, \bibinfo {author}
  {\bibfnamefont {F.}~\bibnamefont {Sala}}, \bibinfo {author} {\bibfnamefont
  {A.}~\bibnamefont {Salvio}},\ and\ \bibinfo {author} {\bibfnamefont
  {A.}~\bibnamefont {Strumia}},\ }\bibfield  {title} {\bibinfo {title}
  {{Investigating the near-criticality of the Higgs boson}},\ }\href
  {https://doi.org/10.1007/JHEP12(2013)089} {\bibfield  {journal} {\bibinfo
  {journal} {JHEP}\ }\textbf {\bibinfo {volume} {12}},\ \bibinfo {pages}
  {089}},\ \Eprint {https://arxiv.org/abs/1307.3536} {arXiv:1307.3536 [hep-ph]}
  \BibitemShut {NoStop}%
\bibitem [{\citenamefont {Workman}\ \emph {et~al.}(2022)\citenamefont {Workman}
  \emph {et~al.}}]{ParticleDataGroup:2022pth}%
  \BibitemOpen
  \bibfield  {author} {\bibinfo {author} {\bibfnamefont {R.~L.}\ \bibnamefont
  {Workman}} \emph {et~al.} (\bibinfo {collaboration} {Particle Data Group}),\
  }\bibfield  {title} {\bibinfo {title} {{Review of Particle Physics}},\ }\href
  {https://doi.org/10.1093/ptep/ptac097} {\bibfield  {journal} {\bibinfo
  {journal} {PTEP}\ }\textbf {\bibinfo {volume} {2022}},\ \bibinfo {pages}
  {083C01} (\bibinfo {year} {2022})}\BibitemShut {NoStop}%
\bibitem [{\citenamefont {Staub}(2014)}]{Staub:2013tta}%
  \BibitemOpen
  \bibfield  {author} {\bibinfo {author} {\bibfnamefont {F.}~\bibnamefont
  {Staub}},\ }\bibfield  {title} {\bibinfo {title} {{SARAH 4 : A tool for (not
  only SUSY) model builders}},\ }\href
  {https://doi.org/10.1016/j.cpc.2014.02.018} {\bibfield  {journal} {\bibinfo
  {journal} {Comput. Phys. Commun.}\ }\textbf {\bibinfo {volume} {185}},\
  \bibinfo {pages} {1773} (\bibinfo {year} {2014})},\ \Eprint
  {https://arxiv.org/abs/1309.7223} {arXiv:1309.7223 [hep-ph]} \BibitemShut
  {NoStop}%
\bibitem [{\citenamefont {Thomsen}(2021)}]{Thomsen:2021ncy}%
  \BibitemOpen
  \bibfield  {author} {\bibinfo {author} {\bibfnamefont {A.~E.}\ \bibnamefont
  {Thomsen}},\ }\bibfield  {title} {\bibinfo {title} {{Introducing RGBeta: a
  Mathematica package for the evaluation of renormalization group
  $\beta$-functions}},\ }\href
  {https://doi.org/10.1140/epjc/s10052-021-09142-4} {\bibfield  {journal}
  {\bibinfo  {journal} {Eur. Phys. J. C}\ }\textbf {\bibinfo {volume} {81}},\
  \bibinfo {pages} {408} (\bibinfo {year} {2021})},\ \Eprint
  {https://arxiv.org/abs/2101.08265} {arXiv:2101.08265 [hep-ph]} \BibitemShut
  {NoStop}%
\bibitem [{\citenamefont {Alloul}\ \emph {et~al.}(2014)\citenamefont {Alloul},
  \citenamefont {Christensen}, \citenamefont {Degrande}, \citenamefont {Duhr},\
  and\ \citenamefont {Fuks}}]{Alloul:2013bka}%
  \BibitemOpen
  \bibfield  {author} {\bibinfo {author} {\bibfnamefont {A.}~\bibnamefont
  {Alloul}}, \bibinfo {author} {\bibfnamefont {N.~D.}\ \bibnamefont
  {Christensen}}, \bibinfo {author} {\bibfnamefont {C.}~\bibnamefont
  {Degrande}}, \bibinfo {author} {\bibfnamefont {C.}~\bibnamefont {Duhr}},\
  and\ \bibinfo {author} {\bibfnamefont {B.}~\bibnamefont {Fuks}},\ }\bibfield
  {title} {\bibinfo {title} {{FeynRules 2.0 - A complete toolbox for tree-level
  phenomenology}},\ }\href {https://doi.org/10.1016/j.cpc.2014.04.012}
  {\bibfield  {journal} {\bibinfo  {journal} {Comput. Phys. Commun.}\ }\textbf
  {\bibinfo {volume} {185}},\ \bibinfo {pages} {2250} (\bibinfo {year}
  {2014})},\ \Eprint {https://arxiv.org/abs/1310.1921} {arXiv:1310.1921
  [hep-ph]} \BibitemShut {NoStop}%
\bibitem [{\citenamefont {Alwall}\ \emph {et~al.}(2011)\citenamefont {Alwall},
  \citenamefont {Herquet}, \citenamefont {Maltoni}, \citenamefont {Mattelaer},\
  and\ \citenamefont {Stelzer}}]{Alwall:2011uj}%
  \BibitemOpen
  \bibfield  {author} {\bibinfo {author} {\bibfnamefont {J.}~\bibnamefont
  {Alwall}}, \bibinfo {author} {\bibfnamefont {M.}~\bibnamefont {Herquet}},
  \bibinfo {author} {\bibfnamefont {F.}~\bibnamefont {Maltoni}}, \bibinfo
  {author} {\bibfnamefont {O.}~\bibnamefont {Mattelaer}},\ and\ \bibinfo
  {author} {\bibfnamefont {T.}~\bibnamefont {Stelzer}},\ }\bibfield  {title}
  {\bibinfo {title} {{MadGraph 5 : Going Beyond}},\ }\href
  {https://doi.org/10.1007/JHEP06(2011)128} {\bibfield  {journal} {\bibinfo
  {journal} {JHEP}\ }\textbf {\bibinfo {volume} {06}},\ \bibinfo {pages}
  {128}},\ \Eprint {https://arxiv.org/abs/1106.0522} {arXiv:1106.0522 [hep-ph]}
  \BibitemShut {NoStop}%
\bibitem [{\citenamefont {Peskin}\ and\ \citenamefont
  {Takeuchi}(1990)}]{Peskin:1990zt}%
  \BibitemOpen
  \bibfield  {author} {\bibinfo {author} {\bibfnamefont {M.~E.}\ \bibnamefont
  {Peskin}}\ and\ \bibinfo {author} {\bibfnamefont {T.}~\bibnamefont
  {Takeuchi}},\ }\bibfield  {title} {\bibinfo {title} {{A New constraint on a
  strongly interacting Higgs sector}},\ }\href
  {https://doi.org/10.1103/PhysRevLett.65.964} {\bibfield  {journal} {\bibinfo
  {journal} {Phys. Rev. Lett.}\ }\textbf {\bibinfo {volume} {65}},\ \bibinfo
  {pages} {964} (\bibinfo {year} {1990})}\BibitemShut {NoStop}%
\bibitem [{\citenamefont {Peskin}\ and\ \citenamefont
  {Takeuchi}(1992)}]{Peskin:1991sw}%
  \BibitemOpen
  \bibfield  {author} {\bibinfo {author} {\bibfnamefont {M.~E.}\ \bibnamefont
  {Peskin}}\ and\ \bibinfo {author} {\bibfnamefont {T.}~\bibnamefont
  {Takeuchi}},\ }\bibfield  {title} {\bibinfo {title} {{Estimation of oblique
  electroweak corrections}},\ }\href {https://doi.org/10.1103/PhysRevD.46.381}
  {\bibfield  {journal} {\bibinfo  {journal} {Phys. Rev. D}\ }\textbf {\bibinfo
  {volume} {46}},\ \bibinfo {pages} {381} (\bibinfo {year} {1992})}\BibitemShut
  {NoStop}%
\bibitem [{\citenamefont {Grinstein}\ and\ \citenamefont
  {Wise}(1991)}]{Grinstein:1991cd}%
  \BibitemOpen
  \bibfield  {author} {\bibinfo {author} {\bibfnamefont {B.}~\bibnamefont
  {Grinstein}}\ and\ \bibinfo {author} {\bibfnamefont {M.~B.}\ \bibnamefont
  {Wise}},\ }\bibfield  {title} {\bibinfo {title} {{Operator analysis for
  precision electroweak physics}},\ }\href
  {https://doi.org/10.1016/0370-2693(91)90061-T} {\bibfield  {journal}
  {\bibinfo  {journal} {Phys. Lett. B}\ }\textbf {\bibinfo {volume} {265}},\
  \bibinfo {pages} {326} (\bibinfo {year} {1991})}\BibitemShut {NoStop}%
\bibitem [{\citenamefont {Quiros}(1999)}]{Quiros:1999jp}%
  \BibitemOpen
  \bibfield  {author} {\bibinfo {author} {\bibfnamefont {M.}~\bibnamefont
  {Quiros}},\ }\bibfield  {title} {\bibinfo {title} {Finite temperature field
  theory and phase transitions}\ }(\bibinfo  {publisher} {ICTP Summer School in
  High-Energy Physics and Cosmology},\ \bibinfo {year} {1999})\ pp.\ \bibinfo
  {pages} {187--259},\ \Eprint {https://arxiv.org/abs/hep-ph/9901312}
  {arXiv:hep-ph/9901312} \BibitemShut {NoStop}%
\bibitem [{\citenamefont {Weinberg}(1973)}]{Weinberg:1973am}%
  \BibitemOpen
  \bibfield  {author} {\bibinfo {author} {\bibfnamefont {E.~J.}\ \bibnamefont
  {Weinberg}},\ }\href@noop {} {\bibinfo {title} {Radiative corrections as the
  origin of spontaneous symmetry breaking}} (\bibinfo {year} {1973}),\ \Eprint
  {https://arxiv.org/abs/hep-th/0507214} {arXiv:hep-th/0507214} \BibitemShut
  {NoStop}%
\bibitem [{\citenamefont {Weinberg}(1974)}]{Weinberg:1974hy}%
  \BibitemOpen
  \bibfield  {author} {\bibinfo {author} {\bibfnamefont {S.}~\bibnamefont
  {Weinberg}},\ }\bibfield  {title} {\bibinfo {title} {{Gauge and Global
  Symmetries at High Temperature}},\ }\href
  {https://doi.org/10.1103/PhysRevD.9.3357} {\bibfield  {journal} {\bibinfo
  {journal} {Phys. Rev. D}\ }\textbf {\bibinfo {volume} {9}},\ \bibinfo {pages}
  {3357} (\bibinfo {year} {1974})}\BibitemShut {NoStop}%
\bibitem [{\citenamefont {Curtin}\ \emph {et~al.}(2018)\citenamefont {Curtin},
  \citenamefont {Meade},\ and\ \citenamefont {Ramani}}]{Curtin:2016urg}%
  \BibitemOpen
  \bibfield  {author} {\bibinfo {author} {\bibfnamefont {D.}~\bibnamefont
  {Curtin}}, \bibinfo {author} {\bibfnamefont {P.}~\bibnamefont {Meade}},\ and\
  \bibinfo {author} {\bibfnamefont {H.}~\bibnamefont {Ramani}},\ }\bibfield
  {title} {\bibinfo {title} {{Thermal Resummation and Phase Transitions}},\
  }\href {https://doi.org/10.1140/epjc/s10052-018-6268-0} {\bibfield  {journal}
  {\bibinfo  {journal} {Eur. Phys. J. C}\ }\textbf {\bibinfo {volume} {78}},\
  \bibinfo {pages} {787} (\bibinfo {year} {2018})},\ \Eprint
  {https://arxiv.org/abs/1612.00466} {arXiv:1612.00466 [hep-ph]} \BibitemShut
  {NoStop}%
\end{thebibliography}%

\end{document}